\documentclass[iop,numberedappendix,twocolappendix]{emulateapj}

\bibpunct[; ]{(}{)}{;}{a}{,}{,}

\newcommand{\secpoint}{\mbox{$''\mskip-7.6mu.\,$}}
\newcommand{\secnopoint}{\mbox{$''\mskip-7.6mu\,$}}
\newcommand{\lya}{Ly$\alpha$}
\newcommand{\minpoint}{\mbox{$'\mskip-4.7mu.\mskip0.8mu$}}

\shorttitle{A High-Resolution $HST$ Study of Apparent LyC Leakers at $z\sim3$}
\shortauthors{Mostardi, R.~E., et al.}

\begin{document}

\title{A High-Resolution Hubble Space Telescope Study of Apparent Lyman Continuum Leakers at $z\sim3$}

\author{R.~E. Mostardi\altaffilmark{1}} \altaffiltext{1}{Department of Physics \& Astronomy, University of California, Los Angeles, 430 Portola Plaza, Los Angeles, CA 90095}

\author{A.~E. Shapley\altaffilmark{1}}

\author{C.~C. Steidel\altaffilmark{2}} \altaffiltext{2}{Cahill Center for Astrophysics, California Institute of
Technology, MS 249-17, Pasadena, CA 91125}

\author{R.~F. Trainor\altaffilmark{3}} \altaffiltext{3}{Miller Fellow; Astronomy Department, 501 Campbell Hall, Berkeley, CA 94720-3411}

\author{N.~A. Reddy\altaffilmark{45}} \altaffiltext{4}{Department of Physics and Astronomy, University of California, Riverside, 900 University Avenue, Riverside, CA 92521} \altaffiltext{5}{Alfred P. Sloan Research Fellow}

\author{B. Siana\altaffilmark{4}}

\begin{abstract}
We present $U_{336}V_{606}J_{125}H_{160}$ follow-up $HST$ observations of 16 $z\sim3$ candidate LyC emitters in the HS1549+1919 field.  With these data, we obtain high spatial-resolution photometric redshifts of all sub-arcsecond components of the LyC candidates in order to eliminate foreground contamination and identify robust candidates for leaking LyC emission.  Of the 16 candidates, we find one object with a robust LyC detection that is not due to foreground contamination.  This object (MD5) resolves into two components; we refer to the LyC-emitting component as MD5b.   MD5b has an observed 1500\AA\ to 900\AA\ flux-density ratio of $(F_{UV}/F_{LyC})_{obs}=4.0\pm2.0$, compatible with predictions from stellar population synthesis models.  Assuming minimal IGM absorption, this ratio corresponds to a relative (absolute) escape fraction of $f_{esc,rel}^{MD5b}=75-100$\% ($f_{esc,abs}^{MD5b}=14-19$\%).  The stellar population fit to MD5b indicates an age of $\lesssim50$Myr, which is in the youngest 10\% of the $HST$ sample and the youngest third of typical $z\sim3$ Lyman break galaxies, and may be a contributing factor to its LyC detection.  We obtain a revised, contamination-free estimate for the comoving specific ionizing emissivity at $z=2.85$, indicating (with large uncertainties) that star-forming galaxies provide roughly the same contribution as QSOs to the ionizing background at this redshift.  Our results show that foreground contamination prevents ground-based LyC studies from obtaining a full understanding of LyC emission from $z\sim3$ star-forming galaxies.  Future progress in direct LyC searches is contingent upon the elimination of foreground contaminants through high spatial-resolution observations, and upon acquisition of sufficiently deep LyC imaging to probe ionizing radiation in high-redshift galaxies.
\end{abstract}


\section{Introduction}    \label{sec:Intro}

The sources responsible for cosmic reionization are still not well understood.  Evidence that quasars (QSOs) cannot be solely responsible for reionization \citep{fontanot12,glikman11,siana08} has prompted many searches for ionizing Lyman continuum (LyC) emission from star-forming galaxies.  While the IGM at $z\gtrsim6$ is opaque to LyC photons and prevents direct observations of LyC-emitting galaxies during the epoch of reionization, many studies have attempted to detect lower-redshift analogs to galaxies responsible for reionization.  Although IGM transmission is highest in the local universe, studies at $z<2$ \citep[e.g.,][]{grimes07,grimes09,cowie09,bridge10,siana07,siana10} have yielded very few detections of LyC emission, with only three objects identified to date \citep{leitet11,leitet13,borthakur14}.  At redshift $z \sim 3-4$, the search for LyC-emitting galaxies has appeared to be more fruitful.  However, even though the examination of hundreds of galaxies \citep[in works such as][]{steidel01,shapley06,iwata09,nestor13,mostardi13,siana15,vanzella10dec,vanzella12,vanzella15} has yielded many promising LyC-emitting candidates, there exist only two robust detections \citep[][]{vanzella12,vanzella15}.  

Amassing large samples of LyC detections in high-redshift star-forming galaxies has been difficult for several reasons.  First, large parent samples of high-redshift galaxies must be identified and confirmed spectroscopically, requiring extensive galaxy surveys (often ground-based) and time-consuming spectroscopic follow-up.  Second, it is necessary to probe flux in the LyC spectral region for these galaxies, either with deep spectroscopy, also very time-intensive, or through narrowband imaging in a filter just blueward of the Lyman limit, in which it is difficult to match a single narrowband filter to the LyC region for many galaxies at once.  Even after potential high-redshift LyC-emitting candidates are identified, there remains the possibility that apparent LyC emission is actually due to a lower-redshift interloper along the line of sight, which cannot be distinguished in ground-based, seeing-limited data.

One method that has proven successful at identifying potential LyC-emitting galaxies is narrowband LyC imaging of galaxy protoclusters.  Large ground-based surveys of UV-selected star-forming galaxies at $z \sim 2 - 3$ \citep{steidel03,steidel04,steidel11,reddy08} have identified and spectroscopically confirmed thousands of high-redshift star-forming galaxies.  These surveys have also located galaxy protoclusters, areas on the sky with large overdensities of galaxies at similar redshift.  A very effective way to simultaneously probe the LyC of large samples of galaxies at the same redshift is to perform deep imaging through a narrowband filter tuned to the LyC spectral region at the protocluster redshift \citep[e.g.,][]{iwata09,nestor11,nestor13,mostardi13}.

Initially, these protocluster studies were entirely based upon ground-based data with seeing FWHMs of $0\secpoint7 - 1\secpoint0$, and thus suffered from the possibility of foreground contamination.  \citet{vanzella10} demonstrated that statistical simulations modeling the distribution of foreground galaxies result in high rates of foreground contamination for high-redshift objects in ground-based studies.  While simulations can account for contamination statistically in LyC-emitting samples \citep[as in][]{nestor11,mostardi13}, contaminants cannot be eradicated on an individual basis.  As two of the main goals of LyC studies are to determine the mechanism of LyC photon escape from the interstellar medium (ISM), and to identify additional features of LyC-emitting galaxies that may enable their identification through other means, it is crucial to identify robust, individual candidates for LyC emission where foreground contamination has been ruled out.

Eliminating contaminants is a complex process.  High-resolution imaging shows that the majority of high-redshift galaxies are not morphologically simple, but are composed of multiple compact clumps and/or diffuse emission \citep[e.g.,][]{law07}.  Contamination can only be firmly ruled out if the redshifts of individual galaxy clumps are measured, and if the clump associated with LyC emission is confirmed to be at the redshift of the target galaxy.  In order to address the issue of contamination in the narrowband LyC survey of the $z=3.1$ SSA22a protocluster \citep{nestor11,nestor13}, \citet{siana15} obtained near-IR spectroscopy with Keck/NIRSPEC to measure the spectroscopic redshifts of the sub-arcsecond components of 5 LyC candidates.  These authors found two foreground contaminants, one galaxy with a misidentified redshift, and two galaxies that could not be definitively confirmed as LyC-emitters.  For galaxies at slightly higher redshifts ($z\sim3.7$), \citet{vanzella12} used photometric redshifts obtained through the high-resolution imaging in the CANDELS survey \citep{grogin11,koekemoer11} to analyze the sub-arcsecond clumps of 19 candidate LyC-emitters, finding 18 contaminants and one bona-fide LyC emitter.  These two studies have shown that both methods $-$ high-resolution spectroscopy and photometric redshifts $-$ are effective ways to locate foreground contaminants.

In \citet{mostardi13} (hereafter M13), we performed a narrowband LyC imaging survey of a galaxy protocluster in the HS1549+1919 field \citep{steidel11}.  In this work, we present follow-up observations with $HST$ for the sample of candidate LyC emitters in M13 with the goal of using photometric redshifts to eliminate contaminants from the LyC emitter sample.  The HS1549 protocluster has a redshift-space overdensity of $\delta_{gal} \sim 5$ at $z = 2.85 \pm 0.03$, and this ``spike'' redshift coincides with that of a hyperluminous QSO \citep{trainor12}.  More than 350 UV-selected galaxies have been identified in the HS1549 field, $\sim160$ of which have been spectroscopically confirmed at $1.5 \leq z \leq 3.5$. Additionally, narrowband imaging with a 4670\AA\ filter tuned to the wavelength of \lya\ at the redshift spike has revealed $\sim$300 potential \lya\ Emitters (LAEs) and several \lya\ ``blobs" \citep{steidel00,steidel11}.  In M13, we used a narrowband filter (NB3420) tuned to wavelengths just below the Lyman limit at $z\geq2.82$, thus observing the LyC spectral region for hundreds of Lyman break galaxies (LBGs) and LAEs at $z\geq2.82$, including 49 LBGs and 91 LAEs with spectroscopic confirmation.  We identified putative LyC emitters in the NB3420 imaging, and also performed a stacking analysis of objects undetected in the NB3420 filter (measuring no signal).  Although we found an NB3420 detection rate of $\sim9$\% in both the LBG and LAE samples, simulations indicated that 40-75\% of the individual NB3420 detections may have resulted from foreground contamination, highlighting the need for further work to disentangle true LyC emitters from low-redshift contaminants.

Our aims in the current work are two-fold.  Our primary goal is to address the question of foreground contamination in the M13 LyC-emitter sample.  Ideally, as in \citet{siana15}, we would obtain spatially-resolved spectroscopy in the vicinity of each putative LyC detection, with resolution of $\leq0$\secpoint5, in order to measure the redshifts of all components.  However, ground-based optical spectroscopy probing the rest-frame UV provides insufficient spatial resolution, and ground-based near-IR spectroscopy of rest-frame optical nebular emission lines (with or without the assistance of adaptive optics) is not feasible because at $z\sim2.85$ the strongest nebular emission lines (H$\alpha$, [OIII]$\lambda$5007, H$\beta$, and [OII]$\lambda$3727) are lost either in the thermal background or gaps in atmospheric transmission.  Therefore, we have identified high-resolution, multi-band $HST$ imaging as the best method for estimating spatially resolved photometric redshifts for the individual galaxy subcomponents associated with apparent LyC emission.  Our second goal is to analyze the properties of galaxies we have verified to be true sources of LyC emission (such as their morphologies, stellar populations, and the ratio of their ionizing to non-ionizing flux densities) with respect to properties of star-forming galaxies without LyC detections.  Such an analysis will help determine whether galaxies with high escape fractions of ionizing photons have different intrinsic properties from those of galaxies without detectable leaking ionizing radiation, and may provide insight into star formation and the structure of the ISM in high-redshift galaxies.

In addition to presenting high-resolution, multiwavelength follow-up $HST$ observations of high-redshift LyC-emitting candidates at $z\sim2.85$ from M13, we discuss the implications for continuing searches for ionizing radiation in star-forming galaxies.  The paper is organized as follows.  Section \ref{sec:methods} describes our methodology, while the galaxy sample and $HST$ observations are presented in Section \ref{sec:obs}.  In Section \ref{sec:DataPhot} we describe the techniques used to reduce the $HST$ imaging data and perform photometry on the sub-arcsecond components of each galaxy.  Section \ref{sec:Photoz} contains a discussion of our methods of fitting photometric redshifts to the data, as well as an analysis of each candidate LyC emitter and any sources of foreground contamination.  In Section \ref{sec:MD5} we analyze the properties of our best candidate for true LyC emission (MD5) with respect to the larger population of star-forming galaxies.  In Section \ref{sec:disc}, we consider the broader implications of these results, and the prospects for future direct searches for LyC radiation at high redshift.  We summarize our results in Section \ref{sec:summary}.  Throughout the paper we employ the AB magnitude system and assume a cosmology with $\Omega_m=0.3$, $\Omega_{\Lambda}=0.7$, and $H_0=70$~km~s$^{-1}$~Mpc$^{-1}$. At $z=2.85$, 1\secnopoint\ corresponds to 7.8 proper kpc.

\section{Methodology}\label{sec:methods}

\begin{figure}
\epsscale{1.0}
\plotone{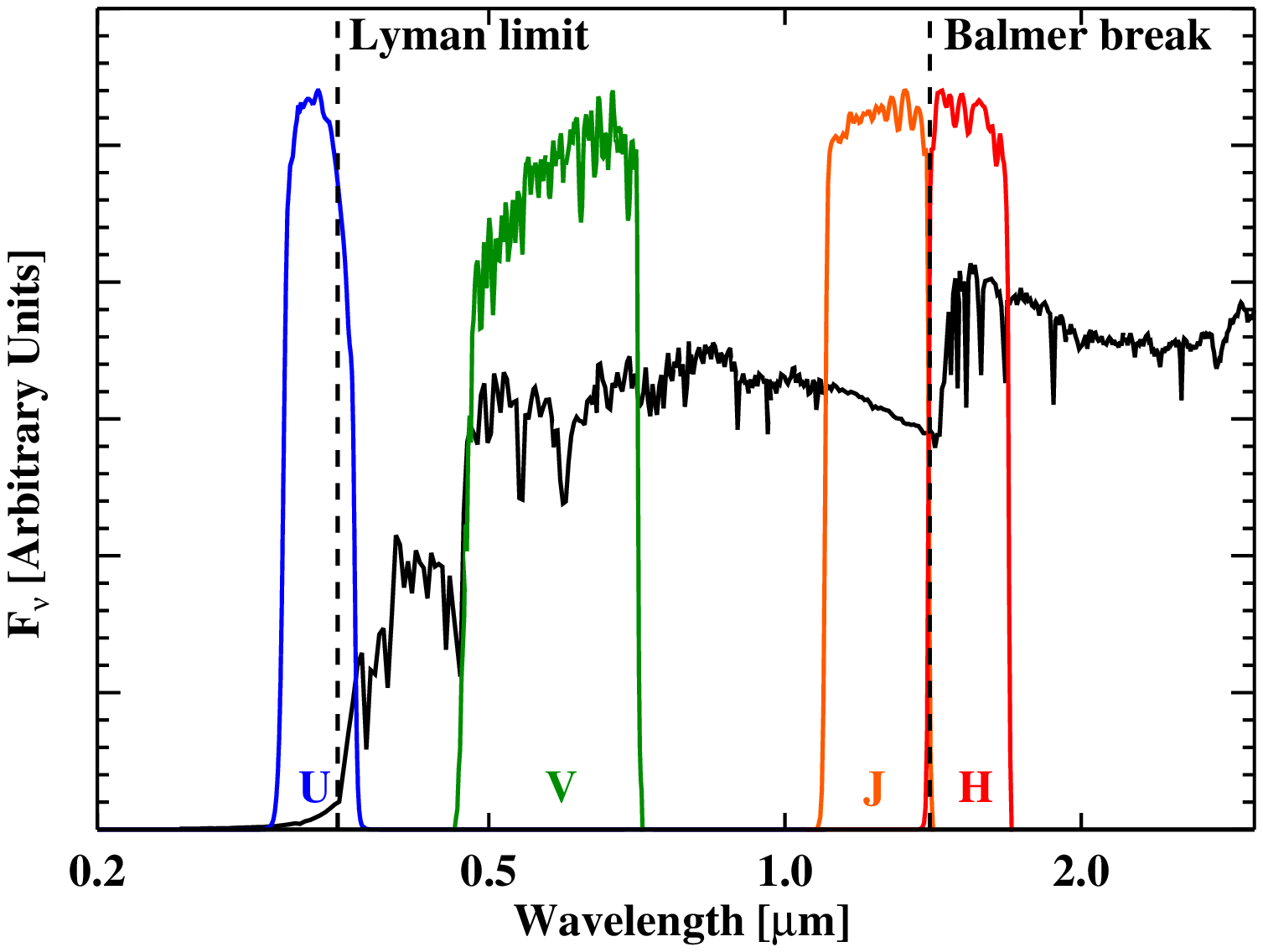}
\caption{
\small
$U_{336}$, $V_{606}$, $J_{125}$, and $H_{160}$ filters superimposed over a model $z=2.85$ galaxy spectrum.  The model spectrum is a solar-metallicity 100 Myr constant star-formation rate model from \citet{bc03}, reddened to $E(B-V)$ = 0.15 with the \citet{calzetti00} attenuation curve.  The $U_{336}$ filter probes the LyC spectral region, but with some contamination ($\sim20$\%) redwards of the Lyman limit.  The $V_{606}$ filter probes the rest-frame non-ionizing UV continuum, and the $J_{125}$ and $H_{160}$ filters probe flux on either side of the Balmer break.  With both the Lyman and Balmer breaks sensitively probed at $z = 2.85$, we can determine photometric redshifts for individual sub-arcsecond components of galaxies within our LAE and LBG samples.  These photometric redshifts enable us to determine whether the NB3420 detections associated with these galaxies are true LyC emission, or contamination from foreground galaxies.
\label{fig:methods} }
\end{figure} 

\begin{figure*}
\epsscale{1.0}
\plotone{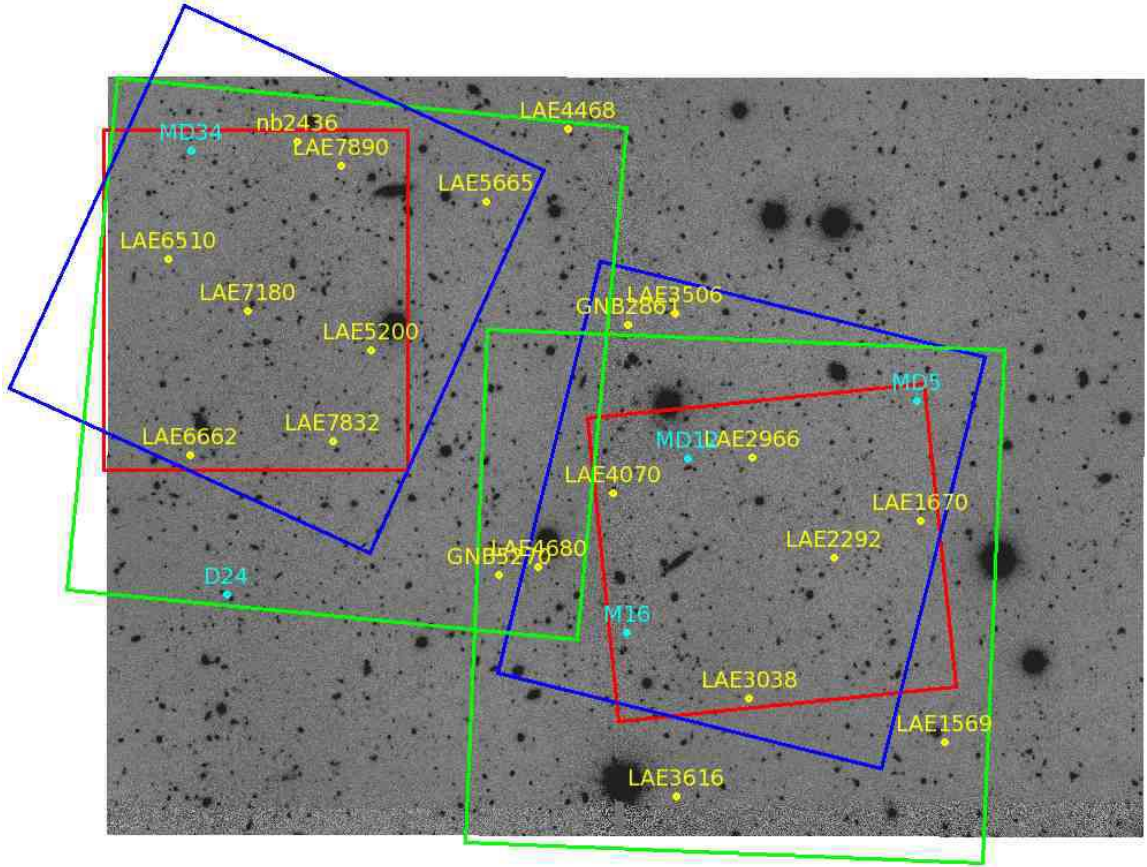}
\caption{
\small
Footprints of HST pointings superimposed on the $5'\times7'$ LRIS NB3420 (LyC) image.  Blue squares indicate the footprint of $U_{336}$, green squares indicate the footprint of $V_{606}$, and red squares indicate the footprints of $J_{125}$ and $H_{160}$.  Cyan and yellow circles indicate the positions of LBGs and LAEs with NB3420 detections that lie within the footprints of the $HST$ images.  The naming scheme for LBGs is presented in \citet{steidel03}.
\label{fig:pointings} }
\end{figure*}

\begin{deluxetable*}{lccc} 
\tablewidth{0pt} 
\footnotesize
\tablecaption{Description of Samples \label{tab:sample}}
\tablehead{
\colhead{Sample\tablenotemark{a}} &
\colhead{$N_{tot}$\tablenotemark{b}} &
\colhead{$N_{\mbox{NB3420}}$ [$N_{UVJH}$]\tablenotemark{c}} &
\colhead{$N_{\mbox{NB3420}}^{\mbox{no}}$ [$N_{UVJH}$]\tablenotemark{d}}
}
\startdata 
LBGs (with $z_{spec}$)		& 49 & 5 [4] & 44 [12] \\
LAEs (with $z_{spec}$)		& 82\tablenotemark{e} & 7 [6] & 75 [18] \\
LAEs (without $z_{spec}$)	& 33 & 10 [4] & 23 [7] \\
Faint LAEs (with $z_{spec}$)    					& --- & 8 [2] & --- \\
GNBs (with $z_{spec}$)   							& --- & 3 [0] & ---
\enddata
\tablenotetext{a}{The samples described in M13.}
\tablenotetext{b}{The total number of galaxies in the M13 sample.}
\tablenotetext{c}{The number of galaxies with NB3420 detections.  In brackets, we indicate the number of these galaxies for which we have obtained $U_{336}V_{606}J_{125}H_{160}$ imaging.}
\tablenotetext{d}{The number of galaxies without NB3420 detections.  In brackets, we indicate the number of these galaxies for which we have obtained $U_{336}V_{606}J_{125}H_{160}$ imaging.}
\tablenotetext{e}{The number of LAEs that are not part of the LBG sample, i.e., 91 LAEs minus the 9 overlap objects (which are listed here as part of the LBG sample).}
\end{deluxetable*}

\begin{deluxetable*}{ccccccc} 
\tablewidth{0pt} 
\footnotesize
\tablecaption{$HST$ Imaging Observations \label{tab:Obs}}
\tablehead{
\colhead{Filter} &
\colhead{$\lambda_{eff}$} &
\colhead{PSF FWHM} &
\colhead{Depth\tablenotemark{a}} &
\colhead{Exposure} &
\colhead{Pixel Scale\tablenotemark{b}} &
\colhead{\texttt{Pixfrac}\tablenotemark{b}}
\cr \colhead{} & \colhead{(\AA)} & \colhead{($''$)} & \colhead{(mag)} & \colhead{(s)} & \colhead{($''$/pixel)} & \colhead{}
}
\startdata 
$U_{336}$   & 3355 & 0.081 & 29.20 & 14176 & 0.025 & 0.7 \\
$V_{606}$   & 5921 & 0.092 & 30.22 & 11848 & 0.03 & 0.7 \\
$J_{125}$    & 12486 & 0.178 & 28.93 & 7835 & 0.075 & 0.7 \\
$H_{160}$   & 15369 & 0.186 & 28.63 & 7835 & 0.075 & 0.7
\enddata
\tablenotetext{a}{The 3$\sigma$ limiting depth obtained in a circular aperture with a diameter of 1.5 times the PSF FWHM.}
\tablenotetext{b}{The $AstroDrizzle$ parameters for pixel scale (\texttt{pixscale}) and \texttt{pixfrac} listed here represent the parameters used to attain optimal resolution.}
\end{deluxetable*}

To evaluate the amount of foreground contamination in M13 using photometric redshifts, we selected four $HST$ filters (WFC3/UVIS $U_{336}$, ACS/WFC $V_{606}$, and WFC3/IR $J_{125}$ and $H_{160}$) designed to probe the strengths of the Lyman and Balmer breaks at $z\sim2.85$.  Figure \ref{fig:methods} shows the locations of the $U_{336}$, $V_{606}$, $J_{125}$, and $H_{160}$ filters superimposed over a model \citet{bc03} galaxy spectrum redshifted to $z=2.85$.  The choice of $J_{125}$ and $H_{160}$ filters is particularly powerful for this test, due to the observed wavelength of the Balmer break at $z = 2.85$ ($\lambda_{obs,BB} = 1.4 \micron$). This wavelength corresponds exactly to the red cut-off of $J_{125}$ and the blue cut-off of $H_{160}$. Therefore, the $J_{125}-H_{160}$ color is very sensitive to the presence of the Balmer break, and provides information about the age of the stellar population.  Additionally, the $V_{606}-J_{125}$ color probes the rest-frame UV slope at $z\sim2.85$, providing information about the stellar populations and dust extinction.  At $z < 2$, where most of the contaminants found by \citet{vanzella12} are located, the $J_{125}$ filter falls entirely on the red side of the Balmer break, and therefore, extremely flat $J_{125}-H_{160}$ colors are expected, with the Balmer break falling instead between $V_{606}$ and $J_{125}$.  At the other end of the spectrum, the $U_{336}$ filter does not lie entirely bluewards of the Lyman limit at $z = 2.85$ (only 80\% of its wavelength range falls below the Lyman limit) and thus does not exclusively probe the LyC spectral region. However, given that the Lyman break passes through this filter at redshifts $z = 2.40 - 2.95$, the $U_{336}-V_{606}$ color should be sensitive to the magnitude of the Lyman break at $z = 2.85$, and to its absence at significantly lower redshift.  With both Lyman and Balmer breaks sensitively probed at $z = 2.85$ using $U_{336}V_{606}J_{125}H_{160}$ photometry, we can distinguish between true sources of LyC emission, and those NB3420 detections attributable to non-ionizing radiation at lower redshift.

Several complications may arise from the method of estimating the photometric redshifts using only these four filters.  First, galaxies with leaking LyC emission may have intrinsically high ratios of escaping ionizing to non-ionizing radiation, and may not exhibit as strong a Lyman break as expected from normal $z=2.85$ galaxies.  We keep this caveat in mind during our subsequent analysis, with the understanding that the $U_{336}-V_{606}$ color may not be accurately represented by the models.  Second, young galaxies ($\lesssim20$ Myr) may not have a significant Balmer break.  The combination of these two scenarios may result in a young, high-redshift, LyC emitter with a relatively featureless spectrum, making it difficult to distinguish such galaxies from low-redshift contaminants.  In our photometric redshift analysis, we highlight cases of galaxies with ambiguous SED shapes and the possible stellar populations these SEDs may indicate.

\begin{figure*}
\epsscale{1.15}
\plotone{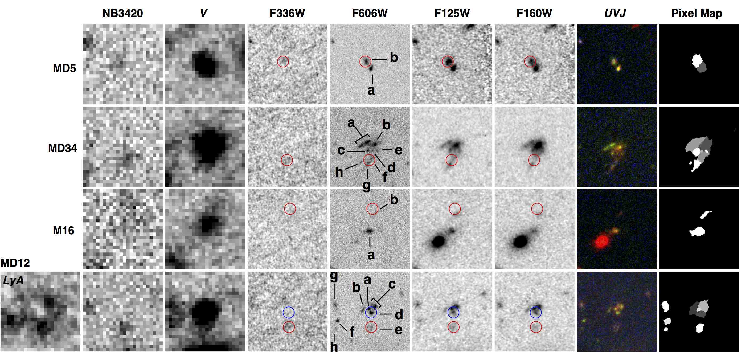}
\caption{
\small
$5''\times5''$ postage stamp images of spectroscopically-confirmed LBGs with NB3420 detections and imaging in all four $HST$ filters.  From left to right, objects are displayed in the LRIS NB4670$-V$ continuum-subtracted image (indicating \lya\ emission; MD12 only), LRIS NB3420 (LyC emission), LRIS $V$ (non-ionizing UV continuum), $HST$ $U_{336}$ (a combination of LyC and non-ionizing UV), $HST$ $V_{606}$ (non-ionizing UV continuum), $HST$ $J_{125}$ (optical, bluewards of the Balmer break), and $HST$ $H_{160}$ (optical, redwards of the Balmer break).  The penultimate column shows a color-composite image of $HST$ $U_{336}$ (blue), $V_{606}$ (green), and $J_{125}$ (red).  The final column shows SExtractor segmentation maps of the pixels used for the photometry of each labeled galaxy component; the arbitrary color scale indicates component edges when two components are adjacent.  Only pixel maps for the labeled objects are shown.  Red circles (1\secpoint0 diameter) indicate the centroid of the NB3420 emission, and blue circles indicate the centroid of the \lya\ emission for MD12.  Photometry was performed individually on sub-arcsecond components associated with each LBG, and all components are labeled in the $V_{606}$ image.  Postage stamps follow the conventional orientation, with north up and east to the left.
\label{fig:LBG_stamps} }
\end{figure*}

\begin{figure*}
\epsscale{1.15}
\plotone{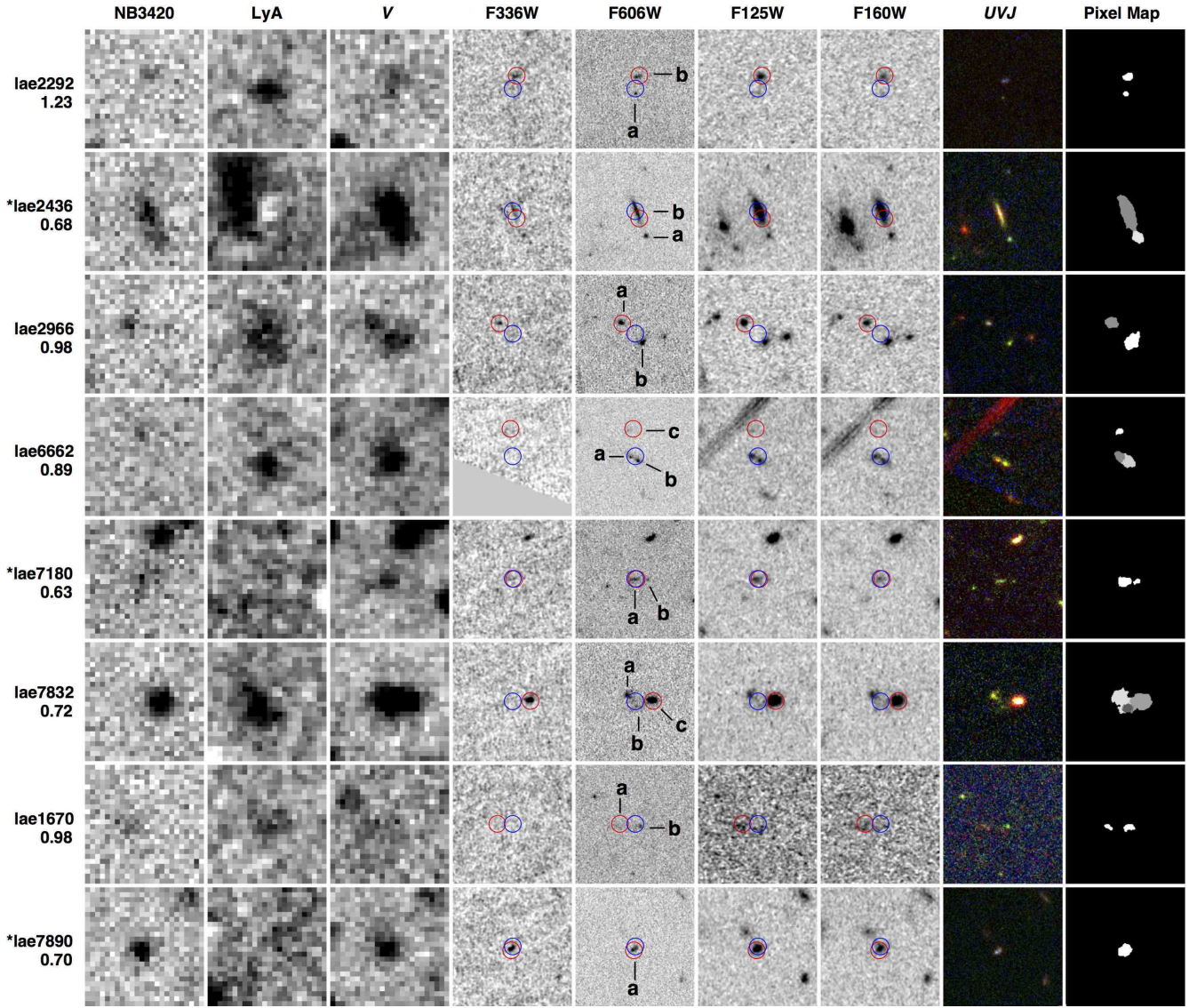}
\caption{
\small
$5''\times5''$ postage stamp images of spectroscopically-confirmed LAEs with NB3420 detections and imaging in all four $HST$ filters.  From left to right, objects are displayed in LRIS NB3420 (LyC emission), LRIS NB4670$-V$ (indicating \lya\ emission), LRIS $V$ (non-ionizing UV continuum), $HST$ $U_{336}$ (a combination of LyC and non-ionizing UV), $HST$ $V_{606}$ (non-ionizing UV continuum), $HST$ $J_{125}$ (optical, bluewards of the Balmer break), and $HST$ $H_{160}$ (optical, redwards of the Balmer break).  The penultimate column shows a color-composite image of $HST$ $U_{336}$ (blue), $V_{606}$ (green), and $J_{125}$ (red).  The final column shows SExtractor segmentation maps of the pixels used for the photometry of each labeled galaxy component; the arbitrary color scale indicates component edges when two components are adjacent.  Only pixel maps for the labeled objects are shown.  Red (blue) circles (1\secpoint0 diameter) indicate the centroid of the NB3420 emission (\lya\ emission).  Photometry was performed individually on sub-arcsecond components associated with each LAE, and all components are labeled in the $V_{606}$ image.  Postage stamps follow the conventional orientation, with north up and east to the left.  The $V-$NB4670 color of each LAE is indicated below the object name, and objects marked by an asterisk (*) were found to have misidentified spectroscopic redshifts (see Section \ref{sec:lyc}).
\label{fig:LAEz_stamps} }
\end{figure*}

\begin{figure*}
\epsscale{1.15}
\plotone{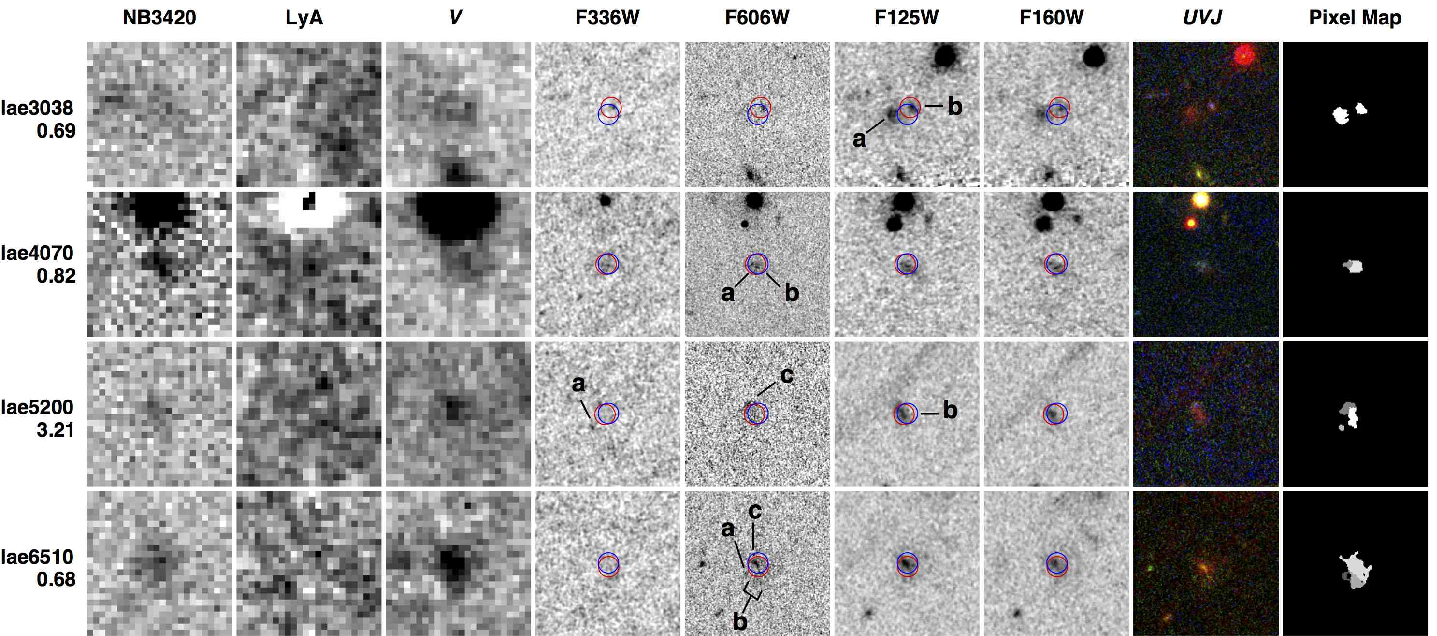}
\caption{
\small
$5''\times5''$ postage stamp images of LAE photometric candidates (no spectroscopic confirmation) with NB3420 detections and imaging in all four $HST$ filters.  Objects are displayed and labeled as in Figure \ref{fig:LAEz_stamps}.
\label{fig:LAEphot_stamps} }
\end{figure*}

\section{Sample and Observations}\label{sec:obs}

The HS1549 galaxy sample discussed in M13 consists of 49 LBGs and 91 LAEs, all spectroscopically confirmed at $z \geq 2.82$.  There are 9 objects part of both the LAE and LBG samples; for simplicity in displaying our data, we group these objects with the LBGs.\footnote{One of the objects in both the LAE and LBG samples is the putative LyC-emitter MD12/\emph{lae3540}, which is discussed in detail in M13.  In this work, we refer to this object simply as MD12.}  The redshift limit of $z \geq 2.82$ ensures that the NB3420 filter is sensitive only to LyC emission, with no contamination from flux redward of the Lyman limit.  Out of these galaxies, 5 LBGs and 7 LAEs have NB3420 detections within 1\secpoint9 of the non-ionizing UV ($\sim1500$\AA) centroid of the galaxy, indicating potential LyC emission if there is no lower-redshift galaxy contaminant along the line of sight.  

In M13 we also present an additional sample of 33 photometric LAE candidates (no spectroscopic confirmation) in the HS1549 field whose magnitudes in the narrowband filter used to select LAEs (NB4670) are in the same range as those of the spectroscopically-confirmed LAEs\footnote{All LAEs within the spectroscopic LAE sample of M13 had $m_{4670} \leq 26$.  Thus, the photometric LAE sample of M13 was defined to be any additional LAEs with $m_{4670} \leq 26$ that did not have spectroscopic confirmation.}.  While these objects are not included in the analysis of M13 because the lack of spectroscopic redshifts increases the possibility of contamination, 10 of them are reported to have NB3420 detections.  

Finally, in M13 we present 8 spectroscopically confirmed LAEs with NB3420 detections that are either fainter than the magnitude limit of the main LAE sample ($m_{4670} > 26$; 5 objects) or were selected by their $G-$NB4670 color rather than $V-$NB4670 (3 objects).  We did not include these objects in the analysis of M13 because we had not assembled a complete and unbiased spectroscopic sample of LAEs with $m_{4670} > 26$ or LAEs selected by their $G-$NB4670 color.  

In the follow-up observations presented in this work, we considered all objects with NB3420 detections presented in M13 and chose $HST$ pointings oriented to maximize the number of these potential LyC-emitting targets on the image footprints.  As the fields of view attained by the various $HST$ instruments employed for these observations (ACS/WFC: 3\minpoint5$\times$3\minpoint5;  WFC3/UVIS: 2\minpoint9$\times$2\minpoint7; WFC3/IR: 2\minpoint3$\times$2\minpoint1) are much smaller than the Keck/LRIS field of view (5$'\times$7$'$), we were unable to acquire imaging for all of the potential LyC-emitters.  However, with two separate $HST$ pointings in each of the 4 filters (see Figure \ref{fig:pointings}), we obtained $U_{336}V_{606}J_{125}H_{160}$ photometry for all but two of the spectroscopically confirmed galaxies in the main sample: 4/5 LBGs and 6/7 LAEs.  We also obtained $U_{336}V_{606}J_{125}H_{160}$ photometry for 4/10 photometric LAE candidates, and 2/6 of spectroscopically confirmed LAEs with $m_{4670} > 26$, totaling 16 galaxies with NB3420 detections covered in all four filters.  Eight additional objects with NB3420 detections are covered by at least one $HST$ filter (usually $V_{606}$, which has the largest field of view), and may be examined morphologically, although it is not possible to obtain photometric redshifts for galaxies without imaging in all four filters.  These objects are presented in Appendix B.  In addition to the 16 galaxies with NB3420 detections, we acquired $U_{336}V_{606}J_{125}H_{160}$ imaging for 30 spectroscopically confirmed galaxies at $z \geq 2.82$ without NB3420 detections (12 LBGs and 18 LAEs), which allows us to calibrate our photometric redshift fitting methods on galaxies without LyC detections and facilitates the differential analysis of the stellar populations of galaxies with and without LyC detections.  Finally, 50 additional spectroscopically confirmed galaxies at $z \geq 2.82$ without NB3420 detections (30 LBGs and 20 LAEs) were partially covered by our suite of $HST$ imaging.  Table \ref{tab:sample} summarizes the $HST$ coverage of the samples.
 
In total, we obtained 5 orbits for each of the WFC3/UVIS $U_{336}$ and ACS/WFC $V_{606}$ pointings and 3 orbits for each of the WFC3/IR $J_{125}$ and $H_{160}$ pointings as part of $HST$ Program ID 12959 (PI: A. Shapley) between 2012 December and 2013 August.  Table \ref{tab:Obs} lists details of the observations.  Individual exposures were half-orbit ($\sim$1400 seconds) for $U_{336}$ and quarter-orbit ($\sim$600 seconds) for $V_{606}$, $J_{125}$, and $H_{160}$, and total exposure times per pointing were 14 ks ($U_{336}$), 12 ks ($V_{606}$), and 8 ks ($J_{125}$ \& $H_{160}$).  We used a combination of the WFC3/UVIS DITHER-BOX and DITHER-LINE patterns for $U_{336}$, the ACS/WFC DITHER-BOX pattern for $V_{606}$, and the WFC3/IR DITHER-LINE pattern for $J_{125}$ and $H_{160}$.  In order to mitigate charge transfer efficiency (CTE) losses in our WFC3/UVIS $U_{F336}$ exposures, we used the ``post-flash" capability with FLASH=8 \citep{biretta13}.  The final $3\sigma$ surface-brightness sensitivities and PSF FWHMs of the $U_{336}$, $V_{606}$, $J_{125}$, and $H_{160}$ images are, respectively, 24.53, 25.71, 25.79, and 25.56 mag arcsec$^{-2}$ and 0\secpoint081, 0\secpoint092, 0\secpoint178, and 0\secpoint186.  The $3\sigma$ depths obtained in circular apertures with a diameter of 1.5 times the PSF FWHM are, respectively, 29.20, 30.22, 28.93, and 28.63 magnitudes.

\section{Data Reduction and Photometry} \label{sec:DataPhot}

Data reduction was performed on calibrated, flat-fielded, and CTE-corrected (in the case of WFC3/UVIS and ACS/WFC) images with $DrizzlePac$ \citep{fruchter10,koekemoer03}.   The task $TweakReg$ was used to align all exposures within each visit, and the $AstroDrizzle$ pipeline was used to perform sky subtraction, mask cosmic rays and bad pixels, and combine the exposures in the final, drizzled image.  Final images were drizzled onto two scales: one optimized for the highest resolution in each filter (for analysis of galaxy morphologies), and one where the pixel scale was consistent across all filters (for matched-aperture photometric analysis).  For the images drizzled to optimum resolution, the $AstroDrizzle$ parameters \texttt{pixscale} and \texttt{pixfrac} are indicated in Table \ref{tab:Obs}.  For the images used in photometric analysis, all filters were drizzled to a pixel scale of 0.03 $''$/pixel, using a pixfrac value of 0.7 for $U_{336}$ and 0.8 for $V_{606}$, $J_{125}$, and $H_{160}$.  The \texttt{pixfrac} values were chosen in order to achieve the optimum balance between the signal-to-noise ratio and visibility of low-surface-brightness features.  Final drizzled images were registered to each other using the tasks $TweakReg$ and $TweakBack$, achieving alignment between the $V_{606}$, $J_{125}$, and $H_{160}$ filters with an rms of 0\secpoint003 $-$ 0\secpoint006, and between $V_{606}$ and $U_{336}$ with an rms of 0\secpoint01.  The $HST$ images were also aligned to the world coordinate system of the Keck/LRIS $V-$band image from M13 (the image to which all other Keck/LRIS images were registered) in order to map where the ground-based NB3420 (LyC) detections fell relative to emission in the $HST$ images.  After registration with $TweakReg$, residual astrometric distortions between the LRIS and $HST$ images were corrected to a precision of 0\secpoint09 (less than half the size of an LRIS pixel) using the IRAF task CCMAP.  Figures \ref{fig:LBG_stamps} $-$ \ref{fig:LAEphot_stamps} display postage stamp images of the 16 galaxies with NB3420 detections and $U_{336}V_{606}J_{125}H_{160}$ imaging.

For objects where imaging in all four filters ($U_{336}V_{606}J_{125}H_{160}$) was available, the widest PSF was that of $H_{160}$ (0\secpoint186).  Accordingly, in order to perform matched-aperture photometry, we smoothed the higher-resolution $HST$ images to match the PSF of the $H_{160}$ image.  When infrared data were unavailable because of the smaller field of view of the WFC3/IR instrument, the widest PSF was that of $V_{606}$ (0\secpoint092) and we smoothed the $U_{336}$ data to match this PSF.  In order to perform the PSF-matching, we first created an empirical PSF from 10$-$30 (depending on the filter) bright, isolated, and unsaturated stars using the IDL routine \texttt{psf\_extract} from $StarFinder$ \citep{diolaiti00}.  These empirical PSFs were then input into the IRAF routine PSFMATCH, which outputs the convolution kernel and the PSF-matched image.  The curves of growth of the stellar profiles in the $H_{160}$ and $V_{606}$ images agree with those of the PSF-matched images to $\leq$2\% for the majority of the stellar profile, and agree within $\leq$10\% at small radii ($\leq$3 pixels).

For objects with $U_{336}V_{606}J_{125}H_{160}$ coverage, matched-aperture photometry was performed with $SExtractor$ \citep{bertin96} in dual-image mode, using the PSF-matched $V_{606}$ image to detect sources and define isophotes, and applying these isophotes (which can be examined using the SExtractor segmentation image) to the $U_{336}$, $J_{125}$, and $H_{160}$ images.  Because the $V_{606}$ image was already smoothed to the $H_{160}$ PSF, no filtering was used, and the $SExtractor$ detection threshold was set to 4.0$\sigma$.  When photometry was extracted for galaxies without $J_{125}$ and $H_{160}$ imaging, the unsmoothed $V_{606}$ image was filtered with a Gaussian kernel of $\sigma$=4 pixels before source extraction.  As most of the galaxies in our $z\sim2.85$ sample have clumpy morphologies, $SExtractor$ was run with maximum deblending (DEBLEND\_MINCONT = 0.0) in order to separate clumps within the galaxies for individual analysis.  All galaxy clumps defined by $SExtractor$ were examined visually in the segmentation image to guarantee that the subregions visible by eye were properly identified.  For some galaxies, the detection threshold parameter was slightly modified ($\pm1\sigma$) to achieve the best isophote.  There were also instances in which substructure was not visible in the $V_{606}$ image; in these cases, the relevant isophotes were defined in the $U_{336}$ or $J_{125}$ images where the substructure was visible.  The final SExtractor segmentation maps used for photometry are displayed in the right-most column of Figures \ref{fig:LBG_stamps} $-$ \ref{fig:LAEphot_stamps}.  The background subtraction algorithm in SExtractor was set to LOCAL, which defines a square sky annulus around the object in question.  However, in cases where multiple adjacent galaxy clumps were deblended, the annulus defined by $SExtractor$ was often contaminated by nearby sources to the extent that the resulting background estimation was biased.  In such cases, we created our own uncontaminated sky annulus around the object and estimated the background using the sigma-clipped mode, the procedure employed by $SExtractor$.

In order to estimate photometric uncertainties for objects in the PSF-matched images, we followed the methods of \citet{fs06} and computed photometric errors as a function of isophotal aperture size.  First, we identified 1000 blank regions that avoided objects and image edges in each of the PSF-matched $U_{336}$, $V_{606}$, $J_{125}$, and $H_{160}$ images used for isophotal photometry.  We then performed photometry on these blank regions with circular apertures of various sizes corresponding to the isophotal areas of our LBGs and LAEs.  We defined the photometric uncertainty for a given object with isophotal area, $A$, as the standard deviation of the number of counts in the 1000 blank apertures of area $A$.  The relationship between aperture size and background rms in our images is qualitatively similar to that found in \citet{fs06}.  From this analysis, we estimated 3$\sigma$ limiting magnitudes in apertures with a diameter of 1.5 times the PSF FWHM for each of the unsmoothed $U_{336}$, $V_{606}$, $J_{125}$, and $H_{160}$ images, and list them in Table \ref{tab:Obs}.  Photometric data for individual objects are listed in Table \ref{tab:phot}.

\begin{deluxetable*}{lcccccccccc} 
\tablewidth{0pt} 
\tabletypesize{\footnotesize} 
\tablecolumns{11} 
\tablecaption{LBG and LAE Photometry.\label{tab:phot}}
\tablehead{
\colhead{ID\tablenotemark{a}} &
\colhead{R.A.\tablenotemark{b}} &
\colhead{Dec.\tablenotemark{b}} &
\colhead{$U_{336}$\tablenotemark{c}} &
\colhead{$V_{606}$\tablenotemark{c}} &
\colhead{$J_{125}$\tablenotemark{c}} &
\colhead{$H_{160}$\tablenotemark{c}} &
\colhead{$z_{spec}$\tablenotemark{d}} &
\colhead{$z_{phot}$} &
\colhead{$z_{phot}$} & 
\colhead{SED Type\tablenotemark{e}}
\cr \colhead{} & \colhead{} & \colhead{} & \colhead{} & \colhead{} & \colhead{} & \colhead{} & \colhead{} & \colhead{P\'{E}GASE} & \colhead{BPASS} & \colhead{}
}
\startdata   
lae1670a & 15:51:45.176 & 19:10:15.261 & $>28.69$ & $29.39^{+0.28}_{-0.22}$ & $28.24^{+0.20}_{-0.17}$ & $>28.64$ & --- & 1.19 & 1.67 & c \\
lae1670b & 15:51:45.110 & 19:10:15.225 & $>28.11$ & $28.25^{+0.16}_{-0.14}$ & $>28.53$ & $>28.18$ & 2.846 & 2.28 & 1.82 & r \\
lae2292a & 15:51:47.633 & 19:10:00.319 & $>29.01$ & $29.14^{+0.16}_{-0.14}$ & $28.97^{+0.32}_{-0.25}$ & $28.65^{+0.33}_{-0.25}$ & 2.851 & 1.61 & 1.44 & r \\
lae2292b & 15:51:47.626 & 19:10:01.034 & $27.31^{+0.20}_{-0.17}$ & $27.73^{+0.10}_{-0.09}$ & $26.71^{+0.07}_{-0.07}$ & $26.76^{+0.11}_{-0.10}$ & --- & 1.15 & 1.17 & c \\
lae2436a & 15:52:03.209 & 19:12:51.261 & $27.33^{+0.32}_{-0.25}$ & $26.98^{+0.08}_{-0.07}$ & $26.92^{+0.10}_{-0.09}$ & $26.32^{+0.08}_{-0.07}$ & 2.04\tablenotemark{f} & 2.29 & 2.33 & c \\
lae2436b & 15:52:03.239 & 19:12:52.296 & $26.13^{+0.29}_{-0.23}$ & $25.16^{+0.04}_{-0.04}$ & $24.46^{+0.03}_{-0.02}$ & $24.33^{+0.03}_{-0.03}$ & 0.44\tablenotemark{f} & 0.42 & 1.65 & c \\
lae2966a & 15:51:50.037 & 19:10:42.064 & $>27.54$ & $26.82^{+0.07}_{-0.07}$ & $26.23^{+0.07}_{-0.07}$ & $26.20^{+0.10}_{-0.09}$ & --- & 1.09 & 1.42 & c \\
lae2966b & 15:51:49.975 & 19:10:41.263 & $>27.14$ & $26.29^{+0.06}_{-0.06}$ & $26.11^{+0.09}_{-0.08}$ & $25.74^{+0.09}_{-0.08}$ & 2.841 & 3.03 & 2.90 & r \\
lae3038a & 15:51:50.131 & 19:09:02.336 & $>27.47$ & $28.02^{+0.25}_{-0.20}$ & $26.17^{+0.07}_{-0.07}$ & $26.08^{+0.09}_{-0.08}$ & --- & 1.18 & 3.10 & a \\
lae3038b & 15:51:50.077 & 19:09:02.483 & $27.93^{+0.36}_{-0.27}$ & $28.20^{+0.15}_{-0.13}$ & $27.03^{+0.09}_{-0.09}$ & $27.08^{+0.14}_{-0.12}$ & --- & 1.20 & 1.21 & c \\
lae4070a & 15:51:54.072 & 19:10:26.789 & $28.24_{-0.24}^{+0.31}$ & $28.48_{-0.11}^{+0.13}$ & $28.30_{-0.19}^{+0.23}$ & $27.75_{-0.16}^{+0.19}$ & --- & 1.25 & 1.21 & a \\
lae4070b & 15:51:54.061 & 19:10:26.731 & $27.38_{-0.22}^{+0.28}$ & $27.51_{-0.10}^{+0.11}$ & $26.85_{-0.09}^{+0.10}$ & $26.81_{-0.12}^{+0.14}$ & --- & 1.24 & 1.23 & c \\
lae5200a & 15:52:01.125 & 19:11:25.487 & $28.87^{+0.30}_{-0.23}$ & $30.29^{+0.43}_{-0.31}$ & $>29.74$ & $>29.34$  & --- & 1.10 & 0.99 & c \\
lae5200b & 15:52:01.097 & 19:11:25.994 & $>27.73$ & $27.90^{+0.17}_{-0.15}$ & $26.69^{+0.07}_{-0.07}$ & $26.50^{+0.09}_{-0.08}$  & --- & 1.23 & 1.51 & a \\
lae5200c & 15:52:01.111 & 19:11:26.206 & $>27.84$ & $27.93^{+0.16}_{-0.14}$ & $27.05^{+0.09}_{-0.09}$ & $27.43^{+0.19}_{-0.16}$  & --- & 1.11 & 1.52 & c \\
lae6510a & 15:52:07.004 & 19:12:02.999 & $>28.68$ & $29.25^{+0.26}_{-0.21}$ & $29.26^{+0.43}_{-0.31}$ & $>28.90$  & --- & 2.03 & 1.71 & c \\
lae6510b & 15:52:06.976 & 19:12:02.822 & $>27.71$ & $27.95^{+0.18}_{-0.16}$ & $27.61^{+0.18}_{-0.15}$ & $27.58^{+0.25}_{-0.21}$  & --- & 1.31 & 1.35 & c \\
lae6510c & 15:52:06.981 & 19:12:03.350 & $>26.81$ & $26.59^{+0.12}_{-0.11}$ & $25.73^{+0.06}_{-0.06}$ & $25.58^{+0.08}_{-0.08}$  & --- & 1.24 & 1.38 & a \\
lae6662a & 15:52:06.369 & 19:10:42.590 & $>27.82$ & $27.31^{+0.09}_{-0.08}$ & $27.01^{+0.09}_{-0.08}$ & $26.64^{+0.09}_{-0.08}$  & 2.833 & 2.82 & 2.28 & r \\
lae6662b & 15:52:06.350 & 19:10:42.450 & $>27.49$ & $26.73^{+0.07}_{-0.07}$ & $26.68^{+0.09}_{-0.08}$ & $26.22^{+0.08}_{-0.08}$  & 2.833 & 2.85 & 2.80 & r \\
lae6662c & 15:52:06.376 & 19:10:43.706 & $>28.55$ & $28.84^{+0.19}_{-0.16}$ & $28.63^{+0.24}_{-0.20}$ & $28.60^{+0.35}_{-0.26}$  & --- & 1.26 & 1.33 & c \\
lae7180a & 15:52:04.669 & 19:11:42.083 & $>27.62$ & $27.29^{+0.11}_{-0.10}$ & $26.72^{+0.08}_{-0.08}$ & $26.67^{+0.11}_{-0.10}$  & --- & 1.07 & 1.54 & a \\
lae7180b & 15:52:04.631 & 19:11:42.088 & $>28.66$ & $29.19^{+0.24}_{-0.20}$ & $28.88^{+0.29}_{-0.23}$ & $>28.89$  & --- & 1.75 & 1.74 & --- \\
lae7832a & 15:52:02.222 & 19:10:48.807 & $>26.88$ & $26.08^{+0.07}_{-0.07}$ & $26.08^{+0.08}_{-0.08}$ & $25.62^{+0.08}_{-0.07}$  & 2.829 & 2.77 & 2.70 & r \\
lae7832b & 15:52:02.199 & 19:10:48.379 & $>27.91$ & $27.65^{+0.11}_{-0.10}$ & $27.68^{+0.16}_{-0.14}$ & $27.37^{+0.17}_{-0.15}$  & 2.829 & 2.68 & 1.86 & r \\
lae7832c & 15:52:02.155 & 19:10:48.614 & $25.57^{+0.13}_{-0.11}$ & $25.16^{+0.03}_{-0.03}$ & $24.48^{+0.02}_{-0.02}$ & $24.51^{+0.03}_{-0.03}$  & --- & 0.72 & 1.72 & c \\
lae7890a & 15:52:01.957 & 19:12:42.255 & $26.61^{+0.19}_{-0.16}$ & $26.52^{+0.06}_{-0.06}$ & $25.73^{+0.04}_{-0.04}$ & $25.86^{+0.06}_{-0.06}$ & --- & 0.95 & 1.38 & c \\
M16a & 15:51:53.648 & 19:09:29.392 & $>27.05$ & $26.15^{+0.06}_{-0.06}$ & $25.32^{+0.04}_{-0.04}$ & $25.19^{+0.06}_{-0.05}$ & 2.954 & 3.92 & 2.50 & r \\
M16b & 15:51:53.619 & 19:09:30.486 & $>27.89$ & $28.45^{+0.25}_{-0.20}$ & $27.26^{+0.14}_{-0.13}$ & $27.21^{+0.19}_{-0.16}$ & --- & 1.25 & 1.36 & c \\
MD5a & 15:51:45.206 & 19:11:04.887 & $>27.21$ & $25.87^{+0.04}_{-0.04}$ & $25.68^{+0.06}_{-0.05}$ & $25.32^{+0.06}_{-0.05}$ & 3.143 & 3.04 & 2.88 & r \\
MD5b & 15:51:45.226 & 19:11:05.300 & $>27.05$\tablenotemark{g} & $25.85^{+0.05}_{-0.04}$ & $25.57^{+0.06}_{-0.05}$ & $25.53^{+0.08}_{-0.07}$ & 3.143 & 3.50 & 1.94 & a \\
MD12a & 15:51:51.887 & 19:10:41.313 & $>28.70$ & $27.38^{+0.04}_{-0.04}$ & $26.92^{+0.05}_{-0.05}$ & $26.51^{+0.05}_{-0.05}$ & 2.852 & 3.17 & 2.88 & r \\
MD12b & 15:51:51.915 & 19:10:41.281 & $>27.20$ & $26.67^{+0.09}_{-0.08}$ & $26.59^{+0.13}_{-0.12}$ & $26.10^{+0.12}_{-0.11}$ & 2.852 & 2.83 & 2.56 & r \\
MD12c & 15:51:51.866 & 19:10:41.457 & $>27.15$ & $25.95^{+0.05}_{-0.04}$ & $25.64^{+0.06}_{-0.05}$ & $25.21^{+0.05}_{-0.05}$ & 2.852 & 3.17 & 2.91 & r \\
MD12d & 15:51:51.879 & 19:10:41.091 & $>27.56$ & $26.41^{+0.05}_{-0.05}$ & $26.31^{+0.07}_{-0.07}$ & $25.97^{+0.08}_{-0.07}$ & 2.852 & 3.00 & 2.94 & r \\
MD12e & 15:51:51.880 & 19:10:40.126 & $27.29^{+0.30}_{-0.23}$ & $27.26^{+0.10}_{-0.09}$ & $26.62^{+0.09}_{-0.09}$ & $26.57^{+0.12}_{-0.11}$ & --- & 1.25 & 1.25 & c \\
MD12f & 15:51:52.028 & 19:10:40.582 & $>27.81$ & $27.10^{+0.07}_{-0.07}$ & $27.35^{+0.17}_{-0.14}$ & $27.25^{+0.21}_{-0.18}$ & 2.852 & 2.93 & 2.74 & r \\
MD12g & 15:51:52.036 & 19:10:41.535 & $27.26^{+0.35}_{-0.26}$ & $27.03^{+0.09}_{-0.09}$ & $26.50^{+0.10}_{-0.09}$ & $26.15^{+0.10}_{-0.09}$ & --- & 1.35 & 1.26 & c \\
MD12h & 15:51:52.035 & 19:10:40.060 & $>28.86$ & $29.11^{+0.18}_{-0.15}$ & $>29.14$ & $>28.79$ & --- & 1.95 & 1.60 & --- \\
MD34a & 15:52:06.336 & 19:12:48.673 & $>26.70$ & $25.49^{+0.05}_{-0.05}$ & $25.21^{+0.04}_{-0.04}$ & $24.67^{+0.04}_{-0.04}$ & 2.852 & 3.15 & 2.87 & r \\
MD34b & 15:52:06.307 & 19:12:48.550 & $>26.83$ & $25.67^{+0.05}_{-0.05}$ & $24.84^{+0.03}_{-0.03}$ & $24.15^{+0.02}_{-0.02}$ & 2.852 & 3.37 & 2.66 & r \\
MD34c & 15:52:06.334 & 19:12:48.205 & $>27.60$ & $26.93^{+0.08}_{-0.07}$ & $26.15^{+0.05}_{-0.05}$ & $25.67^{+0.04}_{-0.04}$ & 2.852 & 3.20 & 2.65 & r \\
MD34d & 15:52:06.314 & 19:12:48.195 & $>29.54$ & $28.97^{+0.08}_{-0.08}$ & $28.06^{+0.06}_{-0.06}$ & $27.46^{+0.05}_{-0.05}$ & 2.852 & 2.71 & 2.63 & r \\
MD34e & 15:52:06.295 & 19:12:48.191 & $>29.34$ & $28.94^{+0.10}_{-0.09}$ & $27.98^{+0.07}_{-0.06}$ & $27.38^{+0.06}_{-0.05}$ & 2.852 & 2.76 & 2.65 & r \\
MD34f & 15:52:06.318 & 19:12:47.708 & $>27.27$ & $26.92^{+0.10}_{-0.09}$ & $25.88^{+0.05}_{-0.05}$ & $25.45^{+0.05}_{-0.05}$ & 2.852 & 2.12 & 2.60 & r \\
MD34g & 15:52:06.333 & 19:12:47.380 & $>27.50$ & $27.25^{+0.11}_{-0.10}$ & $26.27^{+0.06}_{-0.06}$ & $26.23^{+0.08}_{-0.08}$ & --- & 1.21 & 1.33 & c \\
MD34h & 15:52:06.365 & 19:12:47.446 & $>29.43$ & $29.90^{+0.23}_{-0.19}$ & $29.54^{+0.29}_{-0.23}$ & $29.32^{+0.36}_{-0.27}$ & 2.852 & 1.73 & 1.59 & a
\enddata
\tablenotetext{a}{Objects are listed by their IDs from M13.  The final letter of the object name indicates the sub-arcsecond component of the galaxy, according to the labels in Figures \ref{fig:LBG_stamps}, \ref{fig:LAEz_stamps}, and \ref{fig:LAEphot_stamps}.}
\tablenotetext{b}{Object centroids calculated from $V_{606}$.}
\tablenotetext{c}{Isophotal magnitudes and photometric errors as a function of aperture size (see Section \ref{sec:DataPhot}).  Limits are $3\sigma$.}
\tablenotetext{d}{Spectroscopic redshifts listed were obtained via ground-based spectroscopy, and thus in general it is not possible to distinguish between individual sub-arcsecond components of galaxies within the spectrum.  We list spectroscopic redshifts only when the SED fit indicates that the object is not a foreground contaminant.  Spectroscopic redshifts are not listed for the following types of objects: LAEs without spectroscopic follow-up, contaminants identified via photometric redshifts, and objects where $U_{336}V_{606}J_{125}H_{160}$ photometric data were insufficient to determine a photometric redshift.}
\tablenotetext{e}{This column indicates whether the SED of the object implies a real $z\sim2.85$ galaxy (r) or a foreground contaminant (c).  The letter (a) indicates an ambiguous SED shape, defined in Section \ref{sec:sed}.}
\tablenotetext{f}{The original \lya\ redshift associated with \emph{lae2436} from M13 was $z=2.832$.  Subsequent reanalysis of the available spectra near this object indicates that the $z=2.832$ emission was associated with a nearby \lya\ blob (see Section \ref{sec:fcontam}).}
\tablenotetext{g}{Although MD5b is formally undetected in $U_{336}$ at $3\sigma$, emission at the location of MD5b is visible by eye in the $U_{336}$ imaging.  This emission corresponds to a 2.25$\sigma$ detection in $U_{336}$ ($m_{336}=27.37^{+0.64}_{-0.40}$), which is consistent within errors of the detection in NB3420 ($m_{NB3420}=26.89^{+0.43}_{-0.31}$).}
\end{deluxetable*}

\section{Estimation of Photometric Redshifts} \label{sec:Photoz}

One important challenge in isolating LyC emission from high-redshift galaxies is that the majority of these systems have complex morphologies.  In our sample of $z\sim2.85$ LBGs and LAEs, roughly 80\% of the objects have complex morphology in the $U_{336}V_{606}J_{125}H_{160}$ imaging, whether it be due to diffuse emission or multiple sources of nucleated emission.  This fraction of objects with complex morphology is similar to that found in \citet{law07} for a sample of 66 $z\sim3$ LBGs with rest-frame UV $HST$ imaging ($\sim85$\%), and demonstrates that for most high-redshift galaxies, high-resolution images reveal significant substructure.  For the NB3420-detected galaxies, the NB3420 flux may be due to LyC emission from the high-redshift galaxy, or contamination from a lower redshift contaminant.  Therefore, we must obtain photometric redshifts for individual subcomponents in order to identify possible contaminants.

In this section, we begin by discussing the expected SED shapes of $z\sim2.85$ galaxies and low-redshift contaminants, and present the range of properties exhibited by galaxies in our sample.  Next, we explain our procedures for fitting photometric redshifts with EAZY \citep{brammer08} to the objects in our $HST$ sample with imaging in all four of the $U_{336}$, $V_{606}$, $J_{125}$, and $H_{160}$ filters.  Finally, we consider each NB3420-detected galaxy individually, discussing the photometric redshifts of each clump and the implications for the source of the galaxy's NB3420 detection.

\subsection{Empirical Analysis of SEDs}\label{sec:sed} 

\begin{figure*}
\epsscale{1.0}
\plotone{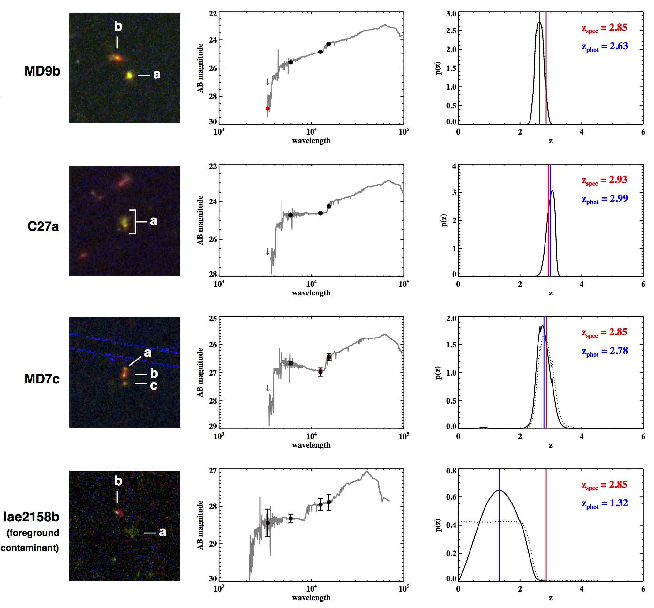}
\caption{
\small
Representative SED shapes found within our $z\sim2.85$ galaxy sample.  The first three rows display $z\sim2.85$ galaxies without NB3420 detections.  From top to bottom, the objects shown are typical examples of SED shapes for objects with a red rest-frame UV slope ($V_{606}-J_{125}>0.1$), a flat rest-frame UV slope ($-0.1 \leq V_{606}-J_{125} \leq 0.1$), a blue rest-frame UV slope ($V_{606}-J_{125}<-0.1$).  The final row shows a typical example of the SED shape of a foreground contaminant.  The left-most panel displays $HST$ $U_{336}V_{606}J_{125}$ composite color images, and indicates the sub-arcsecond components associated with each galaxy.  The middle column shows the $U_{336}V_{606}J_{125}H_{160}$ photometry for the relevant component (black circles with 1$\sigma$ error bars; downward-pointing arrows for 3$\sigma$ limits), the best-fit SED using the SMC-reddened BPASS models in EAZY (gray line), and the expected location of the photometric points based on the best-fit model (red circles).  The right-most panel shows the redshift probability distribution for that component, with the solid black curve indicating the probability distribution after a magnitude-based prior has been applied \citep[see discussion of priors in][]{brammer08} and the dotted black curve indicating the probability distribution before applying the prior.  Blue vertical lines indicate the best-fit photometric redshift ($z_{phot}$) and red lines indicate the observed spectroscopic redshift ($z_{spec}$).  The top three objects demonstrate relatively narrow probability distributions that, while they do not have extremely high redshift precision, do encompass the true spectroscopic redshift.  In general, the precision of the redshift probability distributions decrease with increasing photometric errors.  The final object shows a clear example of a foreground contaminant with the photometric colors described in Section \ref{sec:sed}; accordingly, the EAZY redshift probability distribution does not align with the spectroscopic redshift.  For the case of \emph{lae2158}, the actual LAE is \emph{lae2158a}.
\label{fig:SED_samples} }
\end{figure*}

\begin{figure}
\epsscale{1.0}
\plotone{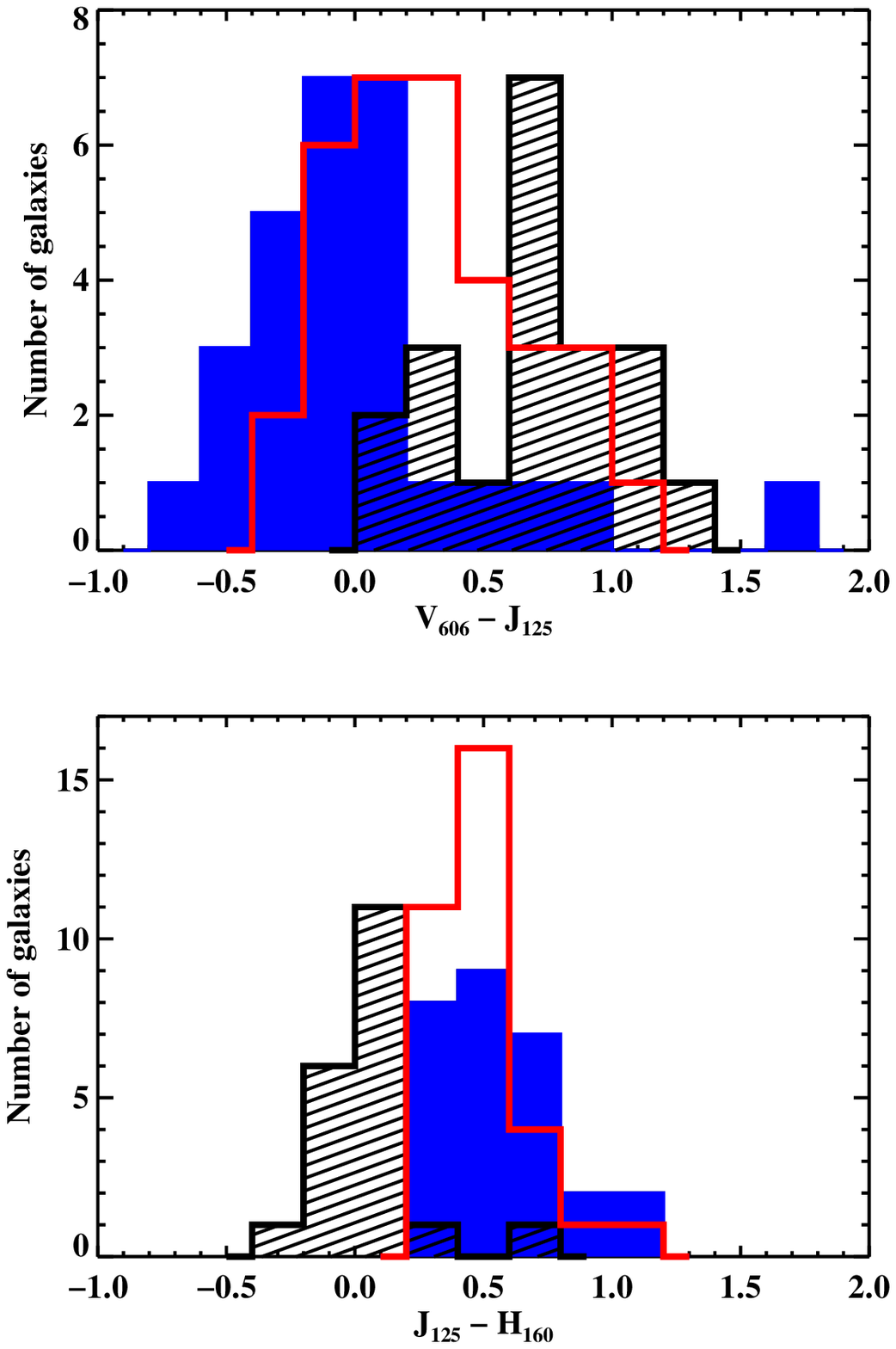}
\caption{
\small
Histograms of $V_{606}-J_{125}$ and $J_{125}-H_{160}$ color for objects with $U_{336}V_{606}J_{125}H_{160}$ photometry that do not have ambiguous SED shapes.  LBGs are plotted in red, LAEs are plotted in blue, and contaminants are plotted in black.  The top panel shows the wide range of $V_{606}-J_{125}$ colors present in both the LAE and LBG samples, although on average LAEs have a bluer rest-frame UV slope than LBGs.  The bottom panel demonstrates how important $J_{125}-H_{160}$ color is in distinguishing typical $z\sim2.85$ galaxies from contaminants, although there are some exceptions (e.g., objects with ambiguous SEDs; see open circles in Figure \ref{fig:colorcolor}).
\label{fig:colorhist} }
\end{figure}

In order to describe the SED shapes of ``typical'' $z\sim2.85$ star-forming galaxies, we first present SEDs from the sample of LBGs and LAEs without LyC detections.  This sample is described in more detail in Section \ref{sec:eazy}.  Figure \ref{fig:SED_samples} shows $U_{336}V_{606}J_{125}H_{160}$ photometry for several galaxies in the sample that span the range of typical SED shapes.  As expected, the Lyman break is present in the SEDs of all galaxies without LyC detections.  Furthermore, every object in the LyC non-detection sample is undetected in $U_{336}$, which is consistent with the NB3420 non-detections in M13.  The presence of the Lyman break is one of the most important features for distinguishing low- and high-redshift galaxies in the sample of galaxies without LyC detections.  It must be kept in mind, however that the strength of the Lyman break may not be an effective way to identify the redshifts of LyC-emitting galaxies \citep[see, e.g.,][]{cooke14}.  While the galaxies in the LyC non-detection sample all exhibit a non-detection in $U_{336}$, there was a large range of $V_{606} - J_{125}$ colors in both the LAE and LBG samples, indicating the corresponding range in rest-frame UV slopes among the galaxies in our sample (see the upper panel of Figure \ref{fig:colorhist}).  In the LBG sample, the $V_{606} - J_{125}$ color ranged from $-0.32 \leq V_{606} - J_{125} \leq 1.03$ (median $V_{606} - J_{125} = 0.24$), with most galaxies displaying red rest-frame UV slopes.  In contrast, the LAE sample had $-0.66 \leq V_{606} - J_{125} \leq 1.7$ (median $V_{606} - J_{125} = -0.02$), with most galaxies displaying blue rest-frame UV slopes.  While on average the LAEs had bluer UV slopes than the LBGs, the LAE sample also contained the galaxy with the reddest $V_{606} - J_{125}$ color (\emph{lae1843}, $z_{spec}=2.847$, which contains clumps with $V_{606} - J_{125} = 0.95$ and $V_{606} - J_{125} = 1.7$).  The $J_{125}-H_{160}$ colors of galaxies ranged from $0.20 \leq J_{125}-H_{160} \leq 1.19$ (mean $ J_{125}-H_{160} = 0.53$), and did not differ significantly between the LBG and LAE samples (see the lower panel of Figure \ref{fig:colorhist}).

\begin{figure}
\epsscale{1.0}
\plotone{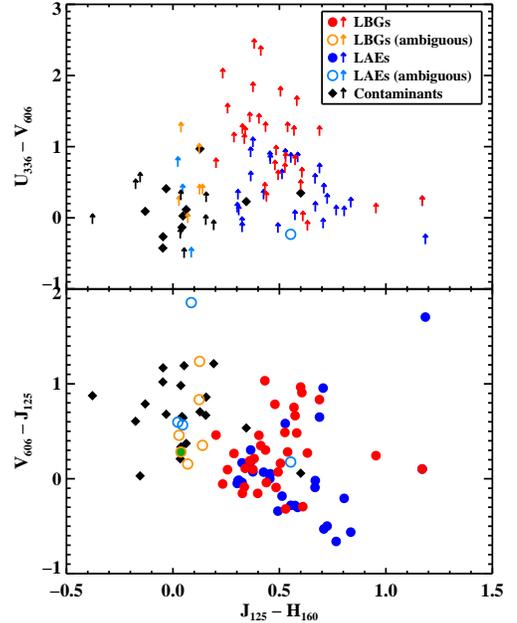}
\caption{
\small
$U_{336}-V_{606}$ vs. $J_{125}-H_{160}$ and $V_{606}-J_{125}$ vs. $J_{125}-H_{160}$ color-color plots of all objects with $U_{336}V_{606}J_{125}H_{160}$ photometry.  LBGs with typical SED shapes are indicated by solid red circles, and LBGs with ambiguous SED shapes are indicated by open orange circles.  LAEs with typical SED shapes are indicated by solid blue circles, and LAEs with ambiguous SED shapes are indicated by open light-blue circles.  Foreground contaminants are indicated by black diamonds.  Lower limits for all objects are indicated by upward-pointing arrows following the same color scheme.  In the lower panel, the best candidate for true LyC emission (MD5b) is indicated by an orange open circle filled with green.
\label{fig:colorcolor} }
\end{figure}

\begin{figure*}
\epsscale{1.0}
\plotone{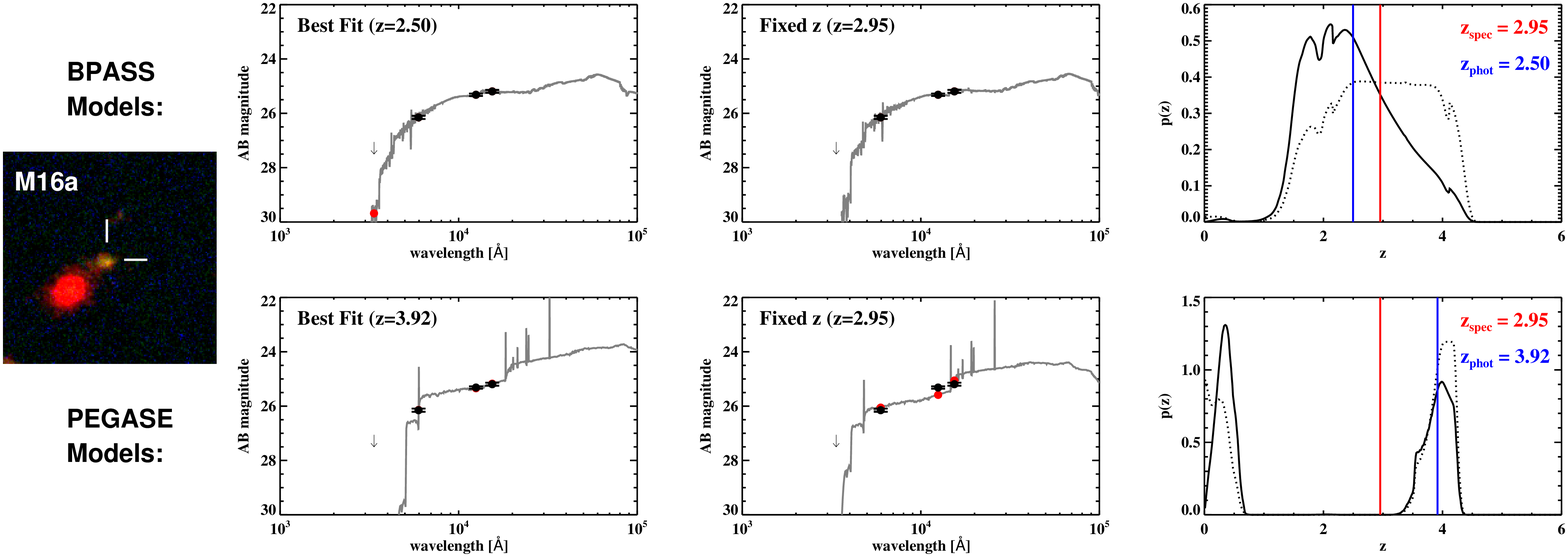}
\caption{
\small
M16a demonstrates the typical shape of an ``ambiguous'' SED, one that can be fit with a wide range of redshifts that encompass both the spectroscopic redshift and lower redshifts typical of foreground contaminants.  Results from EAZY are plotted in the three right hand panels, and both the SMC-reddened BPASS models (top row) and the P\'{E}GASE models that include nebular emission (bottom row) are shown to demonstrate the differences between models.  In the left-most panel, the redshift is allowed to float during SED fitting, while in the middle panel the redshift is fixed to the spectroscopic redshift.  The right-most panel shows the redshift probability distribution.  Colors and symbols are as in Figure \ref{fig:SED_samples}.  The plot showing the P\'{E}GASE fit at fixed redshift demonstrates the limitations of the P\'{E}GASE models in terms of plotting young galaxies with dust attenuation; even with the redshift known, the P\'{E}GASE templates could not provide a satisfactory fit.  Analysis of both the BPASS and P\'{E}GASE redshift probability distributions shows that for an SED of this shape, there is no way to narrow down the redshift to better than $0<z<4.5$.
\label{fig:SED_ambig} }
\end{figure*}

Additionally, there were a few objects in the LyC non-detection sample that demonstrated the typical SED shapes of low-redshift contaminants (see the bottom panel of Figure \ref{fig:SED_samples}).  We found that the main features that help identify a low-redshift interloper are the presence of flat $U_{336} - V_{606}$ and $J_{125} - H_{160}$ colors, especially when accompanied by a large break between $V_{606}$ and $J_{125}$ (see Figure \ref{fig:colorcolor}).  For contaminants, the break between  $V_{606}$ and $J_{125}$ corresponds to a Balmer break or 4000\AA\ break between $0.5 \leq z \leq 2.2$, and has typical values of $0.3 \leq V_{606} - J_{125} \leq 1.2$.  The only complication, however, is that young, dusty galaxies may also present nearly flat $J_{125} - H_{160}$ colors with red $V_{606} - J_{125}$ colors.  If $U_{336} - V_{606}$ is flat (i.e., no Lyman break), then galaxies with flat $J_{125} - H_{160}$ colors and red $V_{606} - J_{125}$ colors can easily be identified as low-redshift contaminants.  However, if $U_{336} - V_{606} > 0$, we must disentangle the degeneracy between contaminants and young, dusty galaxies.  In our analysis, we label galaxies with such SEDs ``ambiguous cases'', an example of which is shown in Figure \ref{fig:SED_ambig}.  As the redshifts of galaxies with ambiguous SED shapes are uncertain, we do not include them in our analysis of $z \sim 2.85$ galaxies without LyC detections.  We also stress again that employing the $U_{336} - V_{606}$ color to distinguish a contaminant from a $z \sim 2.85$ galaxy may be problematic when applied to galaxies with potential LyC emission, as the magnitude of the Lyman break for LyC galaxies is not well understood \citep[M13;][]{nestor13,steidel14}.  We keep this caveat in mind in Section \ref{sec:lyc} when discussing our targets with NB3420 detections.  Figure \ref{fig:colorcolor} demonstrates the relationships between $U_{336} - V_{606}$, $V_{606}-J_{125}$, and $J_{125}-H_{160}$ colors for galaxies at $z \sim 2.85$, galaxies with ambiguous SED shapes, and foreground contaminants.  In the $V_{606}-J_{125}$, and $J_{125}-H_{160}$ plot, $z \sim 2.85$ galaxies generally occupy a different region of color-color space from that of foreground contaminants, with the ambiguous cases (differentiated from contaminants by having $U_{336} - V_{606} > 0$) straddling both distributions.

\subsection{Modeling SEDs with EAZY}\label{sec:eazy}

While the empirical SED shapes suggest qualitative divisions between low- and high-redshift galaxies, we can also obtain more systematic estimates of the redshifts of our targets.  We used the photometric redshift code EAZY \citep{brammer08} to fit the $U_{336}$, $V_{606}$, $J_{125}$, and $H_{160}$ photometry acquired with $HST$ and estimate the redshifts of each sub-arcsecond component in the vicinity of our LAE and LBG targets.  In order to choose the best input parameters for EAZY and learn how to interpret the output EAZY produces, we first ran the code on a test sample of $z\sim2.85$ LBGs and LAEs in our $HST$ images with known spectroscopic redshifts, unambiguous SED shapes, and no NB3420 detections.  Later, we supplemented this sample with $z\sim2.85$ components of galaxies from the NB3420-detected sample for which there was no LyC detection (i.e., galaxies for which the NB3420 detection was proven to be associated with foreground contamination from another clump).  This test sample of galaxies should have a low rate of contamination by foreground interlopers, and thus help us evaluate whether or not EAZY can accurately identify galaxies known to be at $z\sim2.85$ with the photometry provided.  Additionally, we were able to analyze galaxies in this sample that exhibited complex morphologies and evaluate the SED fits for each galaxy clump separately.  In this way, we developed a procedure for differentiating between clumps that belonged to the spectroscopically confirmed LBG or LAE and those that were lower redshift interlopers.  As the galaxies in the test sample do not have NB3420 detections, we were able to make such a distinction without the complication of possible LyC emission, which might be associated with ``non-standard'' SEDs.  Finally, we note that we did not include any components where the SED shape was ambiguous (see definition in Section \ref{sec:sed}) because it was not clear from the SED shape if that component was at high or low redshift.

We varied several input parameters to EAZY in order to determine their optimal values.  First, we experimented with fitting our data using different stellar population synthesis models.  EAZY defaults to P\'{E}GASE models \citep{fioc97}, which span a range of star-formation histories, metallicities, ages, and reddenings using the \citet{calzetti00} attenuation curve.  To the five default P\'{E}GASE models, EAZY adds additional templates representing a dusty starburst galaxy \citep[described in][]{brammer08} and two templates for old, dusty galaxies (from \texttt{EAZY\_v1.1\_lines/} and \texttt{DUSTY/} in the online development version\footnote{https://github.com/gbrammer/eazy-photoz/blob/master/ templates/}).  EAZY also includes a set of model templates from \citet{br07} (BR07), which are based on \citet{bc03} models.  Both the P\'{E}GASE and BR07 models provided good, qualitatively similar fits to most of our LBGs and LAEs, but failed to accurately represent galaxies with blue UV slopes ($V_{606}-J_{125}<-0.1$) and galaxies with SEDs younger than $\sim$50 Myr (see Figure \ref{fig:SED_ambig}).  Therefore, we experimented with additional stellar population models that might provide a better fit to the bluest and youngest galaxies in our sample.  These included BPASS models, which have a more accurate treatment of Wolf-Rayet stars and massive stellar binaries \citep{eldridgestanway09}, and Starburst99 (SB99) models with updated treatment of stellar rotation \citep{leitherer14}.  Additional impetus for considering bluer templates is their increased emission in the LyC spectral region, which may provide a method of more accurately modeling galaxies with LyC detections.  In addition to experimenting with the choice of stellar population models, we also used two different extinction curves to redden the BPASS and SB99 models, for which only constant star-formation (CSF) histories are available.  The \citet{calzetti00} attenuation curve has been traditionally used to model extinction in high-redshift star-forming galaxies, but recent work \citep[e.g.,][]{siana09,reddy06jun,reddy10,reddy12} has shown that an SMC extinction curve may be more appropriate for galaxies with ages younger than 100 Myr.  Therefore, we made two sets of CSF templates with each of the BPASS and SB99 models, one template set reddened exclusively with the \citet{calzetti00} attenuation curve, and another reddened with the \citet{calzetti00} attenuation curve for templates with older galactic ages and the SMC extinction curve from \citet{gordon03} for templates with ages less than 100 Myr.

Based on the analysis of objects without LyC detections, we verified that, for the majority of galaxies, EAZY estimates the correct redshift of the galaxy within roughly $\pm0.5$ of the spectroscopic redshift.  Thus, we can successfully use photometric redshifts to determine if most galaxy clumps are at $z\sim2.85$ or are low-redshift contaminants, keeping in mind that we have fairly coarse redshift precision.  As EAZY provides several estimators for the photometric redshift (e.g., $z_{p}$\footnote{Redshift where the likelihood is maximized after applying the magnitude-based prior.}, $z_{m2}$\footnote{Redshift marginalized over the posterior probability distribution.}, $z_{peak}$\footnote{Hybrid between $z_{p}$ and $z_{m2}$ to address the pathological case where there are two widely-separated peaks in the probability distribution that have similar integrated probabilities.}), we investigated each of them while varying the input parameters to the program and determined that $z_{peak}$ provided the best estimate of the spectroscopic redshift.  All photometric redshifts we quote use the $z_{peak}$ estimator.  Of the range of input parameters tested on the sample of objects without LyC detections, we determined that the P\'{E}GASE models result in photometric redshifts that most closely match the spectroscopic redshifts.  These models had both the smallest systematic offset and lowest standard deviation of all the variations of input parameters we tried, yielding $(z_{phot}-z_{spec})/(1+z_{spec}) = -0.03 \pm 0.07$.  We note that while the P\'{E}GASE models work best for the sample as a whole, the BPASS models provide the best fits to galaxies with blue rest-frame UV slopes ($V_{606}-J_{125}<-0.1$).  Also, as the EAZY P\'{E}GASE models only include 5 templates, they did not accurately fit galaxies with younger SEDs.  Better fits were achieved for young galaxies by implementing a fine grid of young BPASS templates, specifically by including SMC-reddened models with ages of 1 Myr, 5 Myr, 10 Myr, 30 Myr, and 50 Myr.  Throughout our analysis of the galaxies with potential LyC detections (Section \ref{sec:lyc}), we employ both the P\'{E}GASE models and the SMC-reddened BPASS models with additional young galaxy templates in order to fully examine the likely photometric redshifts for each galaxy.  All figures in this paper displaying output from EAZY show fits using the SMC-reddened BPASS models, unless otherwise indicated.

\subsection{Results of Photometric Redshift Fits for Potential LyC Emitters}\label{sec:lyc}

In total, we have observations in all four $U_{336}V_{606}J_{125}H_{160}$ filters for 16 galaxies with potential LyC detections.  These 16 galaxies include 4 LBGs and 8 LAEs with spectroscopic confirmation, as well as 4 LAE photometric candidates.  In this section, we discuss the results of our analysis of these potential LyC emitters.  We describe the sources of contamination for 11 targets with obvious contaminants, present 4 galaxies with ambiguous SEDs that may or may not be at high redshift, and argue for MD5 as a true LyC emitter.  We also note 3 objects for which the $HST$ imaging revealed that the spectroscopic redshift was incorrectly assigned in M13.  This analysis is contingent upon the precise alignment between the $HST$ and LRIS NB3420 imaging, as described in Section \ref{sec:DataPhot}.

\subsubsection{Foreground Contaminants in the LyC Sample}\label{sec:fcontam}

Seven objects with Keck/LRIS spectroscopic redshifts (2 LBGs: MD12, M16; 5 LAEs: \emph{lae1670}, \emph{lae2292}, \emph{lae2966}, \emph{lae6662}, and \emph{lae7832}) and one object in the photometric LAE sample (\emph{lae3038}) had contaminants that stood out plainly with the combination of high-resolution $HST$ imaging and SED fitting.  Each of these objects was resolved into several clumps in the $HST$ imaging.  In all cases, at least one of the clumps had an SED fit corresponding to the redshift of the Keck/LRIS spectrum (or consistent with the spike redshift $z=2.85$, in the case of \emph{lae3038}), while the clump associated with the LyC emission had the unambiguous SED of a $0.5 \leq z \leq 2.2$ contaminant (flat $U_{336} - V_{606}$ and $J_{125} - H_{160}$, red $V_{606} - J_{125}$; similar to the example contaminant shown in the bottom panel of Figure \ref{fig:SED_samples}).  For these eight objects, the NB3420 emission had a fairly large offset from the original galaxy coordinates; the offsets between the NB3420 detection and \lya\ emission ranged from 0\secpoint57 to 1\secpoint15 with a median value of 0\secpoint65 (5.0 kpc at $z=2.85$), and offsets between the NB3420 emission and the LRIS $V-$band emission ranged from 0\secpoint12 to 1\secpoint26 with a median value of 0\secpoint72 (5.7 kpc at $z=2.85$).  The morphology of these objects in the $HST$ images supports the evidence from the SED fits that these objects are not physically associated; the clumps are distinct, with no evidence for diffuse emission between them that might indicate interactions between galaxies at the same redshift.

A third LBG in our sample, MD34, has an NB3420 detection coincident with two clumps in the $HST$ imaging, only one of which is from a foreground contaminant.  Figure \ref{fig:LBG_stamps} shows the eight distinct components associated with MD34 (MD34a though MD34h).  Components MD34a through MD34f all have SED shapes that place them at the spectroscopic redshift of MD34 ($z=2.85$), and MD34g and MD34h have SED shapes indicative of foreground contaminants.  The NB3420 emission for MD34 is coincident with two clumps in the $HST$ imaging: MD34f ($z=2.85$) and MD34g ($z<2.5$), separated on the sky by 0\secpoint31.  Both MD34f and MD34g exhibit emission in the $U_{336}$ filter, although both detections are less than 3$\sigma$.  It is unclear whether the NB3420 emission associated with MD34 is due solely to non-ionizing UV radiation from the foreground contaminant MD34g, or also in part to LyC emission from the $z=2.85$ component MD34f.  As the high-resolution $U_{336}$ image does not exclusively probe the LyC spectral region at $z=2.85$, and the foreground contaminant MD34g is too close to MD34f to distinguish in the seeing-limited NB3420 image, we cannot confirm MD34 as having a robust LyC detection.  Because the interpretation of this case is ambiguous, we do not include MD34 in the final sample of LyC emitters.

\begin{figure}
\epsscale{1.0}
\plotone{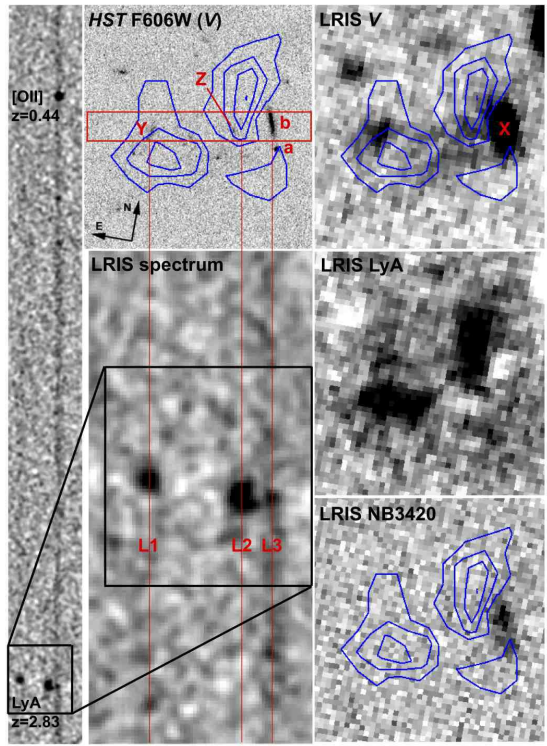}
\caption{
\small
$HST$ and LRIS Imaging for \emph{lae2436} is displayed (9\secpoint5 $\times$ 9\secpoint5), along with the LRIS spectrum probing the rest-frame UV at $z\sim 2.85$.  Along the right-hand column of the figure, the morphology of the \lya-blob in the vicinity of \emph{lae2436} is shown in LRIS $V$ (non-ionizing UV continuum), LRIS continuum-subtracted NB4670$-V$ (indicating \lya\ emission), and LRIS NB3420 (LyC emission).  To the left of the LRIS $V$ image, the higher resolution $HST$ $V_{606}$ image is shown.  Blue contours indicate the location of the \lya\ emission based on the NB4670$-V$ image.  The red rectangle in the $HST$ $V_{606}$ image shows the location of the 1\secpoint2 slit, and the LRIS 2D spectrum below is aligned to match in the orientation and spatial scale of the imaging.  The zoomed-in portion of the 2D spectrum shows the \lya\ emission (L1, L2, L3), and the full 2D spectrum is displayed on the left.  In the original LRIS imaging, only two galaxies were visible near the \lya-blob (X and Y).  These galaxies were identified as LAEs based on their $V-$NB4670 colors and the \lya\ emission in the LRIS spectrum.  As discussed in Section \ref{sec:fcontam}, Galaxy Z was first identified in the $HST$ imaging and is likely associated with the \lya\ emission in the LRIS spectrum.  Galaxy X, meanwhile, can be distinguished as two segments in the $HST$ image (a and b) that were both identified spectroscopically as foreground contaminants through reexamination of the available LRIS spectra.  The [OII] ($z=0.44$) emission from \emph{lae2436b} is visible in the top of the full 2D spectrum.  We conclude that there is no NB3420 emission associated with \emph{lae2436}.
\label{fig:nb2436} }
\end{figure}

For an additional object, \emph{lae2436}, the presence of an extended \lya\ blob \citep{steidel00,steidel11} near the position of the LAE made the original analysis of the LRIS imaging and spectra difficult.  The deep, high-resolution $HST$ imaging helped us clarify the interpretation of this object and determine that the NB3420 detection is associated with a foreground contaminant.  Figure \ref{fig:nb2436} shows the complex morphology of \emph{lae2436} in both the LRIS $V-$band and the continuum-subtracted \lya\ images.  The \lya\ blob extends over more than 5\secnopoint, and in the LRIS $V-$band image two bright galaxies (X and Y) appear to be in the vicinity of the \lya\ blob, with diffuse emission between them.  As shown in Figure \ref{fig:nb2436}, both Galaxy X and Galaxy Y exhibit \lya\ emission in the LRIS spectrum (L3 and L1, respectively) and were identified as LAEs in M13.  Galaxy X was originally identified as the LAE \emph{lae2436} ($z=2.832$), and this galaxy is clearly detected in the NB3420 image.  There are no other NB3420 detections nearby.  Finally, there is a \lya\ emission line (L2) in the LRIS spectrum coincident with the diffuse emission between Galaxies X and Y in the LRIS $V-$band image.  This diffuse emission was originally attributed to the presence of the \lya\ blob and thought to be associated with Galaxy X, as L2 and L3 have nearly identical wavelengths.

In the $HST$ imaging, much additional substructure is revealed in the vicinity of the \lya\ blob.  First, it becomes clear that the diffuse emission in the LRIS $V-$band image is due to several unresolved, faint galaxies in close proximity.  One of these galaxies (Z) is roughly coincident with the L2 \lya\ emission line in the LRIS spectrum, and the SED fit to this galaxy using $HST$ photometry places it near $z=2.832$, the redshift of the emission line.\footnote{We note that Galaxy Z has a unique and extremely red SED, with $V_{606}-J_{125}=2.08$ and $J_{125} - H_{160}=1.18$.  This galaxy is a sub-millimeter source that will be further described in Steidel et al., in prep.}  None of the SEDs for other galaxies in the vicinity of the L2 emission line demonstrate the typical features of $z\sim2.85$ galaxies.  As for object X, originally identified as \emph{lae2436}, it can be distinguished as two separate galaxies in the $HST$ image (a and b), both of which are detected in NB3420.  Subsequent reanalysis of the available spectra near the \lya\ blob allowed us to confirm spectroscopic redshifts for both of these galaxies ($z=2.04$ for \emph{lae2436a}, $z=0.44$ for \emph{lae2436b}), and the SED fits to both objects are consistent with their spectroscopic redshifts.  In particular, we draw attention to the spectrum of \emph{lae2436b}, visible in Figure \ref{fig:nb2436} on the right-hand side of the LRIS spectrum, coincident with L3.  In the zoomed-out version of the spectrum, a spurious emission line is visible at 5375\AA\ which we have identified as [O{\sc II}] at $z=0.44$.  The \lya\ emission line originally identified for this object (L3) must be due to extended \lya\ emission from the \lya-blob.  In conclusion, as both \emph{lae2436a} and \emph{lae2436b} are at $z<2.82$, the NB3420 filter does not probe LyC emission for these galaxies and thus the NB3420 detections associated with both galaxies are foreground contamination.

Finally, we found that one object in the spectroscopic LAE sample, \emph{lae7890}, was misidentified as an LAE.  This object was presented in the Appendix of M13 as a faint LAE with a borderline color excess ($V-$NB4670 $=0.70$) and a possible LyC detection, and was not analyzed with the main LAE sample for which a NB4670 magnitude limit of $m_{4670} \leq 26$ was imposed.  A marginal emission line had been identified for this object, placing it at the spike redshift of $z=2.85$, but the SED (similar to that of the contaminant in Figure \ref{fig:SED_samples}) indicates unequivocally that \emph{lae7890} is at low redshift, in the range of $1<z<2$.  We considered [O{\sc II}] as a possible source of the emission line, but that would place \emph{lae7890} at $z=0.26$, which is also inconsistent with the observed SED.  Thus we conclude that this faint emission line is either spurious, or is possibly consistent with [C{\sc III}] 1907\AA\ / C{\sc III}] 1909\AA\ emission from a galaxy at $z=1.45$.  In either case, the NB3420 emission is not LyC.

In summary, 11/16 candidate LyC emitters in our sample show obvious signs of foreground contamination at the position of the NB3420 detection.  For 9 objects, there is a bona fide $z=2.85$ galaxy at the position of the LAE, with an additional foreground galaxy offset from the LAE and associated with the NB3420 emission.  For 2 objects, $z=2.85$ was erroneously assigned to the candidate LAE, and, again, the NB3420 emission is actually non-ionizing UV flux from a low-redshift contaminant.  In all of these cases, the available evidence suggests that we are not observing LyC emission at $z=2.85$.

\subsubsection{Ambiguous Cases} \label{sec:ambig}

For the four photometric LAE candidates with NB3420 detections, the SED shape was ambiguous and there were no spectroscopic redshifts available to confirm that the objects are indeed at $z\sim2.85$.  These two factors make it impossible to confidently claim a LyC detection for any of these objects.  We discuss the photometry and SED fits for these objects in detail in Appendix A, and summarize the results here.  One object, \emph{lae4070}, has similar $J_{125}-H_{160}$ and $V_{606}-J_{125}$ colors to many $z\sim2.85$ galaxies in our sample and is the most promising photometric LAE candidate for true LyC emission.  The other three objects (\emph{lae5200}, \emph{lae6510}, and \emph{lae7180}) display the ambiguous SED shape described in Section \ref{sec:sed}, which may describe galaxies at many redshifts.  As we cannot unambiguously determine whether or not the four photometric LAE candidates discussed in this section are truly at $z\sim2.85$, we adopt a conservative approach and do not count the NB3420 detections for these objects as secure signatures of leaking LyC radiation.

\begin{figure*}
\epsscale{1.0}
\plotone{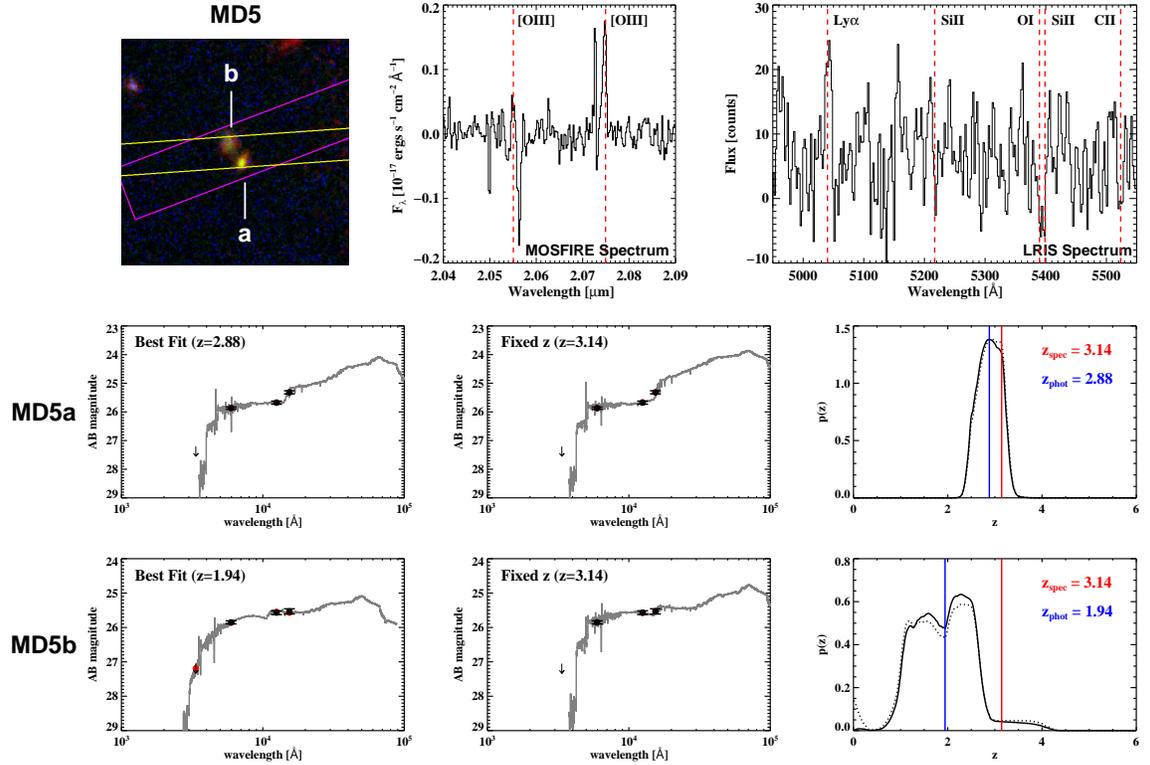}
\caption{
\small
Imaging, spectra, photometry, and SED fits for MD5.  The top left panel shows the $U_{336}V_{606}J_{125}$ color-composite $HST$ image for MD5, indicating the two sub-arcsecond components (MD5a and MD5b).  The locations of the MOSFIRE (yellow) and LRIS (magenta) slits are overlaid on the image.  While both MD5a and MD5b fall within the MOSFIRE and LRIS slits, it is not possible to distinguish between the clumps in the spectra due to the slit orientation and seeing (0\secpoint6 for MOSFIRE, 0\secpoint7-1\secpoint0 for LRIS).  The 1D MOSFIRE and LRIS spectra of MD5 are displayed to the right of the image, showing superimposed emission from both sub-arcsecond components.  We detect the [O{\sc III}] doublet (4959\AA, 5007\AA; $z=3.1426$) in the MOSFIRE $K-$band spectrum, and \lya\ emission ($z=3.147$) along with interstellar absorption lines (1303\AA (O{\sc I} + Si{\sc II}), 1334\AA (C{\sc II}); $z=3.139$) in the LRIS spectrum.  The interstellar absorption lines are blueshifted relative to the \lya\ and [O{\sc III}] emission, indicative of an outflow.  Below the spectra and imaging are shown the EAZY output for MD5a and MD5b.  In the left-most panel, the redshift is allowed to float during SED fitting, while in the middle panel the redshift is fixed to $z=3.14$, the redshift indicated by the spectra.  The right-most panel shows the redshift probability distribution.  Colors and symbols are as in Figure \ref{fig:SED_samples}.  MD5a has the typical SED shape associated with $z\sim3$ galaxies in our sample.  MD5b, which is associated with the NB3420 detection, exhibits the ambiguous SED shape discussed in Section \ref{sec:sed}.  As there is no evidence of foreground contamination in either the MOSFIRE or LRIS spectra, we propose that both of these clumps are at the spectroscopic redshift 3.14 and that the NB3420 emission associated with MD5b is true LyC emission.  We note that the LRIS and MOSFIRE spectra rule out strong emission lines in several redshift ranges, including H$\alpha$ emission between $1.97<z<2.65$, \lya\ emission at $z>1.5$, [O{\sc II}] emission at $z<0.9$, and [O{\sc III}] emission at $z<0.4$.
\label{fig:MD5} }
\end{figure*}

\subsubsection{LyC Emission from MD5}\label{sec:MD5cand}

The best candidate for true LyC emission is the LBG MD5, which has a spectroscopic redshift of $z=3.14$ confirmed by spectra from both LRIS and the MOSFIRE near-IR multi-object spectrograph (see Figure \ref{fig:MD5}).  The LRIS spectrum was taken in May 2011 (M13) and shows \lya\ emission ($z=3.147$), along with multiple absorption features (C{\sc II}, Si{\sc II}, and O{\sc I}; $z=3.139$).  The $K-$band MOSFIRE spectrum was acquired in June 2012, and contains detections of both [O{\sc III}] emission lines (4959\AA, 5007\AA; $z=3.1426$).  Unfortunately, we were unable to measure the $z=3.14$ H$\beta$ emission line in this spectrum because it falls on a sky line.  In the $HST$ imaging, MD5 is composed of two clumps separated by 0\secpoint58 along a direction 32 degrees East of North (MD5a and MD5b; see Figure \ref{fig:MD5}); only MD5b is coincident with the NB3420 emission.  The orientation of the MOSFIRE and LRIS observations (slit PAs of 274 and 111 degrees, respectively; see Figure \ref{fig:MD5}) were such that both clumps fell within the spectroscopic slits.  Although it is not possible to distinguish between the clumps in the spectra due to slit orientation and seeing (0\secpoint6 for MOSFIRE, 0\secpoint7-1\secpoint0 for LRIS), neither spectrum shows evidence for spurious emission or absorption features that would indicate the presence of a low-redshift interloper.

Ideally, as in the case of \emph{Ion1} in \citet{vanzella12}, a candidate LyC-emitting galaxy would have a simple, compact morphology.  In such a case, the probability of a foreground interloper is negligible.  However, it has been shown that high-redshift galaxies typically exhibit clumpy morphologies \citep[see, e.g.,][]{law07,lotz06}.  Of the 35 LBGs in our $z \geq 2.82$ sample with imaging in $V_{606}$, only 20\% have simple, compact morphologies.  Close inspection reveals clumpy morphologies in all other cases, and several LBGs are comprised of clumps with offsets significantly greater than the 0\secpoint58 offset of the MD5 clumps.  For example, LBGs M23, MD9, and C13 are comprised of clumps suggested by SED fits to be at the LBG redshift with offsets of 0\secpoint91, 1\secpoint00, and 1\secpoint19, respectively.  As the majority of LBGs consist of several associated clumps at the same redshift, the fact that MD5 displays multiple clumps does not necessarily indicate that MD5b (the clump associated with the NB3420 detection) is a foreground contaminant.

However, the presence of the second clump opens the possibility that one of these clumps is a low-redshift interloper, and thus we examine the SED fits to MD5a and MD5b for evidence of foreground contamination (see Figure \ref{fig:MD5}).  Both clumps are nearly identical in $V_{606}$ magnitude, and are within 0.1-0.2 magnitudes in $J_{125}$ and $H_{160}$.  MD5a has the SED shape typical of most $z \sim 3$ galaxies in our sample and is almost certainly at the spectroscopic redshift $z=3.14$.  MD5b has the ambiguous SED shape described in Section \ref{sec:sed}.  As the spectroscopic redshift of MD5 is higher than most galaxies in our sample, the $H_{160}$ filter actually falls right at the location of the Balmer break, rather than redwards of the break.  Thus, the $H_{160}$ filter is partially contaminated by flux bluewards of the Balmer break, and $J_{125}-H_{160}$ no longer probes the full strength of the break.  Therefore, the fact that MD5b has a smaller $J_{125}-H_{160}$ color than the lower-redshift LBGs in our sample does not necessarily indicate a young age or low-redshift contaminant.

Figure \ref{fig:colorcolor} shows MD5b plotted with respect to other galaxies in our sample in $V_{606}-J_{125}$ vs. $J_{125}-H_{160}$.  If MD5b is truly at $z=3.14$, it has a very small Balmer break ($J_{125}-H_{160} = 0.04 \pm 0.13$), a red UV slope ($V_{606}-J_{125} = 0.28 \pm 0.10$, equal to the mean of the LBG sample), and a $U_{336}$ detection brighter than expected by standard stellar population synthesis models (unsurprising for a LyC emitter).  If MD5b is a foreground interloper, EAZY estimates that the most likely redshifts for the interloper are $0<z<0.5$ or $2<z<2.5$ using P\'{E}GASE models, and between $1<z<2.8$ using reddened BPASS models.  In both cases, $z\sim 2.3-2.4$ is the most likely contaminant redshift.  

The strongest evidence that MD5b is indeed at $z=3.14$ is the lack of spurious emission or absorption lines in both the LRIS and MOSFIRE spectra.  Because these spectra span different wavelength ranges (3100$-$7000 \AA\ for LRIS and 1.95$-$2.4 $\mu$m for MOSFIRE), we can rule out strong emission lines for several redshift ranges.  These include H$\alpha$ emission between $1.97<z<2.65$, \lya\ emission at $z>1.5$, [O{\sc II}] emission at $z<0.9$, and [O{\sc III}] emission at $z<0.4$.  We note that the lack of spurious emission lines in the $K-$band MOSFIRE spectrum rules out H$\alpha$ emission right in the redshift range predicted by EAZY to be the most probable redshift of a contaminant ($2<z<2.6$).  It might be possible to confirm the redshift of MD5b using MOSFIRE observations with the slit oriented along the axis connecting MD5a and MD5b.  These observations would maximize the distance between the two clumps (separated by 0\secpoint58) and, if taken under conditions of good seeing, potentially distinguish emission from each clump individually.

\section{Properties of the Lyman-Continuum Emitter MD5}\label{sec:MD5}

One of the main goals of this work is to investigate the multiwavelength properties of galaxies with and without LyC emission, in order to better understand the mechanism of LyC photon escape from galaxies.  We are also interested in investigating any systematic differences between galaxies with and without LyC emission, for such differences may facilitate the search for LyC-emitting galaxies both during and after the epoch of reionization.  As our $HST$ data has left us with only one robust candidate for LyC emission (MD5), we here discuss the morphological properties and best-fit stellar population of this object with respect to the properties of typical LBGs.  

\subsection{Morphology of MD5}\label{sec:MD5morph}

Figure \ref{fig:LBG_stamps} displays imaging of MD5 in all available bands.  Morphologically, MD5 is composed of two clumps (MD5a, MD5b) separated by 0\secpoint58 (4.4 kpc at $z=3.14$).  The $V_{606}$ magnitudes for each clump (representing the non-ionizing UV continuum) are $m_{606}^{MD5a} = 25.87 \pm 0.04$ and $m_{606}^{MD5b} = 25.85^{+0.05}_{-0.04}$.  In Section \ref{sec:MD5cand} we present arguments for why both of these clumps are likely at the spectroscopic redshift of $z=3.14$ and why foreground contamination is unlikely.  The clumpy morphology of MD5 is similar to that of many other LBG systems, which commonly exhibit significant substructure.  MD5b, which is more diffuse and lower in surface brightness than MD5a, is the clump associated with the NB3420 detection (i.e., the LyC emission).

Because of the particularly high redshift of MD5, we are able to directly map the LyC emission in the high-resolution $HST$ $U_{336}$ image.  While most galaxies in our sample had redshifts of roughly $z\sim2.85$ and the $U_{336}$ filter was partially contaminated by non-ionizing flux redwards of the Lyman limit, MD5 is at high enough redshift ($z=3.14$) that the $U_{336}$ filter probes the LyC spectral region without any contamination.  While MD5b is formally undetected in $U_{336}$ at 3$\sigma$, emission at the location of MD5b is visible by eye in the $U_{336}$ imaging.  MD5b has a 2.25$\sigma$ detection in $U_{336}$ ($m_{336}=27.37^{+0.64}_{-0.40}$), which is consistent within errors of the detection in NB3420 ($m_{NB3420}=26.89^{+0.43}_{-0.31}$).\footnote{We note that the $U_{336}$ filter is wider than NB3420, and thus the fainter $U_{336}$ magnitude may be due to increased IGM attenuation within the bluer half of the $U_{336}$ filter.}

In light of models in which LyC emission may escape anisotropically from galaxies \citep[e.g.,][]{gnedin08,zackrisson13}, we examined the offset between the centroid of the $U_{336}$ and $V_{606}$ emission for MD5b in order to determine if there was a significant offset between the ionizing and non-ionizing UV emission.  We measured a value for this offset of $\Delta_{UV}$ = 0\secpoint08.   As there are no additional $z\sim3$ galaxies in our sample with $U_{336}$ detections to compare to, we examined the distribution of $U_{336}-V_{606}$ offsets for the foreground contaminants in our sample.  In this way, we measured $\Delta_{UV}$ for objects where $U_{336}$ and $V_{606}$ are both probing the non-ionizing continuum, and thus should not demonstrate significant offsets.  For the contaminant sample, we found a roughly flat distribution of offsets between 0\secpoint0 and 0\secpoint12 with a mean offset of 0\secpoint065 and a standard deviation of 0\secpoint031.  As MD5b has an offset consistent with the mean of the contaminant distribution, we conclude that its measured offset is not significant.  This lack of significant offset implies either that LyC emission is escaping isotropically from MD5b, or that, if LyC emission escapes only from cleared holes in the ISM, MD5b must be oriented such that the opening is along our line of sight.

\subsection{ISM Kinematics of MD5}\label{sec:outflow}

The spectral features of MD5 shed light on the kinematics of its ISM.  The [O{\sc III}] $\lambda 5007$ nebular emission line observed in the MOSFIRE spectrum of MD5, which indicates the systemic redshift, places MD5 at $z=3.1426$.  This line has an intrinsic width of $\sigma_{v}=37$ km s$^{-1}$, typical of LAEs, but half that of typical LBGs \citep[Trainor et al., in prep;][]{pettini01}.  The low-ionization interstellar absorption features observed in the LRIS spectrum of MD5 (O{\sc I} + Si{\sc II} $\lambda 1303$, C{\sc II} $\lambda 1334$) are consistent with a redshift of $z=3.139$, blueshifted relative to the [O{\sc III}] emission.  The magnitude of this blueshift corresponds to a velocity offset of $\Delta v_{IS}\sim280$ km s$^{-1}$, higher than the median $\Delta v_{IS}$ for LBGs \citep[150 km s$^{-1}$;][]{shapley03}, but fairly uncertain due to the low signal-to-noise of the LRIS spectrum.  Additionally, the redshift derived from the centroid of the \lya\ emission line ($z=3.147$) corresponds to a velocity offset of $\Delta v_{Ly\alpha}\sim300$ km s$^{-1}$, which is typical of LBGs \citep[$\Delta v_{Ly\alpha}^{LBG}=360$ km s$^{-1}$;][]{shapley03}.  Thus, in terms of its kinematics, MD5 does not stand out significantly with respect to the full population of LBGs, although the blueshift of its interstellar absorption lines is higher than average.  We note that with the spatial resolution of our spectra, we cannot resolve the two components MD5a and MD5b separately and evaluate their individual kinematics.  Finally, we note that our measured value of $\Delta v_{Ly\alpha}$ is inconsistent with predictions for LyC-leaking galaxies from \citet{verhamme15}, who find small offsets between \lya\ emission and the systemic redshift ($\Delta v_{Ly\alpha} \leq 150$ km s$^{-1}$) in models of galaxies with leaking LyC radiation.

\subsection{Stellar Populations of MD5}\label{sec:MD5sp}

In addition to providing high-resolution imaging of MD5, the multiwavelength $HST$ data also enabled us to fit the photometry for both sub-arcsecond components (MD5a and MD5b) with stellar population synthesis (SPS) models.  While EAZY is a powerful tool for estimating photometric redshifts, the program does not directly provide stellar population information.  In order to examine the stellar populations of the galaxies in our sample, we employed the stellar population fitting code FAST \citep{kriek09}, which accommodates a different set of stellar population synthesis models and dust extinction prescriptions from those of EAZY.  To model the photometry, we fixed the redshift to the spectroscopically measured value and used \citet{bc03} models ranging in age from 50 Myr to 2 Gyr with delayed-$\tau$ star-formation histories (SFR $\propto te^{-t/\tau}$) and a \citet{chabrier03} initial mass function.  We chose delayed-$\tau$ star-formation histories because of their flexibility in accommodating both rising and falling star-formation histories, and we note that we found no significant qualitative differences in the derived stellar population parameters when experimenting with constant or rising star-formation histories.  The lower age limit of 50 Myr is adopted to reflect the LBG dynamical timescale, following \citet{reddy12}.  However, this age limit is conservative and may be larger than necessary given the small sizes of some galaxy subcomponents.  For dust extinction, we employed the \citet{calzetti00} attenuation curve.  We adopted solar metallicity for the models because, with only four photometric points, we did not have enough data to constrain metallicity.  By performing tests with several values for fixed metallicity and with metallicity as a free parameter, we verified that metallicity does not have a significant effect on the final fit.  We note that even when metallicity was allowed to float, the best-fit model for MD5b had solar metallicity.

\begin{figure}
\epsscale{1.0}
\plotone{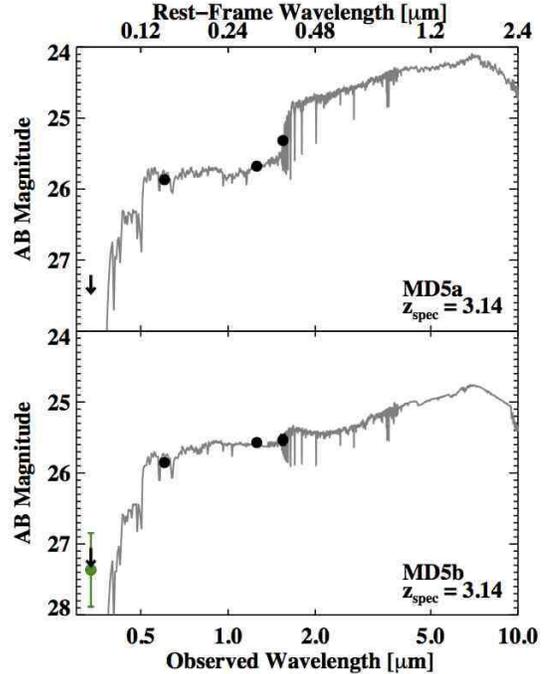}
\caption{
\small
Best-fit \citet{bc03} stellar population synthesis models as computed by FAST (gray line) to $HST$ $U_{336}V_{606}J_{125}H_{160}$ photometry for MD5a and MD5b (black circles), fit at the spectroscopic redshift of $z=3.14$.  One-sigma photometric uncertainties are smaller than the data points.  For MD5b, the $2.25\sigma$ detection in $U_{336}$ (probing the LyC spectral region at $z=3.14$ with no contamination redwards of the Lyman limit) is plotted in green, with the formal $3\sigma$ limit indicated by the black arrow.  We note that while the $U_{336}$ data point for MD5b does not agree with the model prediction, this is to be expected from \citet{bc03} models, which do not have significant emission bluewards of 912\AA.  Parameters of the fits are listed in Table \ref{tab:fast_vals}.
\label{fig:MD5b_fast} }
\end{figure}

We display the best fit models and photometry for MD5a and MD5b in Figure \ref{fig:MD5b_fast}.  We note that while the $U_{336}$ data point for the LyC-emitter MD5b does not agree with the model prediction, this is to be expected from \citet{bc03} models, which do not have significant emission bluewards of 912\AA.  We also report the best-fit values and 68\% confidence intervals for $\tau$, stellar mass, star-formation rate (SFR), dust extinction, and age.  For MD5a, the clump without leaking LyC emission, we found: log($\tau$) = $8.6_{-0.6}^{+1.4}$, log(Mass [M$_{\odot}$]) = $9.70_{-0.40}^{+0.25}$, log(SFR [M$_{\odot}$ yr$^{-1}$]) = $0.51_{-0.14}^{+0.43}$, $ E(B-V) = 0.02_{-0.02}^{+0.10}$, and log(Age [yr]) = $9.10_{-0.67}^{+0.20}$.  In constrast, for MD5b, the clump with leaking LyC emission, we found: log($\tau$) = $8.2_{-0.2}^{+1.8}$, log(Mass [M$_{\odot}$]) = $8.69_{-0.10}^{+0.49}$, log(SFR [M$_{\odot}$ yr$^{-1}$]) = $1.31_{-0.47}^{+0.09}$, $ E(B-V) = 0.17_{-0.09}^{+0.03}$, and log(Age [yr]) = $7.70_{-0}^{+1.05}$\footnote{The confidence interval is bounded by log(Age [yr]) = 7.70 (age = 50 Myr) because that is the minimum age we enforce upon the models.}.

The fits to MD5a and MD5b describe two very different stellar populations.  MD5a has an old stellar population with significant stellar mass, a low star-formation rate, and very little reddening.  MD5b, however, is young and low-mass, with a much higher star-formation rate and a larger $E(B-V)$.  While we have set a minimum LBG age limit of 50 Myr following \citet{reddy12}, the best-fit model to MD5b without a minimum age requirement is even younger (10 Myr).  Because of the differing stellar populations of MD5a and MD5b, it is possible that they are two distinct galaxies in the process of merging.  It is also possible that they are simply two sub-regions of the same galaxy, one of which (MD5b) is undergoing a recent burst of star formation that has greatly increased its non-ionizing UV and LyC fluxes, making it more likely to detect LyC emission.  

In addition to modeling the stellar populations of MD5a and MD5b individually, we used FAST to model the best-fit stellar population to the combined photometry of both components.  While such modeling does not represent a physically meaningful stellar population\footnote{For example, the mass and SFR derived from the best-fit model to the combined photometry of MD5 are less than the summed individual masses and SFRs of MD5a and MD5b.}, it facilitates a comparison to the results of ground-based LBG surveys, recreating the flux that would be measured by instruments lacking the high resolution of $HST$.  For the combined photometry of MD5, we found: log($\tau$) = $8.2_{-0.2}^{+1.8}$, log(Mass [M$_{\odot}$]) = $9.52_{-0.53}^{+0.54}$, log(SFR [M$_{\odot}$ yr$^{-1}$]) = $1.13_{-0.47}^{+0.58}$, $ E(B-V) = 0.1_{-0.1}^{+0.11}$, and log(Age [yr]) = $8.5 \pm 0.8$.

\begin{figure*}
\epsscale{1.0}
\plotone{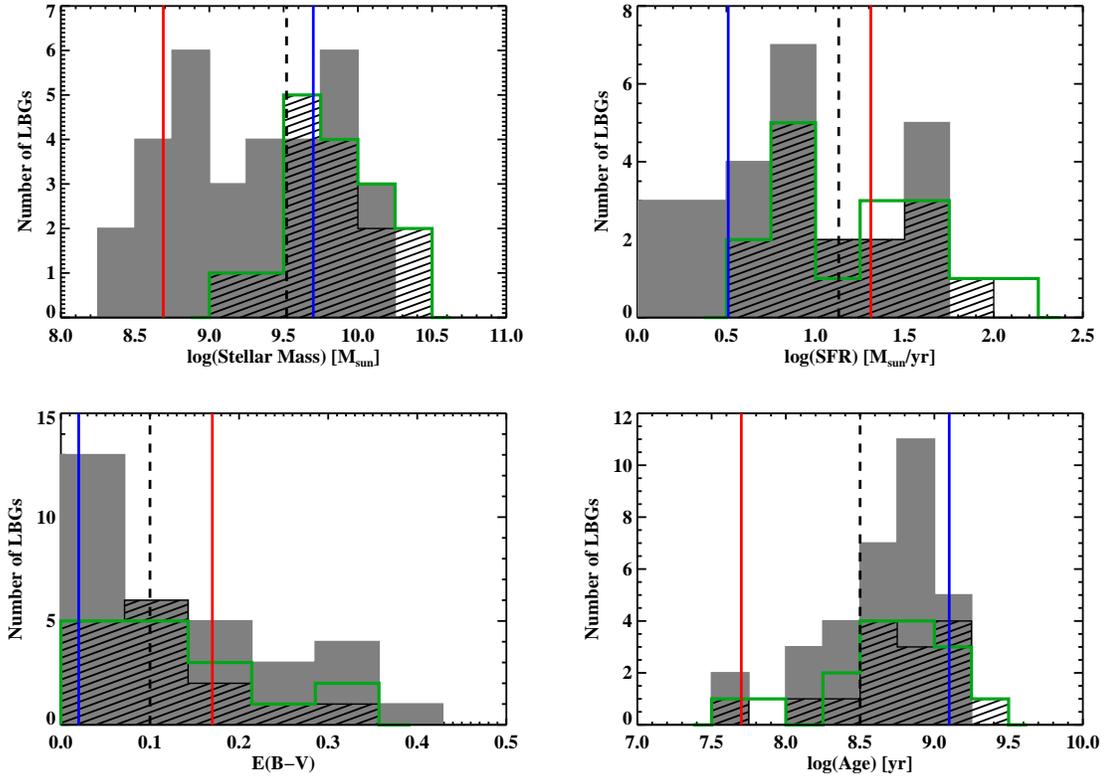}
\caption{
\small
Histograms showing the stellar population parameters of MD5 calculated by FAST with respect to the distribution of parameters for all LBGs with $HST$ $U_{336}V_{606}J_{125}H_{160}$ imaging.  MD5b (the LyC-emitting component) is indicated by the red vertical line, MD5a is indicated by the blue vertical line, and the combined photometry for MD5a and MD5b is indicated by the black dashed vertical line.  Gray filled histograms indicate parameters derived from FAST fits to individual sub-arcsecond components of LBGs.  Black hashed histograms indicate parameters derived from FAST fits to combined photometry, mimicking ground-based studies.  Green open histograms also indicate parameters derived from fits to combined photometry, but include known foreground contaminants in cases where the contaminant was indistinguishable from the LBG in the ground-based LRIS imaging.  The similarities between the green and black histograms show that the inclusion of foreground contaminants does not significantly alter the sample-averaged LBG properties.  MD5b stands out as having an age in the youngest 10\% of the LBG sample.
\label{fig:fast_hist} }
\end{figure*}

\begin{deluxetable*}{lcccc} 
\tablewidth{0pt} 
\footnotesize
\tablecaption{Stellar Population Parameters\label{tab:fast_vals}}
\tablehead{
\colhead{Sample} &
\colhead{log(Stellar Mass)} &
\colhead{log(SFR)} &
\colhead{$E(B-V)$\tablenotemark{a}} &
\colhead{log(Age)}
\cr \colhead{} & \colhead{[M$_{\odot}$]} & \colhead{[M$_{\odot}$ yr$^{-1}$]} & \colhead{[mag]} & \colhead{[yr]} 
}
\startdata 
MD5a\tablenotemark{b} & $9.70_{-0.40}^{+0.25}$ & $0.51_{-0.14}^{+0.43}$ & $0.02_{-0.02}^{+0.10}$ & $9.10_{-0.67}^{+0.20}$ \\
MD5b (LyC detected)\tablenotemark{b} & $8.69_{-0.10}^{+0.49}$ & $1.31_{-0.47}^{+0.09}$ & $0.17_{-0.09}^{+0.03}$ & $7.70_{-0}^{+1.05}$ \\
MD5 combined photometry\tablenotemark{b} & $9.52_{-0.53}^{+0.54}$ & $1.13_{-0.47}^{+0.58}$ & $0.10_{-0.10}^{+0.11}$ & $8.50_{-0.80}^{+0.80}$ \\
LBG sub-arcsecond components\tablenotemark{c} & $9.26^{+0.72}_{-0.53}$ & $0.75^{+0.76}_{-0.76}$ & $0.10^{+0.20}_{-0.10}$ & $8.80^{+0.20}_{-0.50}$ \\
LBG combined photometry\tablenotemark{c} & $9.77^{+0.58}_{-0.26}$ & $1.05^{+0.59}_{-0.22}$ & $0.10^{+0.12}_{-0.07}$ & $8.80^{+0.40}_{-0.40}$ \\
LBG combined photometry + contaminants\tablenotemark{c} & $9.84^{+0.22}_{-0.19}$ & $1.29^{+0.37}_{-0.41}$ & $0.10^{+0.15}_{-0.05}$ & $8.80^{+0.30}_{-0.40}$ \\ 
$2.7<z<3.7$ LBGs from \citet{reddy12}\tablenotemark{e} & $9.78_{-0.39}^{+0.47}$ & $1.65_{-0.49}^{+0.40}$ & $0.18_{-0.10}^{+0.09}$ & $8.05_{-0.35}^{+0.90}$
\enddata
\tablenotetext{a}{Derived from $A_{V}$, assuming $k(V)=k(5500)=4.048$ from \citet{calzetti00}.}
\tablenotetext{b}{Best fit values from FAST assuming a delayed-$\tau$ star-formation history, a minimum age of 50 Myr, and \citet{calzetti00} attenuation.  Uncertainties quoted are 68\% confidence intervals.}
\tablenotetext{c}{Median values for LBG samples are quoted, along with values bracketing the inner 68\% of the distributions.}
\tablenotetext{e}{Best-fit stellar population parameters for $2.7<z<3.7$ LBGs from \citet{reddy12}, assuming a constant star-formation history, a minimum age of 50 Myr, and \citet{calzetti00} attenuation.  Median values are quoted, along with values bracketing the inner 68\% of the distributions.}
\end{deluxetable*}

In order to compare the derived properties of MD5a, MD5b, and the combined photometry with those of typical LBGs, we used FAST to fit SPS models to the remainder of the LBGs in our sample and examined their stellar masses, star-formation rates, dust extinction, and ages.  We performed this analysis both for individual sub-arcsecond components of LBGs, for the combined fluxes from all components of each LBG (simulating ground-based studies that are free of contamination), and for a combined-flux sample that also includes foreground contaminants in cases where the contaminant was indistinguishable from the LBG in the ground-based LRIS imaging (a fair simulation of ground-based studies).  Figure \ref{fig:fast_hist} shows histograms of these three LBG samples, along with the best-fit values for MD5a, MD5b, and the combined photometry for MD5.  MD5a and MD5b clearly have distinct stellar populations from each other, and MD5b, the LyC-emitting component, stands out as being among the youngest 10\% galaxies in the $HST$ LBG sample.

For all the LBGs in our sample, we now compare the stellar population fits to individual clumps with fits to the combined photometry.  As expected, we find that model fits to individual galaxy components generally yield smaller stellar masses and SFRs than fits to the combined photometry.  Also, individual sub-arcsecond components exhibit a wider range of reddening values than do galaxies with combined photometry, although the median reddening value is the same for both samples.  There is no significant difference in the median derived ages between the individual clumps and the combined photometry.  Finally, we find that the addition of foreground contaminants does not significantly alter the sample-averaged LBG properties, as foreground contaminants close to LBGs are rare in the LBG sample without NB3420 detections.  We conclude that occasional foreground contaminants superimposed upon LBGs do not greatly affect the stellar populations derived for galaxies in ground-based LBG surveys.

\begin{figure*}
\epsscale{1.0}
\plotone{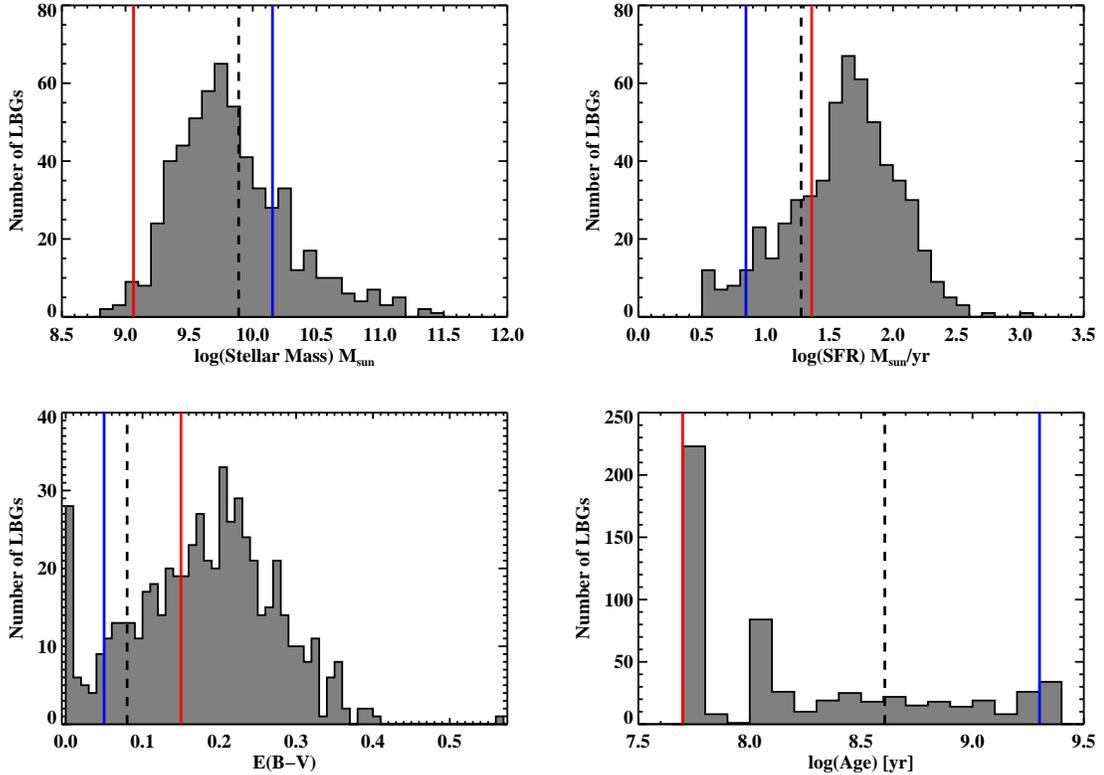}
\caption{
\small
Histograms showing the stellar population parameters of MD5 calculated by FAST with respect to the distribution of parameters for the $2.7<z<3.7$ LBG sample of \citet{reddy12}.  MD5b (the LyC-emitting component) is indicated by the red vertical line, MD5a is indicated by the blue vertical line, and the combined photometry for MD5a and MD5b is indicated by the black dashed vertical line.  All parameters plotted are derived using the stellar population fitting methods described in \citet{reddy12}, with assumptions of constant star formation histories, a minimum age of 50 Myr, and \citet{calzetti00} dust attenuation.  While the properties of the composite object MD5 (containing both MD5a and MD5b) do not stand out among other LBGs in the \citet{reddy12} sample, the LyC-emitting component MD5b stands out as being in the youngest third of the sample.
\label{fig:reddy} }
\end{figure*}

In order to compare the properties of MD5 with those of a much larger parent sample of LBGs, we consider the set of 570 LBGs at $2.7<z<3.7$ from the ground-based survey of \citet{reddy12}.  In Figure \ref{fig:reddy}, we display parameters of the stellar population fit to MD5a, MD5b, and the combined photometry with respect to the LBGs from \citet{reddy12}.  The stellar population parameters we display for the \citet{reddy12} LBGs have been derived from the latest solar metallicity models of S. Charlot \& G. Bruzual, using constant star formation histories and a minimum age limit of 50 Myr.  The median parameters of the \citet{reddy12} LBGs are consistent with those of our combined-photometry LBG sample, although the percentage of young galaxies ($<$100 Myr) in our $HST$ sample is less than that of the \citet{reddy12} sample.\footnote{The fact that we find few young galaxies ($<$100 Myr) among the $HST$-imaged LBGs and LAEs in the HS1549 field may be due to statistical variation inherent to our small sample size (40 galaxies with $U_{336}V_{606}J_{125}H_{160}$ imaging), a peculiarity of the HS1549 field, or possibly a property of all protoclusters \citep[see, e.g., ][]{steidel05}.}  

To facilitate comparison with the \citet{reddy12} LBGs, we have re-fit the stellar populations of MD5a, MD5b, and the combined photometry of MD5 using the methods described in \citet{reddy12}, employing CSF models, a minimum age of 50 Myr, and \citet{calzetti00} extinction.  These fit results for MD5 (plotted in Figure \ref{fig:reddy}) are qualitatively similar to those from our original delayed-$\tau$ fits.  The data in Figure \ref{fig:reddy} show that the fit to the combined photometry of MD5 is unremarkable when compared to the ground-based photometry of the \citet{reddy12} LBG sample: it has a typical stellar mass and age, and slightly below-average values for SFR and $E(B-V)$.  It is only when MD5a and MD5b are fit separately that the young stellar population of MD5b becomes apparent; an age of 50 Myr places it in the youngest third of the \citet{reddy12} LBG sample.  Finally, we also fit the photometry for MD5b using SMC extinction.  \citet{reddy12} found that $\sim$90\% of LBGs with Calzetti-inferred ages of $<$100 Myr had older ages ($>$100 Myr) when fit using an SMC extinction curve.  For MD5b, the best-fit model using SMC extinction had an age of 160 Myr, along with less reddening ($E(B-V) = 0.05$) than the best-fit Calzetti-attenuated model and qualitatively similar values for stellar mass and star-formation rate (log(Mass [M$_{\odot}$]) = 9.16, log(SFR [M$_{\odot}$ yr$^{-1}$]) = 0.95).  \citet{reddy12} model their full LBG sample using a combination of extinction curves, employing \citet{calzetti00} attenuation for the majority of the sample, but using SMC extinction for galaxies with Calzetti-inferred ages younger than 100 Myr.  Compared with the ages derived from these fits, the 160 Myr age estimated for MD5b is still in the youngest third of the sample.  Table \ref{tab:fast_vals} summarizes the stellar population fits to MD5 with respect to those of the non-LyC-emitting galaxies in our $HST$ LBG sample and those of the LBGs from \citet{reddy12}.

\begin{figure*}
\epsscale{1.0}
\plotone{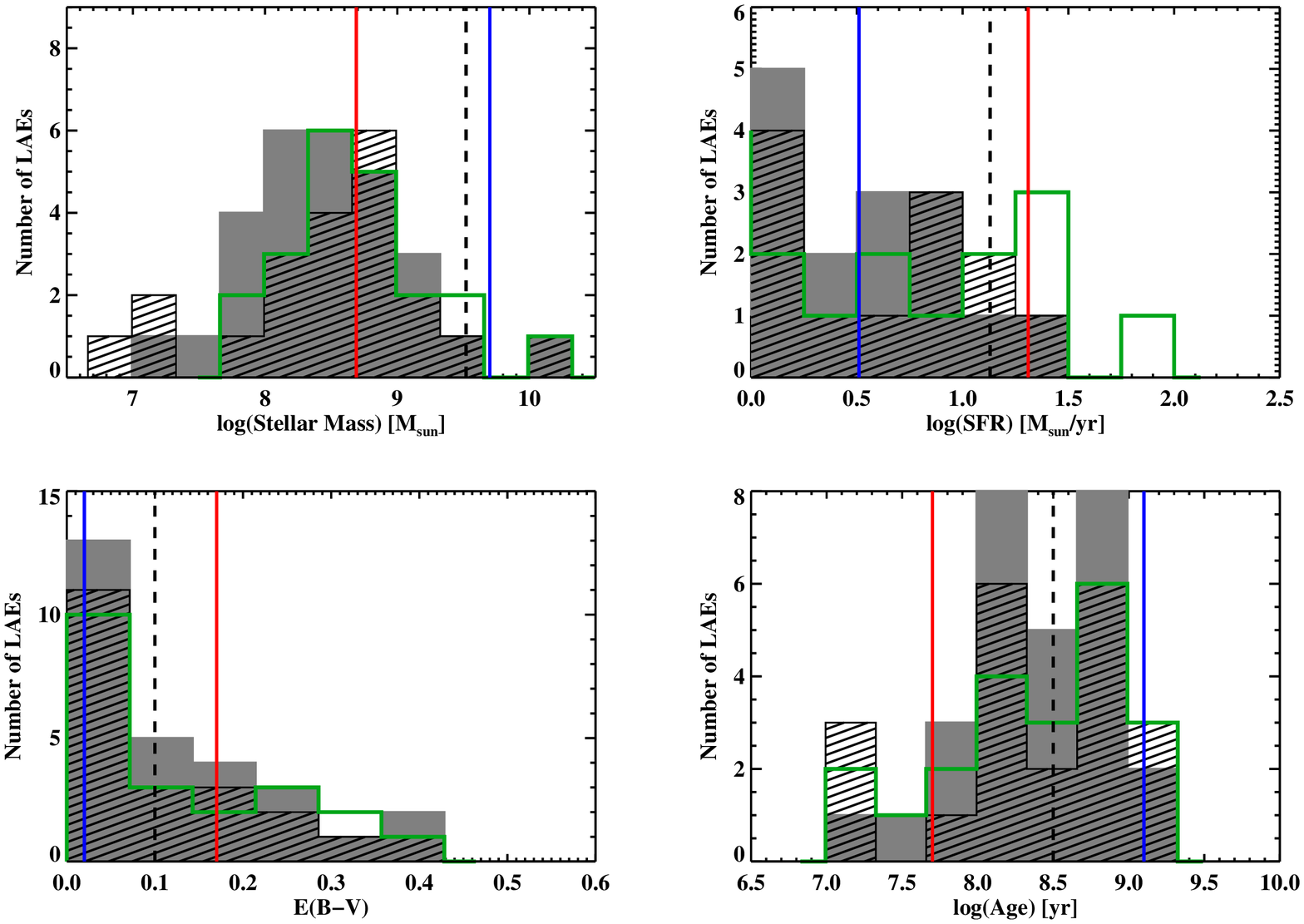}
\caption{
\small
Histograms showing the stellar population parameters of MD5 calculated by FAST with respect to the distribution of parameters for all LAEs with $HST$ $U_{336}V_{606}J_{125}H_{160}$ imaging.  MD5b (the LyC-emitting component) is indicated by the red vertical line, MD5a is indicated by the blue vertical line, and the combined photometry for MD5a and MD5b is indicated by the black dashed vertical line.  Histogram colors are as in Figure \ref{fig:fast_hist}.  The bottom-right plot shows several LAEs with log(age) $<$ 8, similar to MD5, raising the question of whether or not these objects $-$ which do not have NB3420 detections $-$ may also emit LyC radiation.  However, as discussed in Section \ref{sec:MD5sp}, these LAEs are much fainter than MD5.  If their observed ionizing to non-ionizing UV flux density ratios were equivalent to that of MD5, then they would be too faint to detect in the NB3420 image.
\label{fig:fast_lae_hist} }
\end{figure*}

As the model fit to MD5b has a young age and low stellar mass, two properties typical of LAEs \citep[e.g.,][]{gawiser07,guaita11}, we also examined MD5b with respect to the LAEs in our sample, none of which exhibited LyC detections.  We modeled LAE stellar populations with FAST using the same methods as for LBGs, but in order to account for the young ages and lower metallicities associated with LAEs we set the minimum age to 10 Myr and fixed metallicity at 20\% solar.  Figure \ref{fig:fast_lae_hist} shows the distribution of stellar mass, star-formation rate, dust extinction, and age for LAEs, along with values for MD5a, MD5b, and the combined photometry of MD5.  MD5b has a stellar mass more typical of the average LAE in our sample, but a higher than average SFR.  Its age is still young compared to the LAE sample.

While the young age of MD5b is shared by several objects in our $HST$ sample, none of these additional young objects exhibit LyC detections.  In the LBG sample, two object subcomponents in addition to MD5b have ages less than 100 Myr, and five such components exist in the LAE sample.  If the young stellar population of MD5b is responsible for its LyC emission, then we might also expect LyC detections from other LBG and LAE components with similarly young ages.  Both of the young objects in the LBG sample (M16a and MD34f) have small stellar masses (log(Mass [M$_{\odot}$]) $\sim$ 9) and large SFRs (log(SFR [M$_{\odot}$ yr$^{-1}$]) $>$ 1.6) like MD5b, but both of these objects are redder ($E(B-V) \sim 0.4$).  If LyC emission is being produced copiously by the hot stars in these two galaxies, the additional dust extinction might be the reason we do not detect the LyC photons.  We note that MD34f is actually associated with a NB3420 detection (see Section \ref{sec:fcontam}), but its close proximity to a foreground contaminant makes it impossible to distinguish between emission from MD34f and the foreground contaminant in the seeing-limited NB3420 image.  The LAE sample presents several additional young galaxy components that are undetected in the NB3420 image.  These objects have reddening values similar to MD5b ($E(B-V) \sim 0.2$), lower than those of the young LBGs.  One possible explanation for the lack of LyC detections in the LAE sample is that MD5b is located along a fortuitously clear IGM sightline, and these LAEs are not.  Another possibility is that the LAEs are simply too faint to be detected in our LyC imaging.  The $V_{606}$ magnitude of MD5b is $m_{606} = 25.9$, and its LyC detection is near the edge of our detection limit.  The $V_{606}$ magnitudes of the young LAEs in our sample (ages $<$ 100 Myr) are much fainter on average, and range from $26.29 < m_{LyC} < 29.14$, with a median of 27.69.  If these objects have the same ratio of ionizing to non-ionizing radiation as MD5b, the LyC magnitudes of these objects would range from $27.80 < m_{606} < 30.64$, with a median of 29.20.  Such magnitudes are well below the detection limit of the NB3420 filter used for LyC imaging (27.3 mag), and thus these objects would not have been detected in the LyC.

\subsection{No Evidence for AGN}\label{sec:noagn}
\citet{vanzella15} discuss the possibility that some portion of the leaking ionizing radiation from the LyC-emitters \emph{Ion1} and \emph{Ion2} are due to contributions from low-luminosity AGNs within these galaxies.  We considered this possibility in MD5b, but find no evidence for a low-luminosity AGN.  There is no significant variability detected between the $V-$band magnitude of MD5 measured by LRIS in 2007 ($m_{V}=24.96 \pm 0.11$) and the combined $HST$ $V_{606}$ magnitudes of MD5a and MD5b measured in 2013 ($m_{606}=25.11 \pm 0.08$).  Additionally, the available spectra of MD5 do not show any high ionization emission line features (e.g., NV $\lambda1240$ emission), although our spectrum does not cover CIV $\lambda1550$ or HeII $\lambda1640$, nor are we able to examine the OIII / H$\beta$ ratio because H$\beta$ falls on a sky line.  We therefore find no evidence for AGN activity in MD5b with the information available.

\section{Discussion}\label{sec:disc}

In this section, we consider the broader implications of our single robust detection of LyC emission in the HS1549 field.  We discuss the actual rate of foreground contamination with respect to predictions from the contamination simulations of M13, and the LyC emission properties of MD5b, including its intrinsic ionizing to non-ionizing UV flux-density ratio and the implied LyC escape fraction.  With the removal of all foreground contaminants from the M13 sample, we then obtain a revised estimate for the ionizing emissivity due to star-forming galaxies at $z=2.85$.  Finally, we discuss the prospects for future direct searches for LyC radiation in high-redshift galaxies.

\subsection{Comparison to Previous Contamination Estimates}\label{sec:disc_contam}
We wish to address the question of whether or not the simulations of foreground contamination from M13 accurately predicted the number of contaminants in the sample.  The simulations \citep[described in detail in M13;][]{nestor13} employed the surface density of objects in the NB3420 image and the offset of each NB3420 detection to estimate the number of foreground contaminants and a contamination-corrected NB3420 magnitude.  The simulations predicted that $1.5 \pm 1.0$ out of 4 LBGs and $4.3 \pm 1.3$ out of 7 LAEs in the main sample were real LyC-emitters.  The prediction for the LBG sample has held out, as one LBG (MD5) remains a strong candidate for LyC emission.  The prediction for the LAE sample, however, was too high.  We were able to obtain $U_{336}V_{606}J_{125}H_{160}$ imaging for 6 out of 7 LAEs with LyC detections, yet none have proven to be true sources of LyC emission.  We note that 2 of the 7 LAEs had misidentified redshifts (\emph{lae2436} and \emph{lae7180}), and should not have made it into this sample in the first place.  Given the small number of galaxies with true LyC detections, contaminated NB3420 detections from even one or two objects with misidentified redshifts may introduce a non-negligible bias that is not taken into account in the contamination simulations of M13.  Accordingly, we re-ran the same contamination simulations, but only considering the 4 spectroscopically-confirmed LAEs from the main sample of M13 for which we acquired $U_{336}V_{606}J_{125}H_{160}$ imaging.  The revised simulations predict $2.1 \pm 1.0$ out of 4 galaxies to be real LyC-emitters.  As none of the four NB3420 detections proved to be real, the prediction is still too high by $\sim2\sigma$.  With the small sample size of 4 objects, however, such variations may be expected.  We conclude that the contamination simulations still serve as a useful, though blunt, tool for evaluating the likelihood of foreground contamination.

\subsection{Ionizing to Non-ionizing UV Flux-density Ratios}\label{sec:disc_FDR}
One of the intriguing findings reported by all three ground-based LyC studies of galaxy protoclusters \citep[][M13]{iwata09,nestor11} was the high apparent ratio of ionizing to non-ionizing radiation in many of the candidate LyC-emitters, which appeared to be in conflict with results from standard SPS models.  Several models have been proposed to address this question.  \citet{iwata09} proposed a top-heavy IMF.  \citet{nestor13} investigated the intrinsic non-ionizing to ionizing UV flux-density ratios for two sets of stellar population models at varying ages and metallicites.  These authors examined both \citet{bc03} models and BPASS models, which include a more detailed treatment of stellar binaries and Wolf-Rayet stars and result in bluer galaxy spectra.  In this work, we also consider SB99 models with improved treatment of stellar rotation \citep{leitherer14}, as rapidly rotating stars exhibit bluer spectra as well.  

However, our new observations, along with all previous follow-up work aimed at investigating contamination among candidate LyC-emitters \citep{vanzella12,siana15}, have ruled out all objects with high apparent ratios of ionizing to non-ionizing radiation as contaminants.  In our sample of 16 galaxies with putative LyC emission, 10 had anomalous flux-density ratios of $(F_{UV}/F_{LyC})_{obs}<2.0$.  However, the results of our analysis show that MD5 remains the only robust candidate for LyC emission.  The flux-density ratio of MD5, as calculated from the ground-based NB3420 and $V-$band imaging of M13, is $F_{UV}/F_{LyC}=5.9 \pm 2.0$.  Using our $HST$ $U_{336}$ and $V_{606}$ imaging\footnote{We use $U_{336}$ rather than NB3420 to represent flux in the LyC because the matched-PSF, matched-aperture photometry between the $U_{336}$ and $V_{606}$ images allows the most accurate calculation of $F_{UV}/F_{LyC}$.} to revise this flux-density ratio to only include MD5b, the component associated with the ionizing radiation, we obtain a value of $F_{UV}/F_{LyC}=4.0 \pm 2.0$.  At the redshift of MD5 ($z=3.14$), the maximum IGM transmission through the $U_{336}$ filter is $\sim60-70$\% \citep[depending on the absorber statistics used; see, e.g.,][]{rudie13,inoue14}, although IGM transmission varies greatly along the line of sight.  These maximum transmission values give rise to IGM-corrected flux-density ratios of $F_{UV}/F_{LyC}=2.4-2.8$, demonstrating that the flux-density ratio of MD5b is consistent with expectations from BPASS models between the ages of $\sim10-50$ Myr \citep[see Table 7 of][]{nestor13} in the lack of significant IGM absorption.

In similar work investigating contamination in candidate LyC-emitters, \citet{siana15} obtained spatially-resolved spectroscopy of the five LBG candidates for LyC emission presented in \citet{nestor11}.  While unable to confirm any candidates, their data showed that both galaxies in their sample with anomalously high apparent ratios of ionizing to non-ionizing radiation (MD32, aug96m16) are contaminated by lower-redshift objects.  Additionally, neither of the LyC-emitters \emph{Ion1} or \emph{Ion2} \citep{vanzella12,vanzella15} exhibit anomalously high ratios of ionizing to non-ionizing radiation.  While the two dozen LyC-emitter candidates in the literature with extensive follow-up data do not comprise the entire sample of high-redshift candidate LyC-emitters, the fact that all candidates with extreme $(F_{UV}/F_{LyC})_{obs}$ ratios have proven to be contaminants may indicate that the anomalously high ratios of ionizing to non-ionizing radiation originally inferred are simply a result of foreground contamination.

\begin{deluxetable*}{rcccc}
\scriptsize
\tablewidth{0pt} 
\tablecaption{Contributions to the Ionizing Background. \label{tab:emissivity}}
\tablehead{
\colhead{} & \colhead{LF\tablenotemark{a}} & \colhead{$F_{UV}/F_{LyC}$\tablenotemark{b}} & \colhead{Magnitude
range\tablenotemark{c}} & \colhead{$\epsilon_{LyC}$\tablenotemark{d}}
}
\startdata
\multicolumn{5}{c}{LyC Detections:  MD5 only\tablenotemark{e}} \\
\hline
(i)   & LBG & $140^{+70}_{-37}$  & $M_{AB} \leq -19.7$         & $0.8 \pm 0.3$ \\
(ii)  & LAE & $>14$ & $-19.7 < M_{AB} \leq -17.7$ & $<1.7$ \\
(iii) & LBG & $>14$ & $-19.7 < M_{AB} \leq -17.7$ & $<7.3$ \\
(iv) & LBG & $140^{+70}_{-37}$ & $M_{AB} \leq -17.7        $ & $1.5 \pm 0.6$ \\
 (v) & LAE & $>14$ & $M_{AB} \leq -17.7        $ & $<3.6$ \\
& \textbf{Total (lum.-dep.)}\tablenotemark{f} & \nodata & \boldmath{$M_{AB} \leq -17.7$}  & \boldmath{$0.8 \pm 3.7$} \\
& \textbf{Total (LAE-dep.)}\tablenotemark{g} & \nodata & \boldmath{$M_{AB} \leq -17.7$}  & \boldmath{$1.2 \pm 1.9$} \\
\hline
\multicolumn{5}{c}{LyC Detections:  MD5, D24, \emph{lae4680}\tablenotemark{e}} \\
\hline
(i)   & LBG & $74^{+24}_{-16}$  & $M_{AB} \leq -19.7$         & $1.5 \pm 0.4$ \\
(ii)  & LAE & $71^{+34}_{-18}$ & $-19.7 < M_{AB} \leq -17.7$ & $0.3 \pm 0.1$ \\
(iii) & LBG & $71^{+34}_{-18}$ & $-19.7 < M_{AB} \leq -17.7$ & $1.4 \pm 0.5$ \\
(iv) & LBG & $74^{+24}_{-16}$ & $M_{AB} \leq -17.7        $ & $2.9 \pm 0.8$ \\
 (v) & LAE & $71^{+34}_{-18}$ & $M_{AB} \leq -17.7        $ & $0.7 \pm 0.2$ \\
& \textbf{Total (lum.-dep.)}\tablenotemark{f} & \nodata & \boldmath{$M_{AB} \leq -17.7$}  & \boldmath{$3.0 \pm 0.9$} \\
& \textbf{Total (LAE-dep.)}\tablenotemark{g} & \nodata & \boldmath{$M_{AB} \leq -17.7$}  & \boldmath{$2.9 \pm 0.8$} 
\enddata
\tablenotetext{a}{\mbox{Luminosity} function parameters are identical to those in M13.}
\tablenotetext{b}{Sample average flux-density ratio corrected for IGM absorption, described in Section \ref{sec:emis}.}
\tablenotetext{c}{Magnitude range over which the first moment of the luminosity function is determined. $M_{AB} =-19.7$ and $-17.7$ correspond to $0.34L_{*}$ and $0.06L_{*}$, respectively.}
\tablenotetext{d}{Comoving specific emissivity of ionizing radiation in units of
$10^{24}$~ergs~s$^{-1}$~Hz$^{-1}$~Mpc$^{-3}$.}
\tablenotetext{e}{As we do not have the full suite of $U_{336}V_{606}J_{125}H_{160}$ imaging for D24 and \emph{lae4680}, we cannot determine whether their NB3420 detections are true LyC emission or foreground contamination.  Therefore, we perform two calculations of the emissivity in order to determine the full range of its possible values.  In the upper portion of this table we assume that both D24 and \emph{lae4680} are foreground contaminants, and that MD5 is the only true LyC detection.  In the bottom portion of the table, we assume that MD5, D24, and \emph{lae4680} are all true LyC-emitters.}
\tablenotetext{f}{Total for the luminosity-dependent model, determined by summing rows (i) and (iii).  Limits are taken into account using the method described in Section \ref{sec:emis}.}
\tablenotetext{g}{Total for the LAE-dependent model, determined by summing $0.77 \times$ row (iv) and row (v).  Limits are taken into account using the method described in Section \ref{sec:emis}.}
\end{deluxetable*}

\subsection{Escape Fraction for MD5}\label{sec:disc_fesc}
We can estimate the relative and absolute escape fractions of MD5b using the intrinsic luminosity-density ratio $(L_{UV}/L_{LyC})_{intr}$, observed flux-density ratio $(F_{UV}/F_{LyC})_{obs}$, and IGM transmission factor ($t_{IGM}$) with the following equations:

\begin{equation}
f^{LyC}_{esc,\,rel} = \frac{\left(L_{UV}/L_{LyC}\right)_{intr}}{\left(F_{UV}/F_{LyC}\right)_{obs} \times t_{IGM}}
\end{equation}

\begin{equation}
f_{esc,abs}^{LyC} = f^{LyC}_{esc,\,rel}  \times f_{esc}^{UV},
\end{equation}

\noindent 

The intrinsic luminosity-density ratio is highly uncertain, and varies significantly with the stellar population synthesis models used, as well as with the age and metallicity of the stellar population \citep[see Table 7 of][]{nestor13}.  Here, we bracket the full range of possible values for the LyC escape fraction of MD5b by using the lowest value of $(L_{UV}/L_{LyC})_{intr}=2.1$ (from BPASS models of age 10Myr with 20\% solar metallicity) and a maximum IGM transmission of 70\%. From these values, we obtain $f_{esc,rel}=75$\%, which must be interpreted as a lower limit such that the range of allowed relative escape fractions is $f_{esc,rel}^{MD5b}=75-100$\%.  We use the value $f_{esc}^{UV}=0.19$ to estimate the UV escape fraction at 1500\AA, which is calculated from the $E(B-V)$ of the best-fit FAST model to MD5b, assuming the \citet{calzetti00} attenuation curve.   We thus obtain an absolute escape fraction of $f_{esc,abs}^{MD5b}=14-19$\%.  As the $U_{336}$ flux has photometric errors of roughly fifty percent, the uncertainty in $f_{esc}$ is also at minimum fifty percent.

\subsection{Revised LyC Emissivity for Star-forming Galaxies at $z=2.85$}\label{sec:emis}

Here, we present a revised calculation of the emissivity of ionizing photons at $z=2.85$ based on the analysis of the $HST$ data in the HS1549 field.  We estimate the comoving specific emissivity as

\begin{equation}
\epsilon_{LyC} = \left(\frac{F_{UV}}{F_{LyC}}\right)^{-1}_{corr} \, \int^{L_{max}}_{L_{min}} L \, \Phi \; dL
\end{equation}

\noindent following the assumptions of M13 and \citet{nestor13}.  In this expression, $L$ is the non-ionizing UV luminosity, $\Phi$ is the non-ionizing UV luminosity function, and $(F_{UV}/F_{LyC})_{corr}$ is the average flux-density ratio of non-ionizing to ionizing UV radiation for the entire galaxy sample, corrected for the mean IGM attenuation in the LyC spectral region\footnote{To correct for absorption of LyC photons by neutral hydrogen in the IGM, we use the sample-averaged transmission values calculated in M13: $t_{LAE}=0.44\pm0.03$ and $t_{LBG}=0.35\pm0.04$.}.  We perform this emissivity calculation separately for the main sample of spectroscopically confirmed LBGs and LAEs from M13 (using the UV luminosity functions from \citet{reddy08} for LBGs and \citet{ouchi08} for LAEs), and combine the LBG and LAE emissivities to obtain a total emissivity for star-forming galaxies.  As in M13, we use the LRIS $V-$band to represent non-ionizing UV flux and NB3420 to represent LyC flux.  The difference between our calculation and that of M13 lies in our estimation of the average flux-density ratio.  Rather than estimating the average amount of foreground contamination from simulations, we instead know exactly which galaxies are contaminated based on the $HST$ data.  There were only two NB3420-detected galaxies in the M13 spectroscopic sample for which we were unable to acquire $U_{336}V_{606}J_{125}H_{160}$ imaging (D24 and \emph{lae4680}), and for these objects we could not evaluate whether or not the NB3420 detections are due to foreground contamination.  We thus calculate the emissivity twice in order to quote the full range of possible values: in one calculation we assume that MD5 is the only true LyC detection, and in the other calculation we assume that MD5, D24, and \emph{lae4680} are all true LyC-emitters.  In addition to using the $HST$ data to remove the NB3420 flux of foreground contaminants, we also use these measurements to estimate the percentage of contaminated flux in the non-ionizing UV.  All objects with foreground contaminants identified through the $HST$ imaging are blended in the LRIS $V$ imaging, and it is impossible to isolate the uncontaminated $z\sim2.85$ flux in the LRIS $V$ image.  Thus, for each contaminated object, we decrease its LRIS $V-$band flux to match the fraction of uncontaminated $V_{606}$ flux in the $HST$ imaging.  For objects that do not have $HST$ $U_{336}V_{606}J_{125}H_{160}$ imaging and are undetected in NB3420, we decrease their LRIS $V-$band flux to match the average fraction of uncontaminated $V_{606}$ flux in the full sample of $HST$-imaged galaxies without NB3420 detections (99\% for LBGs, 91\% for LAEs).  Finally, we use the same sample-averaged IGM correction to compute $(F_{UV}/F_{LyC})_{corr}$ as described in M13, employing statistics of H{\sc I} absorbers from \citet{rudie13}.  We note that the clustering of Lyman limit systems is not taken into account in these absorber statistics, and thus the true mean IGM transmission may be slightly higher than the values presented in M13 \citep[see, e.g.,][]{prochaska14}.

In order to estimate the total contribution of star-forming galaxies to the ionizing emissivity at $z\sim2.85$, we estimate the emissivity due to LBGs and LAEs separately and use two different models \citep[described in detail in][]{nestor13} to combine these values into a total emissivity of star-forming galaxies.  In the first model, which we refer to as the luminosity-dependent model, the $F_{UV}/F_{LyC}$ values for LAEs are assumed to represent those for star-forming galaxies with faint UV continuum magnitudes ($0.06 L^{*} < L < 0.34 L^{*} $, corresponding to $25.5<V<27.5$) while the $F_{UV}/F_{LyC}$ values for LBGs represent those for brighter star-forming galaxies ($L > 0.34 L^{*}$).  The second model, referred to as the LAE-dependent model, considers the case that LAEs are not simply faint LBG-analogs, but that LBGs and LAEs are actually distinct populations of star-forming galaxies with systematically different $F_{UV}/F_{LyC}$ values on average.  In this model, LAEs are assumed to comprise 23\% of the LBG population \citep{nestor13}, galaxies identified both as LBGs and LAEs are treated as LAEs, and the luminosity function is integrated over the full luminosity range ($0.06L^{*} < L < \infty$) for both LBGs and LAEs. 

In Table \ref{tab:emissivity}, we summarize the contributions to $\epsilon_{LyC}$ as determined from galaxies in the HS1549 field.  By considering MD5 as the only real LyC detection in the entire galaxy sample, we obtain values of the average UV flux-density ratio for LBGs and LAEs the HS1549 field to be $(F_{UV}/F_{LyC})_{corr}^{LBG} = 140^{+70}_{-37}$ and $(F_{UV}/F_{LyC})_{corr}^{LAE} > 14$.  If we include the NB3420 detections for D24 and \emph{lae4680} as well, these values become $(F_{UV}/F_{LyC})_{corr}^{LBG} = 74^{+24}_{-16}$ and $(F_{UV}/F_{LyC})_{corr}^{LAE} = 71^{+34}_{-18}$.  The uncertainties in the flux-density ratios are dominated by the NB3420 photometric errors, and are large because of significant uncertainty in the average NB3420 flux due to our very few NB3420 detections.  We estimated the uncertainties in the flux-density ratios using a Monte Carlo simulation.  The simulation calculates a distribution of average values for $(F_{UV}/F_{LyC})_{corr}$ based on random realizations of our photometric data and IGM correction within their Gaussian uncertainties.  As the resulting distribution is positively skewed, we quote error bars that bracket the inner $68\%$ of the distribution.  Based on the values for $(F_{UV}/F_{LyC})_{corr}$, we infer revised values of the comoving specific emissivity: considering NB3420 emission from MD5 only (or from MD5, D24, and \emph{lae4680}) we obtain $\epsilon_{LyC}= 0.8 \pm 3.7 \; (3.0 \pm 0.9) \times 10^{24}$ ergs s$^{-1}$ Hz$^{-1}$ Mpc$^{-3}$ for the luminosity-dependent model and $\epsilon_{LyC}= 1.2 \pm 1.9 \; (2.9 \pm 0.8) \times 10^{24}$ ergs s$^{-1}$ Hz$^{-1}$ Mpc$^{-3}$ for the LAE-dependent model. The uncertainties in $\epsilon_{LyC}$ reflect only uncertainties in $(F_{UV}/F_{LyC})_{corr}$, which dominate over uncertainties in the luminosity function.  Each total emissivity value represents the linear combination of the LBG and LAE contributions to the emissivity, which are characterized by distinct $(F_{UV}/F_{LyC})_{corr}$ values and integrals over the non-ionizing UV luminosity functions.  To determine the error bar for each contribution to the emissivity, we divided the relevant non-ionizing luminosity function integral by the corresponding random distribution of $(F_{UV}/F_{LyC})_{corr}$ values described above.  We then randomly drew linear combinations of emissivities (a simple sum for the luminosity dependent model, and a weighted linear combination for the LAE-dependent model; see Table 5, notes f and g) from the relevant distributions. To combine the LBG and LAE emissivities in cases where the value of the LAE emissivity is only an upper limit, we treated the LAE emissivity as a normally distributed random variable centered on zero, with a standard deviation equal to its 1$\sigma$ limit. The resulting distribution of emissivities was not skewed, so we report the standard deviation of the distribution as the uncertainty in the emissivity.

The revised values of $\epsilon_{LyC}$ are much lower than those computed in M13: $\epsilon_{LyC}=15.0 \pm 6.7 \times 10^{24}$ ergs s$^{-1}$ Hz$^{-1}$ Mpc$^{-3}$ for the luminosity-dependent model and $\epsilon_{LyC}=8.8 \pm 3.5 \times 10^{24}$ ergs s$^{-1}$ Hz$^{-1}$ Mpc$^{-3}$ for the LAE-dependent model.  The lower emissivity value calculated in the current work is much more compatible with the total ionizing emissivity at $z=2.85$, estimated in M13 from measurements of the \lya-forest opacity \citep{bolton07,fauchergiguere08} to be $\epsilon^{tot}_{LyC} \sim 5 - 10 \times 10^{24}$ ergs s$^{-1}$ Hz$^{-1}$ Mpc$^{-3}$.  As measurements of the contribution of QSOs to the ionizing background at $z=2.85$ range from $\epsilon_{LyC}^{QSO} \sim 1.5 \times 10^{24}$ ergs s$^{-1}$ Hz$^{-1}$ Mpc \citep{cowie09} to $\epsilon_{LyC}^{QSO} \sim 5.5 \times 10^{24}$ ergs s$^{-1}$ Hz$^{-1}$ Mpc \citep{hopkins07}, our data indicate (with large uncertainties) that star-forming galaxies provide roughly the same contribution as QSOs to the ionizing background at this redshift.

\subsection{The Future of LyC Surveys}

The results from this work suggest that identifying true LyC-emitters at high redshift requires an extremely large parent sample of galaxies and/or significantly deeper LyC observations.  With only one confirmed detection out of 49 LBGs, and zero confirmed detections among the 91 LAEs, the detection rate of LyC emitters at high redshift is very small.  While several interesting methods of indirectly identifying LyC-emitting galaxies have been proposed $-$ such as assessing the shape of the \lya\ emission line \citep{verhamme15}, observing reduced flux in nebular emission lines \citep{zackrisson13}, and observing residual flux in the cores of saturated low-ionization absorption lines \citep{jones13,heckman11} $-$ it will not be possible to verify the validity these indirect methods without first obtaining a sample of galaxies with robust detections of LyC emission.  

Given their low detection rate in the HS1549 field, it will likely be very difficult to amass a statistical sample of LyC emitters at $2<z<4$ without a dedicated survey.  As the process of identifying and verifying LyC emission in this redshift range involves several steps, we outline here what we consider to be the most efficient method for doing so.
  
First, there is the question of efficient targeting.  Because the average surface density of LBGs down to $R=25.5$ is roughly $1-2$ galaxies per square arcminute \citep{steidel99,steidel04}, the process of observing LBGs for LyC emission is greatly streamlined by observing galaxy protoclusters, which have an increased density of objects at a particular redshift.  Several galaxy protoclusters have already been identified in the literature at $2<z<4$ \citep[see, e.g.,][]{venemans07,kodama07,hatch11,cucciati14,lemaux14,shimakawa14,diener15}.  At the rate of one LyC detection per protocluster, observations of at least $\sim10$ protoclusters would be necessary to obtain a sample large enough to investigate systematic differences between LyC-leakers and non-leakers.  While the environment in the IGM surrounding protoclusters may not be typical of the universe as a whole, it is unlikely to affect the escape of ionizing photons through the ISM of LyC-emitting galaxies.  The factor that is less well-constrained in protocluster environments is the estimate of the sample-averaged IGM transmission, which enters into the calculation of the global ionizing emissivity.  However, galaxies in protoclusters can still be very useful for studying the multiwavelength properties of LyC emitters.  An additional potential problem with targeting protoclusters for LyC studies is that protoclusters may be composed of galaxies with older stellar populations on average \citep[as we found in this work; also, see][]{steidel05}.  If LyC emission is primarily emitted from galaxies with younger stellar populations (as suggested by the LyC detection for MD5b), then LyC emitters may be less common in protoclusters.

Next, there is the question of the required observations.  Spectroscopic redshifts must be measured for a large sample of protocluster galaxies, as photometric redshifts are not sufficiently precise to determine whether or not apparent LyC emission originates below 912\AA.  Multiwavelength $HST$ imaging should then be obtained to probe the SED shapes and LyC emission for all galaxy components near the high-redshift targets (as in this work).  To allow for direct LyC imaging, the protocluster identified must be at a redshift where one of the currently available $HST$ filters probes the LyC spectral region just bluewards of the Lyman limit (such as $F336W$ for $z>3.06$, or $F275W$ for $z>2.38$).  To obtain useful limits on the amount of escaping ionizing radiation from the faintest galaxies, LyC magnitudes must be probed several times fainter than their non-ionizing UV magnitudes.  The faintest galaxies in our sample, LAEs, have $V_{606}$ magnitudes ranging from $26.08<m_{606}<29.14$ with a median of 27.62.  In order to measure ionizing to non-ionizing flux-density ratios equivalent to that observed for MD5b ($F_{UV}/F_{LyC}\sim4$) for the faintest LAE ($m_{606}=29.14$), the required LyC observations must reach a depth of $\sim30.7$ magnitudes.  Assuming object sizes close to the PSF size, this requires imaging roughly 4 times more sensitive than the $U_{336}$ observations in this current work.  Larger objects, comparable in size to MD5b, would require imaging $\sim20$ times deeper.  In practice, the best way to measure the average $F_{UV}/F_{LyC}$ ratio for the faintest galaxies may be with stacked LyC observations, or with the next-generation UV space telescope \citep[e.g., ATLAST;][]{postman09}.  Finally, if the morphology of the candidates for LyC emission are complex, the last step would be to obtain high-resolution spectroscopic follow-up of the LyC-emitting component of the galaxy.

This plan is streamlined relative to the process we have followed thus far because it skips the time-intensive ground-based LyC imaging and analysis.  The main benefit of the ground-based LyC imaging was that we were able to design a custom, narrowband filter for the exact redshift of the protocluster.  Narrowband filters placed just bluewards of the Lyman limit are the least affected by IGM absorption, and probe LyC emission at wavelengths where the LyC photons are most likely to ionize hydrogen.  Even so, these benefits do not outweigh the cost in time and resources if $HST$ filters are already available at the correct wavelengths to probe LyC emission.  For cluster redshifts where $HST$ filters are not available for LyC imaging, but the Lyman limit falls above the atmospheric cut-off, it would be possible to obtain ground-based narrowband LyC imaging first at $z\sim3$ with, e.g., Keck/LRIS.  The roughly two dozen LyC-emitter candidates that would fall within a single Keck/LRIS pointing (most of which would be contaminants) could be followed up individually with an AO-assisted integral field spectrograph such as Keck/OSIRIS \citep{larkin06}.  In all future LyC searches, it is imperative to obtain high-resolution imaging and redshift confirmation of each galaxy sub-arcsecond component associated with apparent LyC emission in order to rule out foreground contamination.

\section{Summary}\label{sec:summary}

In M13, we identified 30 candidates for LyC emission via detection in the Keck/LRIS NB3420 filter: 5 LBGs and 7 LAEs spectroscopically confirmed at $z \geq 2.82$ (the main sample), 10 photometric LAE candidates, and 8 spectroscopically-confirmed LAEs not part of the main sample.  In this current work, we have presented follow-up $HST$ $U_{336}V_{606}J_{125}H_{160}$ observations of 16 of these objects: 4/5 LBGs and 6/7 LAEs in the main sample, 4/10 photometric LAE candidates, and 2/8 LAEs outside of the main sample.  

In our high-resolution $HST$ imaging, all of the candidates for LyC emission exhibit significant substructure.  We have thus used the $HST$ imaging to obtain photometric redshifts of each galaxy sub-arcsecond component in order to determine if the source of the NB3420 emission is truly at $z \geq 2.82$, or if it is from a lower-redshift contaminant.  Of the sixteen galaxies with NB3420 detections imaged in $U_{336}V_{606}J_{125}H_{160}$, nine were located near foreground contaminants responsible for the NB3420 emission.  Two objects had incorrect redshifts assigned to them, and thus the NB3420 emission was also from a low-redshift galaxy.  Four objects, all LAEs without spectroscopic confirmation, exhibited ambiguous SED shapes consistent with both $z\sim2.85$ galaxies and foreground contaminants.  Lack of spectroscopic redshifts for these objects, combined with their ambiguous SED shapes, makes it impossible to verify their NB3420 detections as true LyC emission.  In the end, only one robust candidate for LyC emission remained: the LBG, MD5.

MD5 has a spectroscopic redshift of $z=3.14$, measured from LRIS and MOSFIRE spectra containing \lya\ and [OIII] emission lines, along with multiple coincident interstellar absorption features, blueshifted with respect to the systematic redshift and indicative of an outflow ($\Delta v_{IS}\sim280$ km s$^{-1}$).  These spectra show no spurious emission or absorption features indicating a foreground contaminant.  In the $HST$ imaging, it becomes apparent that MD5 is composed of two clumps (MD5a and MD5b), which may either be two components of the same galaxy or two separate galaxies in the process of merging.  MD5b is associated with the NB3420 detection.  The best-fit stellar population synthesis model to MD5b indicates that while values for its stellar mass and reddening are typical of LBGs, it has a young stellar population ($\lesssim$50 Myr) and a high SFR (20 M$_{\odot}$ yr$^{-1}$) for such a low-mass object ($5\times 10^{8}$ M$_{\odot}$).  This age places MD5b in the youngest 10\% of the $HST$ sample, and in the youngest third of typical LBGs.  While MD5b is young compared to the full sample regardless of the fitting methods, we caution that the exact value for the best-fit age depends significantly on the dust attenuation curve and star-formation history assumed.

The observed non-ionizing to ionizing UV flux-density ratio of $(F_{UV}/F_{LyC})_{obs} = 4.0 \pm 2.0$ of MD5b is consistent with predictions of the intrinsic flux-density ratio for galaxies of 10$-$50 Myr from stellar population synthesis models \citep{nestor13}, in the absence of significant IGM absorption.  With the assumption of maximum IGM transmission at the redshift of MD5b ($t_{IGM}=70$\%), the observed flux-density ratio results in a relative escape fraction of $f_{esc,rel}^{MD5b}=75-100$\% and an absolute escape fraction of $f_{esc,abs}^{MD5b}=14-19$\%.  We also note that the emission in the $U_{336}$ filter at the location of MD5b, which probes LyC emission at $z=3.14$, shows no spatial offset from the $V_{606}$ emission, supporting the interpretation that the ionizing photons are escaping either isotropically, or through a hole in the ISM directly along our line of sight.

\begin{figure*}
\epsscale{1.0}
\plotone{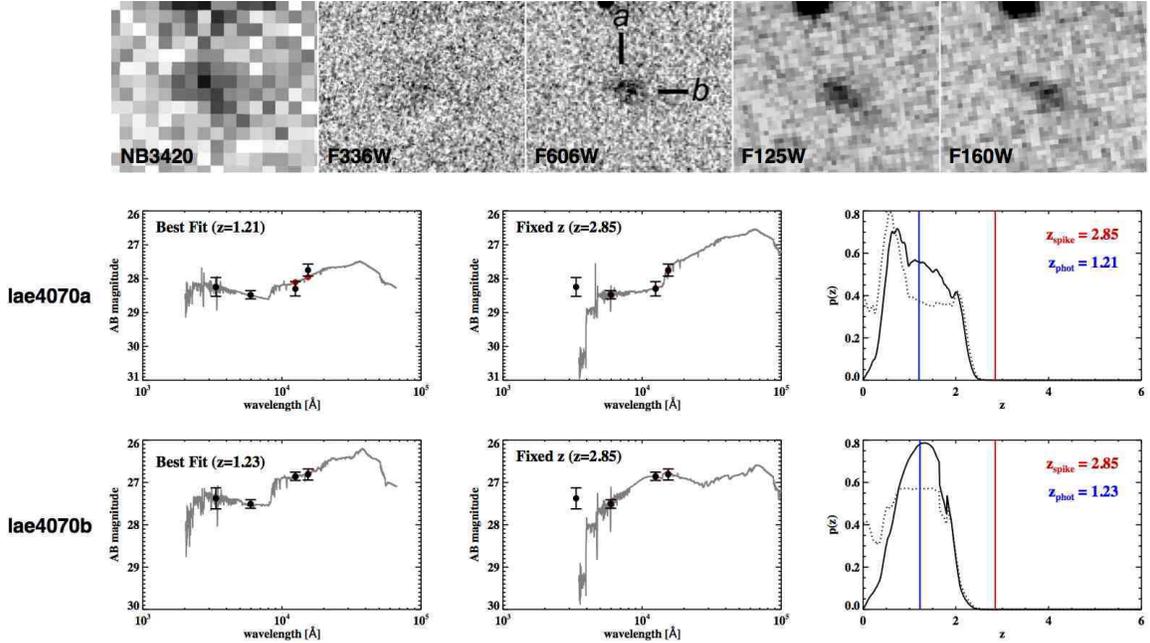}
\caption{
\small
3\secpoint3 $\times$ 2\secpoint8 postage stamp images of the photometric LAE candidate \emph{lae4070}.  From left to right, images displayed include LRIS NB3420 (LyC emission), $HST$ $U_{336}$ (a combination of LyC and non-ionizing UV), $HST$ $V_{606}$ (non-ionizing UV continuum), $HST$ $J_{125}$ (optical, bluewards of the Balmer break), and $HST$ $H_{160}$ (optical, redwards of the Balmer break).  In the high resolution $U_{336}$ and $V_{606}$ images, \emph{lae4070} appears to be composed of two components (a and b, separated by 0\secpoint24; labeled on the $V_{606}$ image).  The photometry and EAZY SED fits for these two components are plotted below the images.  In the left-most panel, the redshift is allowed to float during SED fitting, while in the middle panel the redshift is fixed to $z=2.85$, the redshift corresponding to that of the redshift spike in the HS1549 field.  The right-most panel shows the redshift probability distribution.  Colors and symbols are as in Figure \ref{fig:SED_samples}.  The main qualitative difference between the SEDs of \emph{lae4070a} and \emph{lae4070b} is the magnitude in $J_{125}$.  In the case of \emph{lae4070b}, the resulting SED has the typical shape of a contaminant.  However, in the case of \emph{lae4070a}, the resulting SED cannot be well-fit by model spectra at any redshift.  As discussed in Section \ref{sec:ambig}, it is thus possible that this is a case of a LyC-emitting LAE (\emph{lae4070a}) projected very close to a foreground contaminant (\emph{lae4070b}).
\label{fig:NB4070} }
\end{figure*}

MD5b stands out as one of the youngest galaxies in our $HST$-imaged sample.  The best-fit model to MD5b had the youngest age available (50 Myr), and the best-fit becomes even younger (10 Myr) without the minimum age requirement.  While there are an additional two LBGs and five LAEs in our sample with ages $<$100 Myr, none of these galaxies exhibit NB3420 detections.  The two young LBGs exhibit increased dust extinction ($E(B-V)\sim0.4$), which may be the reason we do not detect the LyC photons.  The young LAEs have less dust extinction ($E(B-V)\sim0.2$), but have very faint UV continuum magnitudes (median $m_{606} = 27.7$).  If the young LAEs have the same ratio of non-ionizing to ionizing radiation as MD5b, then their median LyC magnitude would be $m_{LyC} = 29.20$, well below the detection limit of the NB3420 filter used for LyC imaging (27.3 mag).

After eliminating foreground contaminants from our sample, we obtained a revised estimate for the comoving specific ionizing emissivity ($\epsilon_{LyC}$) at $z=2.85$.  We calculated the emissivity associated with LBGs and LAEs separately, and combined these values using two different models, described in Section \ref{sec:emis}. If we consider MD5 as the only galaxy with a LyC detection, we obtain $\epsilon_{LyC}= 0.8 \pm 3.7 \times 10^{24}$ ergs s$^{-1}$ Hz$^{-1}$ Mpc$^{-3}$ for the luminosity-dependent model and $\epsilon_{LyC}= 1.2 \pm 1.9 \times 10^{24}$ ergs s$^{-1}$ Hz$^{-1}$ Mpc$^{-3}$ for the LAE-dependent model.  If we also add in as true LyC detections the only two galaxies (D24 and \emph{lae4680}) for which we were unable to obtain sufficient $HST$ data to evaluate contamination, we obtain $\epsilon_{LyC}= 3.0 \pm 0.9 \times 10^{24}$ ergs s$^{-1}$ Hz$^{-1}$ Mpc$^{-3}$ for the luminosity-dependent model and $\epsilon_{LyC}= 2.9 \pm 0.8 \times 10^{24}$ ergs s$^{-1}$ Hz$^{-1}$ Mpc$^{-3}$ for the LAE-dependent model.  These revised values of $\epsilon_{LyC}$ are much lower than those computed in M13, and much more compatible with the total ionizing emissivity at $z=2.85$ ($\epsilon_{LyC}=5.6 \pm 1.6 \times 10^{24}$ ergs s$^{-1}$ Hz$^{-1}$ Mpc$^{-3}$; M13).  Within the large photometric uncertainties, and uncertainties due to the small dynamic range in which we can probe LyC emission, our data indicate that star-forming galaxies provide roughly the same contribution as QSOs to the ionizing background at this redshift.

Overall, the rate of foreground contamination for apparent LyC leakers in our $z\sim2.85$ sample was very high.  While the single detection in the LBG sample is consistent with contamination expectations from M13, the contamination rate in the LAE sample was higher than predicted by our simulations.  With this work we have shown that ground-based LyC imaging studies are insufficient for obtaining a full understanding of LyC emission from star-forming galaxies because they are so heavily contaminated by foreground objects.  In order to eliminate cases of foreground contamination, it is essential to obtain high-resolution observations of putative LyC-emitters to confirm the redshifts (either photometrically or spectroscopically) of all substructure associated with the galaxy.  To date, all such observations have shown that candidate LyC-emitters from ground-based studies with anomalously high ratios of ionizing to non-ionizing radiation, both within our sample and in the literature, have proven to be from foreground contaminants.  

Future progress in understanding the physical properties of LyC-emitters and the role of star-forming galaxies in reionization is contingent upon two factors.  First, observations of sufficient depth to probe ionizing radiation in galaxies at the faint end of the luminosity function must be obtained efficiently for a large sample of galaxies.  Second, these observations must be obtained at high spatial resolution, and with redshift information for each galaxy component.  With such observations, it will be possible to distinguish between emission from foreground contaminants and genuine high-redshift LyC emitters, learn more about LyC photon escape by studying the multiwavelength properties of LyC emitters, and place more stringent constraints on the contribution of star-forming galaxies to the ionizing background.

\medskip

We thank Anahita Alavi and Eros Vanzella for helpful discussions about the $HST$ imaging reduction and photometry.  Support for program GO-12959 was provided by NASA through a grant from the Space Telescope Science Institute (STScI), which is operated by the Association of Universities for Research in Astronomy, Inc., under NASA contract NAS 5-26555.  A.E.S. acknowledges additional support from the David and Lucile Packard and Sloan Foundations, C.C.S. acknowledges support from the NSF grants AST-0908805 and AST-1313472, as well as STScI grants GO-11638.01 and GO-11694.02, and R.E.M. and A.E.S. acknowledge the generous support of Mr. Richard Kaplan.  R.F.T. receives funding from the Miller Institute for Basic Research in Science at U.C. Berkeley.  N.A.R. acknowledges support from the Sloan Foundation.  We wish to extend special thanks to those of Hawaiian ancestry on whose sacred mountain we are privileged to be guests. Without their generous hospitality, most of the observations presented herein would not have been possible.

\begin{appendix}
\section{Objects with Ambiguous SEDs}

Here we discuss the interpretation of four SEDs of photometric LAE candidates with NB3420 detections where the SED shape is ambiguous, and there are no spectroscopic redshifts available to confirm that the objects are indeed at $z\sim2.85$.  In total, we present three photometric LAEs from the Appendix of M13 (\emph{lae4070}, \emph{lae5200}, and \emph{lae6510}) and one photometric LAE that was originally in the spectroscopic LAE sample, but for which the $HST$ data showed that the wrong redshift had been assigned (\emph{lae7180}).  For \emph{lae5200}, \emph{lae6510}, and \emph{lae7180}, Keck/LRIS spectroscopy with a total exposure time of 5400 seconds was attempted on clear nights with 0\secpoint5$-$0\secpoint6 seeing, but no redshifts were measured.  For \emph{lae4070}, Keck/LRIS spectroscopy with a total exposure time of 17400 seconds was attempted under suboptimal conditions (intermittent clouds with seeing of 0\secpoint7$-$1\secpoint0 during clear spells), but again, no redshift could be measured.  The lack of spectroscopic redshifts for these LAE candidates makes it impossible to confidently claim a LyC detection for any of them.  Furthermore, the shapes of their SEDs are ambiguous.  Thus, while we can't unequivocally confirm any of these candidates as LyC-emitters, we also cannot rule them out as foreground contaminants.

The most promising LAE photometric candidate is \emph{lae4070}.  LAEs were selected by their $V-$NB4670 colors, as described in M13, and \emph{lae4070} has a $V-$NB4670 color of 0.82, slightly below the median value of the LAE sample and 0.22 mag above the selection threshold of $V-$NB4670$>$0.6.  While \emph{lae4070} is fairly compact, it is still composed of two clumps separated by 0\secpoint24 (\emph{lae4070a} and \emph{lae4070b}, indicated in Figure \ref{fig:NB4070}) for which we analyzed separate SEDs.  Both of these clumps are associated with NB3420 emission, and both clumps are detected individually in $U_{336}$.  \emph{Lae4070b} has the typical SED shape of a foreground contaminant.  \emph{Lae4070a}, however, has an SED shape that could not be well described by any of the stellar population models we used with EAZY.  The feature that most strongly indicates a redshift of $z\sim2.85$ for \emph{lae4070a} is the large break between $J_{125}$ and $H_{160}$, combined with a flat $V_{606} - J_{125}$ color.  In Figure \ref{fig:colorcolor}, \emph{lae4070a} has colors that place it in the same region of color-color space as typical LAEs and LBGs at $z\sim2.85$ ($J_{125}-H_{160}=0.55$, $V_{606}-J_{125}=0.18$).  The emission in the $U_{336}$ filter is anomalously high compared to any model that provides a good fit $V_{606}$, $J_{125}$, and $H_{160}$.  While some of this emission may be contamination from nearby \emph{lae4070b}, there is definitely emission in the $U_{336}$ filter at the location of \emph{lae4070a}, which was not the case for any of the LAEs or LBGs without NB3420 detections.  If \emph{lae4070a} is truly at $z\sim2.85$, then it is a LyC emitter.  However, in this case, measuring the ratio of ionizing to non-ionizing flux for \emph{lae4070a} in our NB3420 image is impossible.  The foreground galaxy \emph{lae4070b} is located so close along the line of sight to \emph{lae4070a} that there is no way to distinguish the NB3420 fluxes of these two objects with the 0\secpoint7 seeing in the LRIS NB3420 image.  Finally, we note that the foreground contaminant \emph{lae2436a} (which spectroscopy proved to be at $z=2.04$) has an SED similar in shape to that of \emph{lae4070a} in $V_{606}$, $J_{125}$, and $H_{160}$, and this object lies near \emph{lae4070a} in Figure \ref{fig:colorcolor} ($J_{125}-H_{160}=0.60$, $V_{606}-J_{125}=0.06$).  The fact that degeneracies still exist in this area of $J_{125}-H_{160}$ and $V_{606}-J_{125}$ color-color space where the majority of $z \sim 2.85$ LAEs and LBGs lie demonstrates the need for spectroscopic redshifts to resolve cases with ambiguous SEDs.

\begin{figure}
\epsscale{0.6}
\plotone{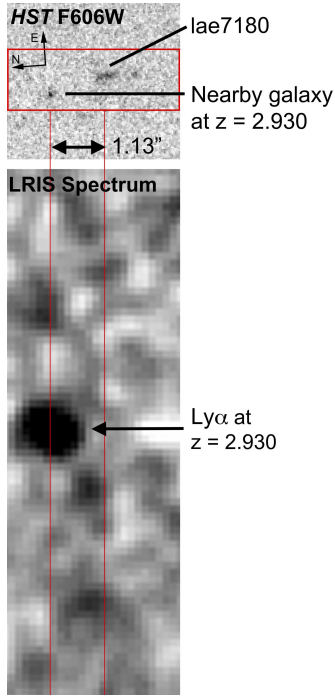}
\caption{
\small
$HST$ $V_{606}$ imaging of \emph{lae7180}, along with the corresponding Keck/LRIS spectrum.  Thick red lines in the $HST$ $V_{606}$ image indicate the location of the 1\secpoint2 slit, and the LRIS 2D spectrum is aligned to match in the orientation and spatial scale of the imaging.  As explained in Section \ref{sec:ambig}, the $z=2.93$ galaxy near to \emph{lae7180} was not visible in the lower resolution LRIS imaging, and thus the bright \lya\ emission line in the spectrum was originally assigned to NB7180.  However, analysis of the spatial distribution of the emission along the LRIS slit and how it corresponds to the $HST$ imaging, along with the resulting SED fits of the sub-arcsecond components near \emph{lae7180}, indicates that the neighboring galaxy is associated with the emission line and no emission line is visible at the location of \emph{lae7180} (which is now only a photometric LAE candidate).
\label{fig:NB7180} }
\end{figure}

We now consider the object \emph{lae7180}, an LAE originally assigned a spectroscopic redshift of $z=2.930$ based on a Keck/LRIS spectrum with a single high signal-to-noise emission line.  The narrowband NB4670 filter used for LAE selection only probes redshifts of $2.80 < z < 2.88$, so the redshift of $z=2.930$ was anomalously high for a NB4670-selected LAE candidate.  However, we considered the possibility that a higher-redshift object with very large \lya\ equivalent width scattered into the LAE sample, and therefore retained \emph{lae7180} for analysis in M13.  The $HST$ imaging for this object, however, indicates that there is another faint galaxy in the vicinity of the $z=2.930$ emission line, in addition to \emph{lae7180}.  Closer examination of the LRIS spectrum with respect to the $HST$ images (see Figure \ref{fig:NB7180}) suggests that \emph{lae7180} is offset by 1\secpoint13 from the location of the $z=2.930$ emission.  With the $HST$ imaging we were able to identify the true galaxy associated with the emission line, an object so faint ($m_{606}=28.85$) that it was undetected in the original LRIS imaging.  The SED fit for this new object matches the redshift $z=2.930$ identified in the spectrum, and we conclude that the emission line belongs to this faint, nearby object and not to \emph{lae7180}.  Now that the $z=2.930$ redshift is no longer associated with \emph{lae7180}, this object can be reevaluated as a photometric LAE candidate with a NB3420 detection.

\emph{Lae7180}, \emph{lae5200}, and \emph{lae6510}, all exhibit ambiguous SED shapes.  Rather than having SED shapes like that of \emph{lae4070}, these objects have SED shapes similar to the one described in Section \ref{sec:sed}, which may represent either young, dusty, high-redshift galaxies or foreground contaminants.  As shown in Figure \ref{fig:SED_ambig} for a spectroscopically confirmed galaxy with an ambiguous SED shape, EAZY gives a wide range of possible values for the redshift using both the P\'{E}GASE and SMC-reddened BPASS models.  However, as Figure \ref{fig:SED_ambig} also shows, the resulting redshift probability distributions may differ when using different sets of models.  Thus, without spectroscopic redshifts, it remains unclear for these LAE photometric candidates with ambiguous SED shapes whether they are low-redshift foreground galaxies with an old stellar population, or high-redshift LAEs with young, dusty stellar populations and LyC detections.  While we cannot absolutely confirm if they are LyC emitters without spectroscopic redshifts, the high-resolution $HST$ imaging and SED fits can help narrow down the possible interpretations for these objects.  

\begin{figure*}
\epsscale{1.0}
\plotone{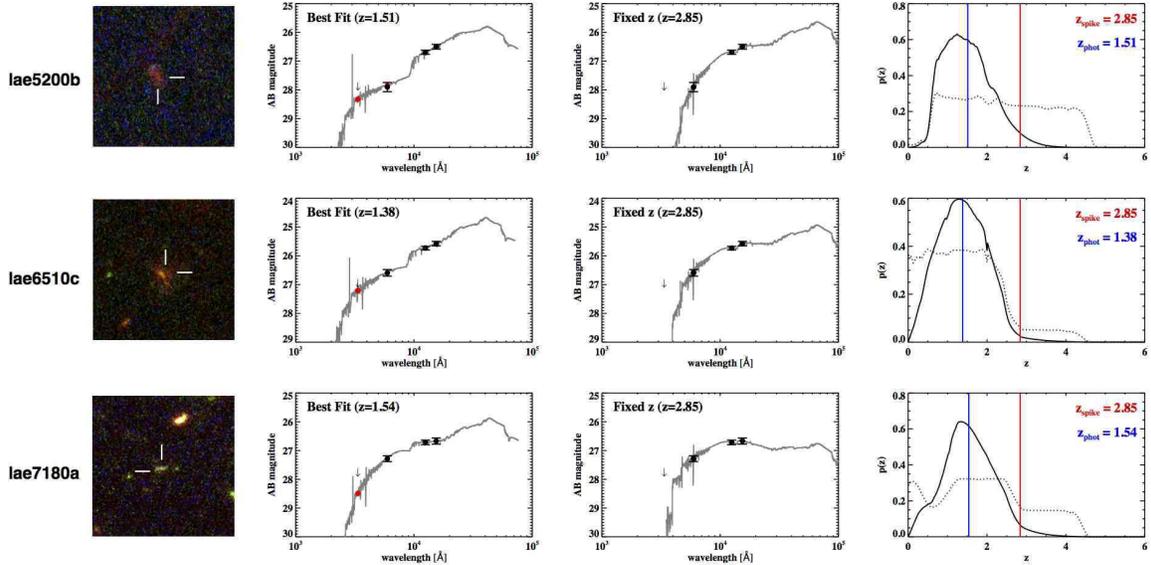}
\caption{
\small
$HST$ $U_{336}V_{606}J_{125}$ color-composite imaging ($5''\times5''$) and SED fits for the three photometric LAE candidates with the ambiguous SED shape described in Section \ref{sec:sed}.  In all cases, the sub-arcsecond component shown is the main component associated with the LAE, and its position is indicated in the imaging.  Results from EAZY are plotted in the three right hand panels.  In the left-most panel, the redshift is allowed to float during SED fitting, while in the middle panel the redshift is fixed to $z=2.85$, the redshift of the spike in the HS1549 field.  The right-most panel shows the redshift probability distribution.  Colors and symbols are as in Figure \ref{fig:SED_samples}.  Fits to these objects using the P\'{E}GASE templates yield slightly different redshift probability distributions, but they are qualitatively similar in that they span a wide redshift range ($0<z<4.5$) for all objects.
\label{fig:amb_lae} }
\end{figure*}

Figure \ref{fig:amb_lae} shows the $HST$ imaging and photometry for the sub-arcsecond component of each LAE that is associated with the NB3420 emission, along with the EAZY redshift probability distributions for these components.  For each component, EAZY predicts a wide range of possible redshifts ($0 < z_{phot} < 4.5$).  Figure \ref{fig:amb_lae} shows BPASS SED fits to the photometry both at low and high redshift, and the fact that both redshifts can fit the data well demonstrates again the difficulties in distinguishing between low and high redshifts for galaxies with these SED shapes.  In all of these cases, the detection of LyC emission depends on whether or not the LAE photometric candidate is truly at $z \sim 2.85$ - something that we cannot confirm for galaxies with ambiguous SED shapes and without spectroscopic redshifts.

Here we consider possible causes for the scenario in which all three of these photometric LAE candidates (\emph{lae5200}, \emph{lae6510}, and \emph{lae7180}) were incorrectly identified as LAEs.  \emph{Lae6510} and \emph{lae7180} have fairly marginal $V-$NB4670 colors (0.68 and 0.63, respectively) when compared to the LAE selection threshold of $V-$NB4670$>$0.6.  These values are in the lowest 15\% of the $V-$NB4670 colors for spectroscopically confirmed LAEs in M13, and indicate that these objects are among the weaker LAE candidates and may have scattered into the LAE sample through photometric errors.  As for \emph{lae5200}, its $V-$NB4670 color is 3.21, an anomalously high value that may be due to contaminating light from a nearby bright star.  Finally, these photometric LAE candidates were chosen for follow-up because of their NB3420 emission.  As our new dataset reveals, true LyC detections in $z\sim3$ galaxies are rare.  Therefore, while many of the LAE photometric candidates from M13 may be true LAEs, singling out objects from the photometric sample with NB3420 detections may result in a higher-than-average selection of foreground contaminants.  

In summary, these four photometric LAE candidates with NB3420 detections all have ambiguous SED shapes that make it difficult to verify their redshifts photometrically and confirm their possible LyC detections.  One object, \emph{lae4070}, has similar $J_{125}-H_{160}$ and $V_{606}-J_{125}$ colors to many $z\sim2.85$ galaxies in our sample and is the most promising photometric LAE candidate for true LyC emission.  The other three objects (\emph{lae5200}, \emph{lae6510}, and \emph{lae7180}) display the ambiguous SED shape described in Section \ref{sec:sed}, which may describe galaxies at many redshifts.  As we cannot unambiguously determine whether or not the four photometric LAE candidates discussed in this section are truly at $z\sim2.85$, we adopt a conservative approach and do not count the NB3420 detections for these objects as secure signatures of leaking LyC radiation.

\section{Objects Without Full $HST$ Coverage}
\begin{figure*}
\epsscale{0.8}
\plotone{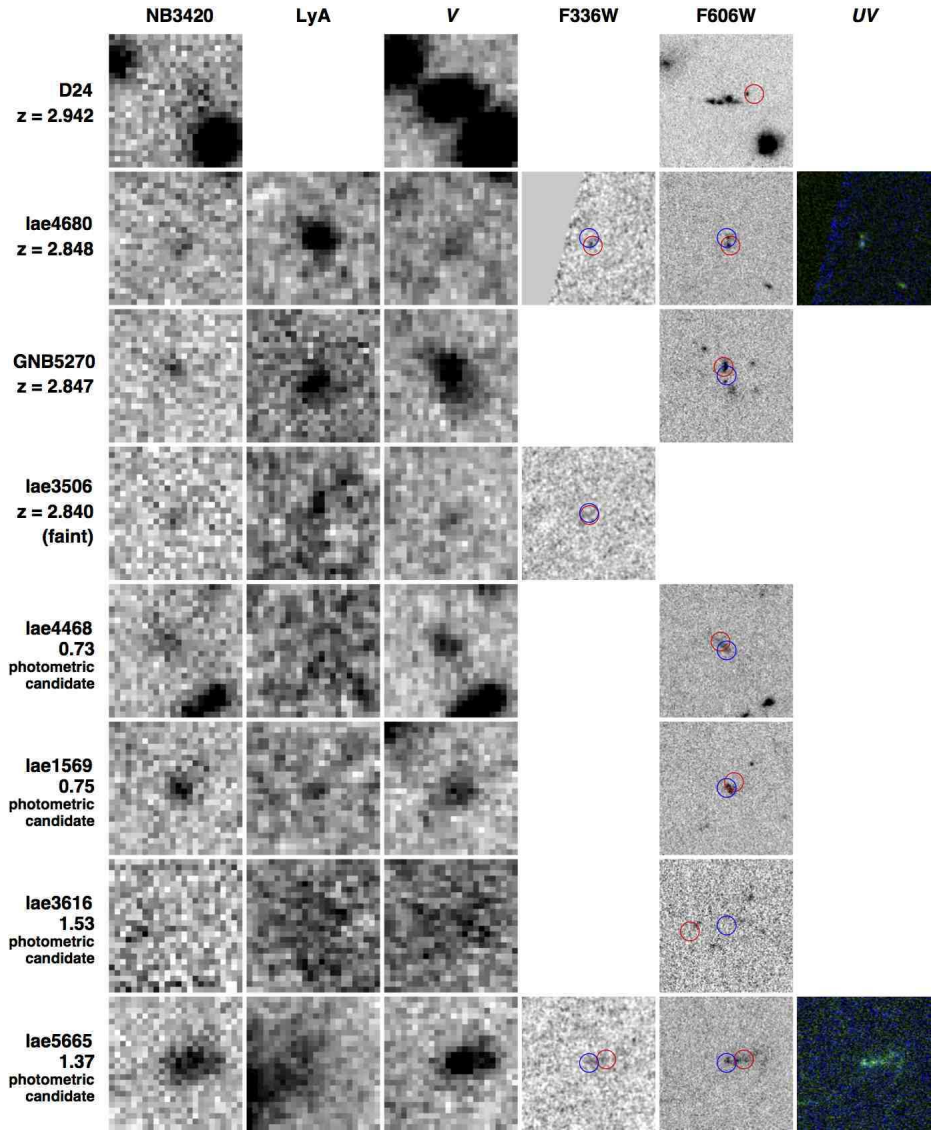}
\caption{
\small
$5''\times5''$ postage stamp images of galaxies with NB3420 detections, but for which imaging was not available in all four $HST$ filters.  From left to right, objects are displayed in LRIS NB3420 (LyC emission), LRIS NB4670$-V$ (indicating \lya\ emission), LRIS $V$ (non-ionizing UV continuum), and (when available) $HST$ $U_{336}$ (a combination of $\sim80$\% LyC and $\sim20$\% non-ionizing UV) and $HST$ $V_{606}$ (non-ionizing UV continuum).  For objects with imaging in both $HST$ $U_{336}$ and $V_{606}$, we show a color-composite image of these two bands.  Red (blue) circles (1\secpoint0 diameter) indicate the centroid of the NB3420 emission (\lya\ emission).  The redshift of each object is indicated below the object name, or, if the object is a photometric LAE candidate, the $V-$NB4670 color is indicated.  Postage stamps follow the conventional orientation, with north up and east to the left.  As there is insufficient photometric data to fit photometric redshifts for these objects and as none of these objects have simple, compact morphology that would lessen the chance of foreground contamination, we cannot draw conclusions about the contamination rate of this sample of objects.
\label{fig:stamps_appendix} }
\end{figure*}

Here we present postage stamp images of galaxies with NB3420 detections, but for which imaging was not available in all four $HST$ filters (see Figure \ref{fig:stamps_appendix}).  While insufficient photometric data prevents us from fitting SEDs and determining photometric redshifts, we attempted to examine the morphologies of these objects in the $V_{606}$ image, when available, to find objects with simple morphology where the possibility of contamination is low.  However, none of the objects shown in Figure \ref{fig:stamps_appendix} have simple, compact morphologies in $V_{606}$.  All objects either break into individual sub-arcsecond components or show extended diffuse emission.  As both clumpy $z\sim3$ galaxies and foreground contaminants may be responsible for these multi-component $V_{606}$ morphologies, we are unable to draw any conclusions about the amount of contamination in this galaxy sample.

\end{appendix}

\end{document}